\documentstyle[aps,epsf,floats,twocolumn,prb]{revtex}
\begin{document}

\twocolumn[\hsize\textwidth\columnwidth\hsize\csname@twocolumnfalse%
\endcsname
\draft

\title{Dynamics and transport in random quantum systems governed by
strong-randomness fixed points}
\author{Olexei Motrunich,$^{1}$ Kedar Damle,$^{1,2}$ and 
David A. Huse$^1$}
\address{$^1$ Department of Physics, Princeton University, 
Princeton, NJ 08544\\
$^2$ Department of Electrical Engineering, Princeton University, 
Princeton, NJ 08544}
\date{June 1, 2000}
\maketitle

\begin{abstract}
{We present results on the low-frequency dynamical and transport 
properties of random quantum systems whose
low temperature ($T$), low-energy behavior is controlled 
by strong disorder fixed points.
We obtain the momentum and frequency dependent dynamic structure 
factor in the Random Singlet (RS) phases of both spin-1/2 and 
spin-1 random antiferromagnetic chains, as well as in the
Random Dimer (RD) and Ising Antiferromagnetic (IAF)
phases of spin-1/2 random antiferromagnetic 
chains.  We show that the RS phases are unusual `spin metals' 
with divergent low-frequency spin conductivity at $T=0$, and 
we also follow the conductivity through novel `metal-insulator' 
transitions tuned by the strength of dimerization or Ising 
anisotropy in the spin-1/2 case, and by the strength of disorder 
in the spin-1 case.  We work out the average spin and energy 
autocorrelations in the one-dimensional random transverse field 
Ising model in the vicinity of its quantum critical point.
All of the above calculations are valid in the frequency dominated 
regime $\omega \agt T$, and rely on previously available
renormalization group schemes that describe these systems 
in terms of the properties of certain strong-disorder
fixed point theories. In addition, we obtain some
information about the behavior of the dynamic structure
factor and dynamical conductivity in the opposite `hydrodynamic'
regime $\omega < T$ for the special case of spin-1/2 chains
close to the planar limit (the quantum x-y model) by analyzing
the corresponding quantities in an equivalent model
of spinless fermions with weak repulsive interactions
and particle-hole symmetric disorder.}
\end{abstract}

\vskip 0.3 truein
]

%%%%%%%%%%%%%%%%%%%%%%%%%%%%%%%%%%%%%%%%%%%%%%%%%%%%%%%%%%%%%%%%%%
%%%%%%%%%%%%%%%%%%%%%%%% INTRO %%%%%%%%%%%%%%%%%%%%%%%%%%%%%%%%%%%
\section{Introduction}
\label{intro} 
Disorder effects arising from quenched randomness are at the
heart of many interesting and novel phenomena observed in
condensed matter systems---examples include Griffiths 
singularities near phase transitions in disordered magnets 
(and the related phenomenon of local-moment formation 
in disordered electronic systems~\cite{ssroysoc}), 
metal-insulator transitions in disordered electronic 
systems,~\cite{revbelkir} and two-dimensional phenomena 
such as weak localization and the quantum Hall plateau 
transitions.~\cite{dassarma}

In particular, the interplay between disorder and quantum 
interference leads to unusual dynamics and transport 
in these systems.  Such effects are well understood 
for disordered quantum systems in which many-body 
correlations are not significant (such as disordered 
Fermi liquids).  In contrast, relatively little is reliably known
about the effects of strong disorder in the presence of strong
correlations (say, due to electron-electron interactions in 
an itinerant electronic system, or due to exchange interactions 
in a system with localized spin degrees of freedom).

However, there does exist one class of systems where 
theoretical tools are available to analyze this interplay 
between strong disorder, correlations and quantum fluctuations; 
important examples include one-dimensional random 
antiferromagnetic spin chains~\cite{fraf,hy} and random quantum 
Ising models in one and two dimensions~\cite{fising,RTFIM2d}.  
In these quantum systems, it is possible to systematically treat 
disorder and correlation using a strong-disorder renormalization 
group (RG) technique that is designed to be accurate when 
the strength of the disorder, as measured by the widths of 
the distributions of the various couplings, is large. 
Such a strong-disorder approach works in these problems
because these systems, 
when studied at ever larger length scales (and correspondingly 
lower energy scales), appear more and more disordered.  
More precisely, the low-energy effective theory obtained from 
the RG has the remarkable property that the widths of the 
distributions of the various couplings in the theory 
{\em grow rapidly} as the energy cutoff is lowered; this means 
that the RG procedure gives reliable results for the effective 
Hamiltonian that governs the low-energy properties of the system. 
Moreover, the extremely strong disorder present at low energies 
in the effective theory actually allows one to straightforwardly 
calculate some thermodynamic properties and ground state 
correlators within the effective theory---this is, in essence, 
because strong disorder implies that some particular terms in
the effective Hamiltonian dominate over all others;
calculations can then be performed by treating these terms 
first and including the effects of the other terms perturbatively.
This approach has been used successfully in the past to obtain 
a wealth of information about the low-temperature thermodynamics 
and ground state correlators in such 
systems.~\cite{fising,fraf,mgj,hybg,RTFIM2d}

Here, we exploit this simplicity that emerges at strong disorder 
to obtain the first analytical results on the low-frequency 
dynamics and transport in these systems at low temperature $T$.  
Most of our results are obtained for $T=0$; these are expected 
to be exact at zero temperature in the low-frequency limit, and 
to remain valid at non-zero temperatures for low frequencies 
$\omega \agt T$.  Moreover, in certain special cases,
we can also access the regime $\omega < T$.

In the remainder of this section, we introduce the various systems 
that are studied in this paper, and describe the organization of 
the rest of the paper.  A brief summary of some of our results 
has already appeared elsewhere.~\cite{prlus}

Our focus is on three model systems.  The first model we consider
is the one-dimensional random antiferromagnetic XXZ spin-1/2
chain with the Hamiltonian
\begin{equation}
{\cal H}_{\rm XXZ} = \sum_j 
\left[ J^\perp_j (s^x_j s^x_{j+1} + s^y_j s^y_{j+1}) + 
       J^z_j s^z_j s^z_{j+1} \right] \, ,
\label{HXXZ}
\end{equation}
where $\vec{s}_j$ are spin-1/2 operators at lattice sites $j$ 
separated by spacing $a$, and both $J^\perp_j$ and $J^z_j$ 
are random positive exchange energies drawn from some probability 
distributions.  Such a Hamiltonian describes the low-energy 
(magnetic) dynamics of insulating antiferromagnetic spin-1/2 
chain compounds~\cite{rafm1,rafm2} with chemical disorder 
that affects the bond strengths.  We will also consider
chains with slightly different probability distributions 
of the even and the odd bonds and study the effects of such 
{\it enforced dimerization}.  The strength of the dimerization 
in the bonds is conveniently characterized by a dimensionless 
parameter $\delta$ defined as
\begin{equation}
\delta=\frac{ \overline{\ln J_{\rm e}} - \overline{\ln J_{\rm o}} }
       { {\rm var}(\ln J_{\rm e}) + {\rm var}(\ln J_{\rm o}) }\, ,
\end{equation}
where $J_{\rm e}$ ($J_{\rm o}$) represents even (odd) bonds, and
the overline and ``var'' denote correspondingly the average 
and variance over the distribution of bonds.  Thus, we have 
$\delta > 0$ ($\delta < 0$) if even (odd) bonds are stronger
on average.  For future reference, we also introduce the basic 
length scale in this system,
\begin{equation}
l_v=\frac{2a}{ {\rm var}(\ln J_{\rm e}) + {\rm var}(\ln J_{\rm o}) }
\, .
\label{lv}
\end{equation}

Detailed information about the spin dynamics in such systems 
can be obtained by inelastic neutron scattering (INS) experiments
that directly probe the frequency and momentum dependent
dynamic structure factor $S^{\alpha \beta}(k,\omega)$.  At $T=0$, 
$S^{\alpha \beta}(k,\omega)$ has the spectral representation
\begin{equation}
S^{\alpha \beta}(k,\omega) =
\frac{1}{L}\sum_m \langle 0| \hat s^\alpha_{-k} |m \rangle
                  \langle m| \hat s^\beta_k |0\rangle
\delta(\omega - E_m) \, ,
\label{struc}
\end{equation}
where $\hat s^\alpha_k = \sum_j e^{ikx_j} s^\alpha_j$,
and $\{ |m\rangle \}$ denote the exact eigenstates of the system 
with excitation energies $E_m$ relative to the ground 
state $|0\rangle$.  
The symmetry of ${\cal H}_{\rm XXZ}$ under rotations about the $z$ 
axis implies that we can restrict our attention to two independent 
components $S^{zz}$ and $S^{+-}$.  The same symmetry also implies 
that the total $s^z_{\rm tot}=\sum_j s^z_j$ is conserved---it then 
makes sense to talk of the spin transport in such a system.  
We characterize the transport of $s^z$ in terms of the dynamical 
spin conductivity $\sigma(\omega)$.  The real part 
$\sigma'(\omega)$ of $\sigma(\omega)$ is defined by the relation 
$P(\omega) = \sigma'(\omega)|\nabla H|^2(\omega)$, 
where $P(\omega)$ is the power absorbed per unit volume by 
the system when magnetic field with a uniform gradient 
$\nabla H(\omega)$ (with the field $H$ always in the $z$ direction) 
oscillating at frequency $\omega$ is applied along the length of 
the chain.  From standard linear response theory, we have the 
following Kubo formula for $\sigma'(\omega)$ at $T=0$:
\begin{equation}
\sigma'(\omega) = \frac{1}{\omega L} \sum_m 
|\langle m| \sum_{j=1}^L \tau_j |0 \rangle|^2 
\delta(\omega - E_m) \, .
\label{sigmakubo}
\end{equation}
In the above, 
$\tau_j = iJ^\perp_j(s^+_j s^-_{j+1} - s^+_{j+1} s^-_j)/2$
is the current operator on link $j$ that transfers one unit of
the $s^z$ from one site to the next. 
Here and everywhere in the following, the frequency $\omega$ 
is taken positive for notational convenience.
Note that both $S^{\alpha \beta}(k,\omega)$ and $\sigma'(\omega)$ 
as defined here are self-averaging in the thermodynamic limit.

The second model we consider is the random antiferromagnetic 
Heisenberg spin-1 chain with the Hamiltonian
\begin{equation}
{\cal H}_{\rm S1} = \sum_j J_j \vec{S}_j \cdot \vec{S}_{j+1} \, ,
\label{Hs1}
\end{equation}
where $\vec{S}_j$ are spin-1 operators on lattice sites $j$, 
and the $J_j$ are random positive nearest-neighbor exchanges;
randomness in the system is characterized by a width $W$ of 
the corresponding distribution of log-exchanges $\ln(J_j)$.
As in the spin-1/2 case, we can characterize spin dynamics and 
transport in terms of the dynamic structure factor and 
the dynamical conductivity; the definitions remain the same 
except for the obvious replacement of all spin-1/2 operators 
with their spin-1 counterparts.  
Experimental realizations of pure Heisenberg spin-1 chains are 
known,~\cite{takigawa} and experimental studies of systems with 
randomness have also been reported in the recent 
literature.~\cite{ran1}
We caution, however, that the degree of disorder needed to 
destroy the gapped Haldane phase of a pure spin-1 chain 
appears to be quite strong,~\cite{yh} and that all our calculations 
are done only in this strong-disorder regime.

The third problem that we consider is the one-dimensional
random transverse field Ising model
\begin{equation}
{\cal H}_{\rm RTFIM} = -\sum_j J_j \sigma_j^z\sigma_{j+1}^z 
-\sum_j h_j\sigma_j^x \, ,
\label{HRTFIM}
\end{equation} 
with random ferromagnetic interactions $J_j$ and positive random 
transverse fields $h_j$; here $\sigma_j$ are Pauli spin matrices.
The strong-disorder RG approach, and its consequences for 
the low-temperature thermodynamics and static correlators,
have been analyzed in greatest detail for this particular 
model.~\cite{foot2dI} 
Also, there are extensive numerical results available for some
dynamical properties.~\cite{numerics} 
This model thus serves as a benchmark to test reliability 
of our approach to the calculation of dynamical properties in 
these strong-disorder systems---we will analyze various 
average autocorrelation functions in considerable detail 
and compare our results with the earlier numerical work. 

The paper is organized as follows: 
We begin in Sec.~\ref{Overview} with a general discussion
of the various types of states that we encounter in these
models, along with an overview of our most important results for 
the dynamics and transport in various regimes; the last part of this
section is devoted to a general outline of the basic approach 
that is used to obtain these results.  The following 
Sections~\ref{XXZ-1/2},~\ref{HS1},~and~\ref{RTFIM}, present careful 
derivations of our results for the zero-temperature dynamical 
properties of the three model systems that we consider, with each 
section starting with a review of the basic RG approach used to 
study the corresponding system. 
In Sec.~\ref{XXZ-1/2} we evaluate the dynamic structure factor
and the dynamical conductivity in the various phases of the 
random XXZ spin-1/2 chain.  This is followed, 
in Sec.~\ref{HS1}, by an analysis of the spin conductivity 
in the strongly-random Heisenberg antiferromagnetic spin-1 chains,
and, in Sec.~\ref{RTFIM}, by an analysis of the average local 
dynamical properties of the random quantum Ising model in the 
vicinity of its critical point.
Section~\ref{FiniteT} is devoted to a qualitative analysis of
the dynamical and transport properties of the XXZ spin-1/2 chains
at non-zero temperatures in the regime $\omega < T$, 
along with some quantitative calculations in the XX spin-1/2 
chain that are possible in this case because of the mapping 
to free fermions.
We conclude, in Sec.~\ref{Experiments}, with a discussion of 
the possible experimental tests of some of our predictions for the 
one-dimensional random-exchange antiferromagnetic spin chains.
Some technical details, as well as some additional developments 
slightly removed from the main thrust of the paper, are relegated 
to the appendices.
%%%%%%%%%%%%%%%%%% END INTRO %%%%%%%%%%%%%%%%%%%%%%%%%%%%%%%%%%%%

%%%%%%%%%%%%%%%%%%%%%%%%%%%%%%%%%%%%%%%%%%%%%%%%%%%%%%%%%%%%%%%%%
%%%%%%%%%%%%%%%%%% OVERVIEW %%%%%%%%%%%%%%%%%%%%%%%%%%%%%%%%%%%%%
\section{Overview}
\label{Overview}
Broadly speaking, our results are for two types of states.
First, there are ground states governed (and therefore
best described by some suitable strong-disorder RG approach)
by infinite-randomness fixed points---examples include 
the random singlet states of the spin-1/2 antiferromagnetic 
chains and the critical point of the random transverse field 
Ising model.  Then, there are the so-called ``Griffiths'' phases
in the immediate vicinity of these critical states;
in these phases, the low-energy renormalized randomness 
is strong, but not infinite.  

In both cases the low-energy excitations are localized, but with 
a characteristic `localization length'---{\em i.e.} the `size of 
the excitation'---that diverges as a power of $\ln \omega$ for 
energy $\omega \rightarrow 0$.  [We emphasize that this is 
the statement about the (rare) low-energy excitations
and is indeed valid in the Griffiths phases, even though 
in this case all equal-time correlators at $T=0$ indicate 
a finite localization length; for details see the main 
body of the paper.]  Apart from this logarithmically divergent 
`localization length', we can also define, from the integrated 
density of states $n_\omega$ for excitations up to 
energy $\omega$, a length $L_{\omega} \equiv n_{\omega}^{-1/d}$ 
that is the typical spacing between these excitations 
in $d$ dimensions [the results we report here are for $d=1$, 
but similar phases do occur for $d>1$~\cite{RTFIM2d}].

For a ground state governed by an infinite-randomness fixed 
point, $L_{\omega}$ diverges at low energies with the same 
power of $\ln \omega$ as the typical size of the excitation.  
This means a strongly-divergent density of states at 
low energy, which allows the system to behave as a conductor
if there is a conserved quantity (e.g., spin or particle number) 
to be transported.  In a Griffiths phase, on the other hand, 
$L_{\omega} \sim \omega^{-1/z}$, with $z$ a nonuniversal
dynamical exponent that varies continuously within the phase.
Here, the low-energy excitations are rare; they are
typically spaced by distance $L_{\omega}$, which diverges as 
a power-law at low energy and thus is much larger than the 
excitation's typical size, which is diverging only 
logarithmically.  In the RG language, the Griffiths phases are 
governed by lines of fixed points ending in the infinite-randomness 
critical fixed point; along such a line, the dynamical exponent 
$z$ varies continuously and diverges near the critical point.

In terms of the original microscopic model, the low-lying 
excitations in the Griffiths phases come from regions
where the local quenched random variables deviate strongly 
from their global averages.  These deviations are such that 
the local averages would put that region in a different phase.
If the system is not at a phase transition, the probability of 
such a rare region occuring and being of linear size $L$ 
behaves as $e^{-c_1 L^d}$ for large $L$, for some constant 
$c_1$.  Such a rare region typically results in a low-lying
mode with a sharply defined (in the sense that $c_2$, 
introduced below, is sharply defined) characteristic frequency 
proportional to $e^{-c_2 L^d}$.  This gives rise to a power-law 
low-energy density of states, with the dynamical exponent $z$ being
determined by the ratio of the constants $c_1/c_2$.
For a disordered Griffiths phase, the rare regions are finite 
``islands'' of either an ordered phase, or a different disordered 
phase.  The resulting low-lying excitations localized on these 
rare regions produce a low-frequency conductivity $\sigma'(\omega)$
or scaled dynamic structure factor $\omega S(k,\omega)$ vanishing
as $\omega^{1/z}$ at low frequencies (apart from possible 
logarithmic factors attributable to singular low-energy behavior
of the relevant matrix elements that may, in some cases,
be sensitive to the logarithmically divergent size of the relevant
excitations).

For one-dimensional systems, there are also power-law Griffiths 
effects in Ising-ordered phases.  These occur because of rare 
regions locally in the disordered phase.  The low-energy excitation
associated with such a region is a domain wall (or ``kink'').
To produce a single such low-energy domain wall requires flipping 
the spontaneous magnetization on one side of the the wall, which 
is tantamount to flipping a semi-infinite piece of the chain.  
Such a flip of an infinite domain cannot occur at a finite 
(non-zero) frequency.  The leading contribution to the 
low-frequency dynamics is then associated with {\it two} nearby 
such rare low-energy domain walls which allow the ordered domain 
between them to flip at a low but nonzero frequency.  
The result of this is that the low-frequency $\sigma'(\omega)$ 
and $\omega S(k,\omega)$ vanish as $\omega^{2/z}$ at low frequency 
in these one-dimensional Ising-ordered Griffiths phases 
(we are again ignoring possible logarithmic factors which
can arise for precisely the same reasons as in the disordered
phase).
Note however that the Griffiths singularities in Ising-ordered 
phases in $d>1$ are of a very different character---in these 
cases, the low-energy density of states vanishes faster 
than any power of $\omega$, as is discussed in 
Ref.~\onlinecite{RTFIM2d}.

In Sections~\ref{XXZ-1/2}-\ref{RTFIM} we will provide a detailed 
justification of these general observations by explicitly 
calculating the low-frequency dynamical properties in a variety of
cases.
In the rest of this Section, we review the phase diagrams of our 
model systems, and highlight our most important results in each case.

%%%%%%%%%%%%%% OVERVIEW::XXZ-1/2 %%%%%%%%%%%%%%%%%%%%%%%%

\subsection{Random antiferromagnetic XXZ spin-1/2 chains}
\subsubsection{Phase diagram}
The phase diagram of the random antiferromagnetic XXZ spin-1/2 
chains is best understood as a product of the competition between 
the transverse part of the coupling $J^{\perp}$, which
favors singlet formation, and the `classical' interaction
$J^{z}$, which favors a ground state with Ising antiferromagnetic 
order.

\narrowtext
\begin{figure}
\epsfxsize=\columnwidth
\centerline{\epsffile{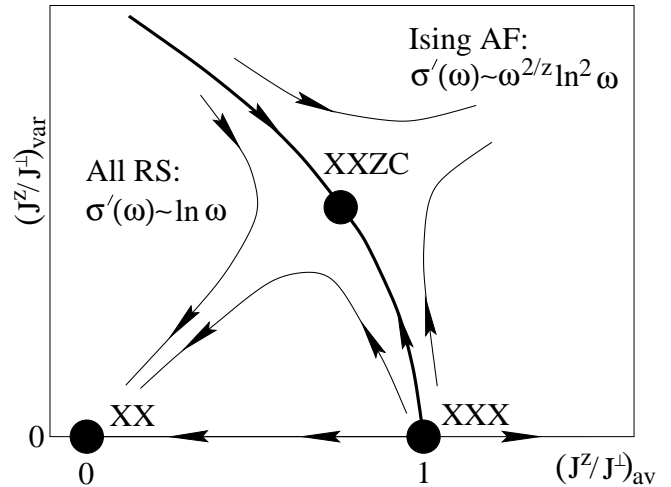}}
\vspace{0.15in}
\caption{Schematic phase diagram of random antiferromagnetic
XXZ spin-1/2 chains obtained in Ref.~{\protect \onlinecite{fraf}},
showing the three different random singlet (RS) fixed points
and RG flows. Our prediction for the low-frequency behavior of the
dynamical conductivity is indicated for each phase.
For details, see Sec.~{\protect \ref{XXZ-1/2}}; 
here $z=z(\delta_{\rm IAF})$ is a (continuously varying)
dynamical exponent in the IAF Griffiths phase.
}
\label{phasesXXZ}
\end{figure}

When the $J^\perp$ dominate, the ground state can
be loosely thought of as being made up of singlet pairs.
In this Random Singlet (RS) state, the interplay of disorder
and quantum fluctuations locks each spin into a singlet pair
with another spin; the two spins in a given singlet pair
can have arbitrarily large spatial separation, with the disorder
determining the particular pattern of the singlet bonds in a given
sample.  On the other hand, when the $J^z$ dominate, the system 
has Ising antiferromagnetic (IAF) order in the ground state 
(with the spins all oriented parallel to the $z$ axis), 
although Griffiths effects can fill in the gap leading to
an IAF ordered Griffiths phase.

These two states are separated by a quantum phase transition
that occurs when the couplings $J^\perp$ and $J^z$ have roughly 
similar distributions (have roughly equal strengths).
A special feature of this system is that the ground state
at any point on the critical manifold is also a Random Singlet 
state, though the details of the excitation spectrum are somewhat 
different.

If we now turn on enforced bond dimerization starting with 
the RS state that obtains for small $J^z$, or the RS state 
of the Heisenberg chain, the system moves into a Griffiths phase 
dubbed the Random Dimer (RD) phase; in this phase the singlet bonds 
in the ground state now preferentially start on one sublattice and 
end on the other.

\begin{figure}
\epsfxsize=\columnwidth
\centerline{\epsffile{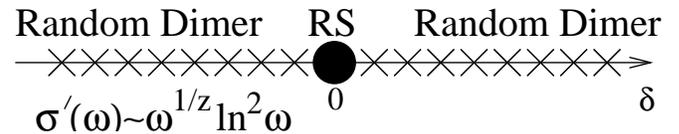}}
\vspace{0.15in}
\caption{The Random Dimer phases for XX or Heisenberg spin-1/2 
antiferromagnetic chains, represented as lines 
of fixed points ending in the critical fixed point labeled RS 
that describes the Random Singlet state at zero 
dimerization;~{\protect \cite{hybg}} here $z=z(\delta)$ is 
a dynamical exponent in the RD Griffiths phase.}
\vspace{0.1in}
\label{phasesRD}
\end{figure}

Schematic phase diagrams summarizing the above are shown 
in Figs.~\ref{phasesXXZ}~and~\ref{phasesRD}.

\subsubsection{Spin transport}

%Figure here so as to force tex to fit it nicely on the page
\narrowtext

We characterize the spin transport properties of the various 
phases in terms of the low-frequency behavior of the dynamical 
conductivity: 
We find that the $T=0$ dynamical conductivity {\it diverges} at low
frequencies in the RS phase as well as at the RS critical points as
\begin{equation}
\sigma'(\omega) =  {\cal K}_{\rm RS} l_v \Gamma_{\omega},
\label{sigmaKRS}
\end{equation}
where we have taken the opportunity to introduce the log-energy scale
\begin{equation}
\Gamma_{\omega} \equiv \ln(\Omega_0/\omega).
\end{equation}
Here and henceforth we use $\Omega_0$ 
to denote the non-universal microscopic energy cutoff, which
corresponds roughly to the energy scale in the bare Hamiltonian
for our various models; also, we use $l_v$ to denote the
non-universal microscopic length scale in the problem.
For the XXZ spin-1/2 system near the RS phase and with sufficiently 
strong disorder, which is what we assume in the following, 
the microscopic length $l_v$ is given by Eq.~(\ref{lv}).
[If, on the other hand, the bare disorder is weak and the system 
flows to strong disorder, then $l_v$ is the length scale at which
the strength of the disorder becomes of order one.] 
${\cal K}_{\rm RS}$ in Eq.~(\ref{sigmaKRS}) is an order one
numerical constant. 
The RS phase and the RS critical points separating it from 
the IAF phase are thus unusual {\it spin conductors}.  

On the other hand, the IAF Griffiths phase is a {\it spin insulator} 
with the low-frequency $T=0$ dynamical conductivity 
\begin{equation}
\sigma'(\omega) = {\cal K}_{\rm IAF}l_{v}\, 
(\omega/\Omega_0)^{2/z_{\rm IAF}} \ln^2(\Omega_0/\omega) \, ,
\end{equation}
where $z_{\rm IAF}(\delta_{\rm IAF})$ is a (continuously varying) 
dynamical exponent diverging at the critical point as
$z_{\rm IAF} \sim \delta_{\rm IAF}^{-(2-\psi)/\lambda}$, 
and ${\cal K}_{\rm IAF}$ is a non-universal amplitude 
vanishing at the transition as 
${\cal K}_{\rm IAF} \sim \delta_{\rm IAF}^{(2-\psi)/\lambda}$.
Here we parametrized the distance from the transition to 
the RS phase by
$\delta_{\rm IAF} \equiv \overline{\Delta}-\overline{\Delta}_c$ 
(where $\Delta \equiv J^z/J^\perp$).  The exponent $\lambda$ is 
the relevant RG eigenvalue controlling the flow away from the 
critical fixed point describing the generic transition between 
the RS phase and the IAF phase, and the exponent $\psi$ 
characterizes the low-energy spectrum 
above the RS ground state at this critical point
(see Ref.~\onlinecite{fraf} and Sec.~\ref{reviewXXZ} for details).
The above result is expected to hold in the frequency regime 
$\omega \ll \Omega_{\delta_{\rm IAF}}$
with the cross-over scale $\Omega_{\delta_{\rm IAF}}$ given
in terms of $\delta_{\rm IAF}$ as
$\ln(\Omega_0/\Omega_{\delta_{\rm IAF}}) \sim 
\delta_{\rm IAF}^{-(2-\psi)/\lambda}$.

Similarly, the RD phases are also {\it spin insulators}, with the
$T=0$ low-frequency dynamical conductivity 
\begin{equation}
\sigma'(\omega) = {\cal K}_{\rm RD} l_{v} 
(\omega/\Omega_0)^{1/z_{\rm RD}} \ln^2(\Omega_0/\omega) \, ;
\end{equation}
the dynamical exponent $z_{\rm RD}(\delta)$ in the RD
phase diverges at the transition as $z_{\rm RD} \sim |\delta|^{-1}$,
and the non-universal amplitude ${\cal K}_{\rm RD}$ vanishes at 
the transition as ${\cal K}_{\rm RD} \sim |\delta|$.
As in the IAF phase, this result is valid at frequencies well
below the corresponding crossover scale $\Omega_\delta$ 
(which can be also viewed as the conductivity pseudo-gap scale); 
in the RD phases $\ln(\Omega_0/\Omega_\delta) \sim 1/|\delta|$.

Thus, in both the IAF phase and the RD phase, the conductivity 
has the functional form
\begin{equation}
\sigma'(\omega) \sim \omega^\alpha \ln^2 \omega \, ,
\end{equation}
with the non-universal exponent $\alpha$ vanishing at the 
corresponding transition.  Note that a similar form but with 
fixed $\alpha=2$ ---the Mott formula--- is obtained via
the usual Mott argument for the $T=0$ dynamical conductivity of 
the one-dimensional Anderson insulator (the fixed value of $\alpha$ 
in this case simply reflects the fact that low-energy density of 
states in the Anderson insulator is {\em constant}, in contrast to 
the situation in the Griffiths phases of interest to us here).

\subsubsection{Spin dynamics}
\begin{figure}
\epsfxsize=\columnwidth
\centerline{\epsffile{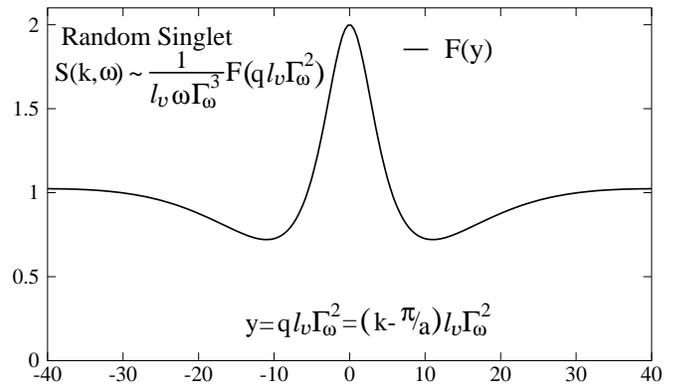}}
\vspace{0.15in}
\caption{Sketch of the dynamic structure factor at fixed 
$\omega \ll \Omega_0$ in the RS states.}
\label{figSXXZ}
\end{figure}

\begin{figure}
\epsfxsize=\columnwidth
\centerline{\epsffile{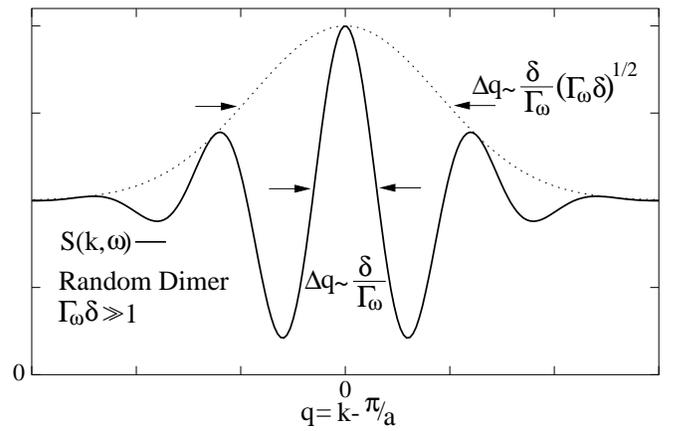}}
\vspace{0.15in}
\caption{Sketch of the dynamic structure factor at fixed 
$\omega \ll \Omega_\delta$ in the RD phases.}
\label{figSRD}
\end{figure}

Turning to the spin dynamics, we find that the $T=0$ dynamic 
structure factor in the RS states in the vicinity of $k=\pi/a$ 
can be written in the following unusual {\it scaling form}
\begin{equation}
S^{\alpha \beta}(k=\frac{\pi}{a}+q,\omega) =
\frac{{\cal A}}{l_v \omega \ln^3(\Omega_0/\omega)}
\Phi\left(|q l_v|^{1/2}\ln(\Omega_0/\omega)\right)
\end{equation}
for $|q| \ll a^{-1}$ and $\omega \ll \Omega_0$; here 
$\alpha \beta \equiv +-$ or $zz$, ${\cal A}$ is an order one 
numerical constant, $l_v$ is the microscopic length defined 
earlier, and the fully {\em universal} function $\Phi(x)$, 
calculated in Sec.~\ref{XXZ-1/2}, interpolates smoothly between 
the limiting forms $\Phi(x) \sim 2-7x^4/90$ for $x \ll 1$ and 
$\Phi(x) \sim 1+2x(\cos x + \sin x) e^{-x}$ for $x \gg 1$.
A plot of the momentum dependence of the dynamic 
structure factor near $k=\pi/a$ (at fixed low frequency) is 
shown in Fig.~\ref{figSXXZ}; an interesting aspect is 
the non-monotonic nature of the lineshape.
We will see in Sec.~\ref{XXZ-1/2} that this oscillatory behavior 
becomes more pronounced and leads to really striking structure in 
the momentum dependence of the dynamic structure factor at (fixed) 
low frequency $\omega \ll \Omega_\delta$ in the Random Dimer 
phases---a plot of the expected $k$ dependence is shown in
Fig.~\ref{figSRD}.  A very similar dependence is also predicted
in the IAF Griffiths phase close to the transition to
the RS state.

As mentioned earlier, these results are expected to remain valid
at small non-zero temperatures so long as the frequency
$\omega$ satisfies $\omega \agt T$.  In Sec.~\ref{FiniteT},
we will see that we can partially overcome even this restriction
in the vicinity of the XX point.

%%%%%%%%%%%%%%%%%%%%% OVERVIEW::SPIN-1 %%%%%%%%%%%%%%%%%%%%%%%%%
\subsection{Spin-1 Heisenberg antiferromagnetic chains}
\subsubsection{Phases}

\begin{figure}
\epsfxsize=\columnwidth
\centerline{\epsffile{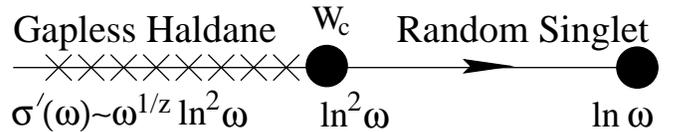}}
\vspace{0.15in}
\caption{Schematic phase diagram of the strongly disordered
Heisenberg spin-1 chain,~{\protect \cite{hy}} along with our 
results for the dynamical conductivity in various regimes. 
Moving to the right along the horizontal axis corresponds to 
increasing disorder.}
\label{phasesH1}
\end{figure}

The effect of randomness on antiferromagnetic Heisenberg spin-1 
chains is even more interesting.  For spin-1, there are, in general
three distinct phases possible in the presence of disorder.  If the
disorder is weak, and the support of the probability distribution
$P(J)$ of the exchanges is confined to a narrow enough region near
the mean, then the system remains in the usual gapped, topologically
ordered Haldane state.  For stronger disorder, or when $P(J)$ has
tails to large or small enough $J$, one has the `Gapless Haldane' 
(GH) phase in which the system still has the topological order that 
characterizes the Haldane state, but becomes gapless due to 
Griffiths effects.  Finally, if the disorder is extremely
strong, with the (bare) distribution of exchanges broad on 
a logarithmic scale, a Random Singlet state completely analogous 
to the one encountered in the spin-1/2 case obtains.  While the GH 
state and the RS state are separated by a quantum critical point 
with universal critical properties (these properties are in fact 
controlled by a strong-disorder fixed point~\cite{hy,mgj}), 
the corresponding transition between the gapped and gapless Haldane 
states is a non-universal feature of the phase diagram, depending 
sensitively on the nature of the initial distribution of couplings 
(see Fig.~\ref{phasesH1} for a summary of the universal
aspects of the phase diagram).

\subsubsection{Overview of results}
In the spin-1 RS state, we obtain the same results for 
the dynamic structure factor and spin conductivity as in 
the spin-1/2 RS state, as the low-energy behavior of the RS 
state does not depend on the spin magnitude except through 
the values of some microscopic scale factors.
Unfortunately, once we move away from the Random Singlet state, 
it is difficult to discuss reliably the momentum dependence of 
the dynamic structure factor of the original spin-1 chain---this 
is because our actual calculations are done in an 
{\em effective model} (see section~\ref{reviewS1} and 
Refs.~\cite{hy,mgj} for details) in which much of the spatial 
information about the original system is missing.

\narrowtext 

However, it is still possible to calculate {\em transport} 
properties, such as the dynamical conductivity, that are insensitive
to the details of the spatial structure (this is, in essence,
a consequence of spin conservation).
At the critical point separating the Gapless Haldane state
from the Random Singlet state, we find for $\omega \ll \Omega_0$
\begin{equation}
\sigma'(\omega) = {\cal K}_{\rm HY} l_v 
\ln^2(\Omega_0/\omega) \, ,
\end{equation}
which is a {\it stronger} divergence than in the strong-disorder 
RS phase; here $l_v$ is the non-universal microscopic 
length scale beyond which the effective model applies, 
$\Omega_0$ is the corresponding microscopic energy scale, 
and ${\cal K}_{\rm HY}$ is an order one numerical constant.  
Thus, the critical point separating the RS phase from 
the GH phase is also an unconventional `spin metal'.  
The GH phase, on the other hand, is a `spin insulator', 
not unlike the RD phase of spin-1/2 chains.  
We find for the conductivity in the GH phase
\begin{equation}
\sigma'(\omega) \sim {\cal K}_{\rm GH} l_v 
(\omega/\Omega_0)^{1/z_{\rm GH}}\ln^2(\Omega_0/\omega) \, .
\end{equation}
The dynamical exponent $z_{\rm GH}$ varies continuously
in the Gapless Haldane phase, diverging at the critical
point as $z_{\rm GH} \sim (W_c -W)^{-\nu/3}$, while the
the non-universal amplitude ${\cal K}_{\rm GH}(W)$ remains
non-zero as one approaches the critical point.  
In the above, $W_c$ is the critical value of the bare disorder 
(the parameter $W$ has already been defined in Sec.~\ref{intro}), 
and the correlation length exponent $\nu=6/(\sqrt{13}-1)$ is known 
from the analysis of Refs.~\onlinecite{hy} and \onlinecite{hPhD}.

%%%%%%%%%%%%%%%%%%% OVERVIEW::RTFIM %%%%%%%%%%%%%%%%%%%%%%%%%%
\subsection{Random quantum Ising spin chains}
\subsubsection{Phases}
The self-dual nature of the random transverse field Ising model 
in one dimension implies that 
the system will be in a critical state if the distributions 
of bonds and fields are identical.  The deviation from 
criticality may be parametrized by
\begin{equation}
\delta=\frac{ \overline{\ln h} - \overline{\ln J} }
            { {\rm var}(\ln h) + {\rm var}(\ln J) } \, ,
\end{equation}
with $\delta > 0$ corresponding to the quantum disordered
paramagnet, and $\delta < 0$ corresponding to the ordered 
ferromagnet.  (Note that we use `$\delta$' both as a dimensionless
measure of dimerization in spin-1/2 chains, and in the present 
context; there is however no cause for confusion and the meaning 
will always be clear from the context in what follows.)

This quantum critical point is flanked, for small $|\delta|$ 
on either side, by paramagnetic and ferromagnetic Griffiths 
phases with gapless excitations.

\subsubsection{Overview of results}
\label{reviewRTFIM}
As mentioned earlier, these Griffiths phases and the quantum 
critical point separating them are among the best 
understood examples of such strong-randomness phenomena. 
However, all previous analyses of the dynamical properties 
relied on numerical results supplemented by scaling ideas.

In contrast, our approach allows us to analytically calculate
the average local autocorrelations of both the spin and 
the energy operators at, and in the vicinity of, the quantum 
critical point, as well as obtain the scaling behavior of 
the dynamic structure factor of the spins.
The main features of the average autocorrelations 
(as well as distributions of autocorrelations, which we do not 
address here) have already been noted in the earlier numerical 
work (Refs.~\onlinecite{numerics}), while our results
on the dynamical structure factor are new.  Here, we only highlight 
some of the subtleties, missed in these numerical studies, that our 
analytical work has uncovered regarding the autocorrelations---a 
complete tabulation of our predictions (and their interpretation in 
terms of Griffiths effects) is given in Sec.~\ref{RTFIM}.

Our results for the $T=0$ imaginary-time off-critical 
spin autocorrelation in the bulk have the form
\begin{equation}
\left[ C_{\rm loc} \right ]_{\rm av}(\tau)
\sim \frac{ |\ln \tau| }{\tau^{n/z(\delta)}} \, ,
\end{equation}
where $z(\delta)$ is the continuously 
varying dynamical exponent characterizing the Griffiths phases 
(from the results of Ref.~\onlinecite{fising}, 
$z^{-1} \approx 2|\delta|$, for small enough $\delta$).
In the above, the parameter $n$ distinguishes between the
disordered and ordered phases with $n=1$ in the disordered phase and $n=2$ in the ordered phase.
Thus, the exponent controlling the power-law decay in
the ordered Griffiths phase is {\em twice} $z^{-1}$,
while the corresponding exponent in the paramagnetic
Griffiths phase is $z^{-1}$.
This reflects the physical distinction between the disordered 
and the Ising ordered Griffiths phases noted in our general 
discussion at the beginning of this Overview.
Moreover, the autocorrelations in the Griffiths phases are 
{\em not} pure power-law, but have a logarithmic correction, 
which reflects the fact that the appropriate `spin' degrees of 
freedom relevant at a time-scale $\tau$ have an effective moment 
of order $|\ln \tau|$.
Both these subtleties have been ignored when extracting 
the dynamical exponent from the numerical results for 
the average spin autocorrelations via the ansatz
$\left[ C_{\rm loc} \right ]_{\rm av}(\tau)
\sim \frac{1}{\tau^{1/z(\delta)}}$, and this could account for 
some of the discrepancies observed in the numerical studies.
Similar remarks apply to other average autocorrelations considered,
and we refer to Sec.~\ref{RTFIM} for details.
%%%%%%%%%%%%%% END QUANTUM ISING OVERVIEW %%%%%%%%%%%%%%%%%%%%%%

%%%%%%%%%%%%%%%% OVERVIEW::BASIC STRATEGY %%%%%%%%%%%%%%%%%%%%%%
\subsection{The basic strategy}
\label{strategy}
We conclude with an overview of the basic strategy
introduced by us in Ref.~\onlinecite{prlus} for the calculation of 
dynamical and transport properties---we will be using
this approach over and over again in what follows, and while 
the details will differ from calculation to calculation, 
the basic approach will remain unchanged.

Consider, for concreteness, the calculation of the dynamic 
structure factor $S^{\alpha \beta}(k,\omega)$ for 
the Hamiltonian ${\cal H}_{\rm XXZ}$.  The basic idea is 
to eliminate high-energy degrees of freedom using 
an appropriate strong-disorder renormalization group procedure 
(in this case, the singlet RG reviewed in Sec.~\ref{reviewXXZ}), 
and trade in the spectral sum~Eq.~(\ref{struc}) for
a sum over the eigenstates of the renormalized Hamiltonian
$\tilde{\cal H}_{\rm XXZ}$, which has fewer degrees of
freedom and renormalized bond strengths.  This renormalized
spectral sum must use the matrix elements of the 
{\it renormalized versions} of the spin operators; these 
renormalized operators are of course defined by the requirement that 
their matrix elements between the eigenstates of the 
renormalized problem reproduce the matrix elements of 
the original operators between the corresponding
eigenstates of the original problem.  In the systems
of interest to us, the low-energy renormalized randomness 
is very large.  In the renormalized problem at the energy
cutoff $\Omega \ll \Omega_0$, the effective bonds thus have a
very broad distribution characteristic of the fixed point 
to which the system flows in the low energy limit.
This allows us to reason as follows:  Focus on pairs of spins
coupled by `strong' bonds in the renormalized problem, with
strengths equal to the cutoff $\Omega$.  The broad distribution 
of bonds implies that these pairs are effectively isolated from 
their neighbors.  It is therefore possible to unambiguously 
identify the excited states of these pairs with excitations of 
the full system at the same energies and work out the matrix
elements connecting these to the ground state using
the renormalized operators.  Thus, to calculate the spectral 
sum~Eq.~(\ref{struc}), the RG is run till the cutoff $\Omega$ 
equals $\Omega_{\rm final}$, and the problem is reduced
to calculating the renormalized spectral sum in this new theory; 
$\Omega_{\rm final}$ is chosen so that the energy of such 
excited states (associated with these strong bonds)
measured from the ground state equals $\omega$.  
The calculation of $S^{\alpha \beta}(k,\omega)$ then becomes 
a counting problem.  One uses the known statistical properties 
of the renormalized bonds in the theory with cutoff 
$\Omega_{\rm final}$ to calculate the number of such strong bonds,
and simply adds up their contributions weighted by 
the corresponding matrix elements to obtain the required 
result.  This result is expected to be asymptotically accurate 
in the limit of small $\omega$, since these contributions 
clearly dominate in the low-frequency limit.  A certain 
simplicity thus emerges when the low-energy effective theory 
has strong disorder, and we will exploit this to the fullest 
in what follows.
%%%%%%%%%%%%%%%%%%%%%%%%%%%%%%%%%%%%%%%%%%%%%%%%%%%%%%%%%%%%%%%%

%%%%%%%%%%%%%%%%%%%%%%%%%%%%%%%%%%%%%%%%%%%%%%%%%%%%%%%%%%%%%%%%
%%%%%%%%%%%%%%%%%%%%%%%%  XXZ S=1/2  %%%%%%%%%%%%%%%%%%%%%%%%%%%
\section{Dynamics and transport in the S=1/2 XXZ chains}
\label{XXZ-1/2}
%%%%%%%%%%%%%%%%%%%%% XXZ-1/2::PHASES %%%%%%%%%%%%%%%%%%%%%%%%%%
\subsection{Detailed characterization of the phases}
\label{reviewXXZ}
\subsubsection{Singlet RG description of the Random Singlet states: 
A review}
\label{singletRG}
We begin by noting that the weak-randomness analysis of 
Doty and Fisher~\cite{doty} implies that randomness is relevant 
for pure antiferromagnetic XXZ spin-1/2 chains for 
$0 \leq J^z/J^\perp \leq 1$; any amount of randomness is 
thus expected to drive the system to strong disorder 
in this entire regime.

In the strong-disorder regime, the singlet RG proceeds as
follows:~\cite{dm,fraf}
We look for the bond with the largest $J^\perp$ coupling, 
say $J^\perp_{23}$ between spins $2$ and $3$; this sets 
the energy cutoff $\Omega \equiv \max\{ J^\perp_j \}$.
We first solve the corresponding two-spin problem 
and introduce the neighboring bonds later as a perturbation.
So long as the $J^z$ couplings are not large compared to 
the $J^\perp$ couplings, the ground state of the two-spin problem 
will always be a singlet separated by a large gap from 
the triplet excited states.  We can then trade
our original Hamiltonian in for another Hamiltonian
(determined perturbatively in the ratio of the neighboring
bonds to the strongest bond) which acts on a truncated Hilbert 
space with the two sites connected by the `strong' bond removed.
To leading order, this procedure renormalizes the Hamiltonian 
${\cal H}_{\rm 4sites} = \sum_{j=1}^3 
[ J^\perp_j (s^x_j s^x_{j+1} + s^y_j s^y_{j+1})+
  J^z_j s^z_j s^z_{j+1} ]$ to
$\tilde{\cal H}_{14} = \tilde{J}^\perp_1 (s^x_1 s^x_4 + s^y_1 s^y_4) 
                     + \tilde{J}^z_1 s^z_1 s^z_4$
with $\tilde{J}^\perp_1 = J^\perp_1 J^\perp_3 / (J^\perp_2 + J^z_2)$
and $\tilde{J}^z_1 = J^z_1 J^z_3 / 2J^\perp_2$; note that the length
of this new bond is $\tilde{l}_1 = l_1 + l_2 + l_3$. 
This procedure, if it remains valid upon iteration, thus
ultimately leads to the Random Singlet state described
in the overview.

A complete understanding of the possible phases then requires an
analysis of the effects of iterating the basic RG procedure.
Such an analysis was performed in Ref.~\onlinecite{fraf} leading 
to the following conclusions (see Fig.~\ref{phasesXXZ}):
So long as the $J^z$ couplings do not dominate over 
the $J^\perp$ couplings and therefore do not produce a state with 
Ising antiferromagnetic order, the ground state is a
Random Singlet state.  In this case, a detailed characterization 
of the low-energy effective Hamiltonian is best couched in terms 
of logarithmic variables as follows: 
Let $\Omega \equiv \max\{ J^\perp_j \}$ at
any given stage of the RG, and define the log-cutoff
$\Gamma \equiv \ln(\Omega_0/\Omega)$.  Also define
log-couplings $\zeta_j \equiv \ln(\Omega/J^\perp_j)$ 
and log-anisotropy parameters $D_j \equiv \ln(\Delta_j)$, 
where $\Delta_j \equiv J^z_j/J^\perp_j$.  As $\Gamma$ increases, 
the fraction of remaining sites $n_\Gamma$ at log-cutoff scale 
$\Gamma$ is given as $n_\Gamma \sim 1/\Gamma^2$.  
When the $J^\perp$ couplings dominate, the system rapidly flows 
to the `XX-RS' fixed point and the probability distribution 
$P(\zeta,\, \Delta,\, l \,|\, \Gamma)$ that determines 
the strengths and lengths of the bonds connecting the remaining 
sites in the effective Hamiltonian quickly converges to the 
following scaling form characteristic of the XX-RS fixed point:
$P(\zeta,\, \Delta,\, l \,|\, \Gamma) = \frac{1}{\Gamma^3}
{\cal P}_1(\frac{\zeta}{\Gamma},\, \frac{l}{\Gamma^2})
\times \delta(\Delta)$.  The function ${\cal P}_1$ has been
characterized in detail in Ref.~\onlinecite{fraf}; here we only
note that $\int\! dy\, {\cal P}_1(x,y) = e^{-x}$.
Between the IAF phase and this XX-RS phase lie 
two kinds of critical points.  If the initial problem has 
full Heisenberg symmetry ($J^z = J^\perp$ for each bond), 
the low-energy effective Hamiltonian preserves this symmetry and 
has bond strengths and lengths drawn from the same probability
distribution:
$P(\zeta,\, l \,|\, \Gamma) = \frac{1}{\Gamma^3}
{\cal P}_1(\frac{\zeta}{\Gamma},\, \frac{l}{\Gamma^2})$. 
In the RG language, the Heisenberg system is critical and
is controlled by the `XXX-RS' critical fixed point.  Finally, 
in this language, the generic critical point between the IAF 
phase and the XX-RS phase is controlled by the `XXZC-RS' fixed 
point---the low-energy effective theory has bond strengths and 
lengths drawn from a distribution 
$P(\zeta,\, D,\, l \,|\, \Gamma) = \frac{1}{\Gamma^{3+\psi}}
{\cal P}_2(\frac{\zeta}{\Gamma},\, \frac{D}{\Gamma^{\psi}}, \,
\frac{l}{\Gamma^2})$ with $\psi <1$ and
$\int\! dy\, {\cal P}_2(x,y,z) = {\cal P}_1(x,z)$.
Notice that these scaling forms imply that the distributions 
of the couplings become infinitely broad as $\Omega \to 0$; 
thus, the RG becomes asymptotically exact at low-energies 
and, in particular, predicts the ground state properties and 
low-temperature thermodynamics correctly.

\subsubsection{Scaling Description of the Ising Antiferromagnet}
\label{scalingIAF}
On the Ising Antiferromagnet side, the singlet RG becomes
invalid at low energies, and the system has a ground state 
with IAF order.  The proper characterization of the system at 
these low energies is in terms of IAF-ordered spin clusters, 
as well as the domain-wall excitations that act to disrupt this 
order.  This section is devoted to providing such a description.
In what follows, we will be considering mainly the IAF phase 
{\em close to the transition to the RS state}.  In this regime, 
the system will `look' IAF ordered only well below a crossover 
energy $\Omega_{\delta_{\rm IAF}}$, while resembling 
a {\em critical} system controlled by the XXZC critical 
point above the crossover scale.  $\Omega_{\delta_{\rm IAF}}$ is 
the scale at which the singlet RG breaks down and is thus
determined by the properties of the RG flows in the vicinity of
the XXZC critical point.  The corresponding log-energy scale 
$\Gamma_{\delta_{\rm IAF}} \equiv 
 \ln(\Omega_0/\Omega_{\delta_{\rm IAF}})$ is given as~\cite{fraf}
$\Gamma_{\delta_{\rm IAF}} \sim \delta_{\rm IAF}^{-\theta}$,
with $\theta=(2-\psi)/\lambda$, where $\lambda$ is the leading 
relevant RG eigenvalue at the XXZC fixed point and $\psi$ has 
already been defined in the previous section.  Below, we construct 
a {\em scaling} description of the IAF phase near criticality by 
combining information obtained from the singlet RG about the nature
of the system at this crossover scale, with a `cluster RG' approach
that is designed to work in the limit of low-energies (well below 
$\Omega_{\delta_{\rm IAF}}$) above the IAF ordered ground state.

We begin with a sketch of our cluster RG approach.  
Consider the Hamiltonian ${\cal H}_{\rm XXZ}$ with $J^z$ couplings 
completely dominating the $J^\perp$ couplings.  Now, spins tend to 
order antiferromagnetically, and we can try formulating a cluster RG 
similar to that for the ordered phase of the random transverse field
Ising model.
Consider combining two such spins, say $s_2$ and $s_3$, coupled 
by a strong bond $J^z_2$ into a new `superspin' $\tilde s_{(23)}$. 
If we identify the two states
$|\! \Uparrow_{(23)} \rangle$ and $|\! \Downarrow_{(23)} \rangle$
of this superspin with the states 
$|\! \uparrow_2 \downarrow_3 \rangle$ and 
$|\! \downarrow_2 \uparrow_3 \rangle$ (which is not a unique 
choice), and treat the $J^\perp$ couplings to second order in 
perturbation theory, the effective Hamiltonian that we obtain is
\begin{eqnarray*}
\tilde{\cal H}_{1(23)4}
& = & \tilde J^z_1 s^z_1 \tilde s^z_{(23)}
    - \tilde J^z_3 s^z_4 \tilde s^z_{(23)}
    + \tilde h_{(23)} \tilde s^x_{(23)} -  \\
&   &
    - \tilde J^\perp_{1(23)4} 
(s^+_1 \tilde s^-_{(23)} s^-_4 + s^-_1 \tilde s^+_{(23)} s^+_4) \, ,
\end{eqnarray*}
where 
$\tilde h_{(23)} \!\!=\!\! J^\perp_2$, 
$\tilde J^\perp_{1(23)4} \!=\! J^\perp_1 J^\perp_3 / J^z_2$,
$\tilde J^z_1 \!=\! J^z_1+(J^\perp_1)^2/J^z_2$, and
$\tilde J^z_3 \!=\! J^z_3+(J^\perp_3)^2/J^z_2$.
Thus, we see that new terms, not present in the original 
Hamiltonian, are generated: an effective transverse field, which 
acts to flip the new spin, and also a three-spin exchange 
interaction.  Before we proceed, a couple of comments regarding 
the new terms:
The effective transverse field appears because the ground state of 
${\cal H}_{23}$ is not exactly a degenerate doublet (the two 
lowest eigenstates, which are the symmetric and antisymmetric 
combinations of
$|\! \uparrow_2 \downarrow_3 \rangle$ and 
$|\! \downarrow_2 \uparrow_3 \rangle$, are actually split 
by a small energy $J^\perp_2$).  Note also that the three-spin 
term does {\em not} violate spin conservation---for example, 
if we consider coupling the conserved total $s^z_{\rm tot}$ 
to a magnetic field, we immediately realize that the superspin 
$\tilde s_{(23)}$ does {\it not} couple to this field. 
 
In principle, we may proceed with such a clustering process, 
keeping track of all additional one- or two- or multi-spin--flip 
terms that are generated.  While this RG is not analytically 
tractable, we do not expect the generated terms to have any 
drastic consequences, since they generally become weaker and 
weaker, while the $J^z$ couplings remain almost unchanged.  
Alternatively, we can remedy this proliferation of new couplings 
by combining an odd number of spins at a time---because of 
the symmetries of the Hamiltonian, any odd length chain will 
have a degenerate pair of ground states with the total 
$s^z_{\rm tot}=\pm \frac{1}{2}$.  In addition, three-spin
terms of the form encountered previously will now be forbidden
by spin conservation.  More explicitly, if we combine three 
spins, say $s_2$, $s_3$, and $s_4$, with relatively strong 
couplings $J^z_2$ and $J^z_3$, into a new superspin 
$\tilde s_2 \equiv \tilde s_{(234)}$, and treat the $J^\perp$ 
couplings perturbatively, the XXZ form of the effective 
Hamiltonian is preserved, with the new couplings  
$\tilde J^z_{12} \!=\! J^z_1+(J^\perp_1)^2/(2 J^z_2)$,
$\tilde J^z_{25} \!=\! J^z_4+(J^\perp_4)^2/(2 J^z_3)$,
$\tilde J^\perp_{12} \!=\! -2 J^\perp_1 J^\perp_3/J^z_2$, and
$\tilde J^\perp_{25} \!=\! -2 J^\perp_2 J^\perp_4/J^z_3$.

Either way, we will have effective spin-half objects
with dominant Ising AF interactions.  Almost always, 
we will be decimating strong $J^z$ couplings making larger
and larger clusters, with the other $J^z$ couplings remaining
essentially unchanged, and the remaining $J^\perp$ couplings
growing weaker and weaker. Only rarely will there be a bond 
with the $J^\perp$ coupling large compared to the neighboring 
couplings, and this will then produce a singlet.
Thus, the picture that emerges is very reminiscent of 
the ordered phase in the RTFIM.

We may now combine this schematic cluster RG description valid
at low energies, with information about
the crossover region obtainable from the singlet RG.
At the crossover scale, the distribution 
of $\zeta^z \equiv \ln(\Omega_{\delta_{\rm IAF}}/J^z)$ is given as 
$P(\zeta^z|\Gamma_{\delta_{\rm IAF}}) \sim 
     \Gamma_{\delta_{\rm IAF}}^{-1} 
     \exp(-\zeta^z/\Gamma_{\delta_{\rm IAF}})$.  
Roughly speaking, beyond the crossover scale, the cluster RG 
merely eliminates the strongest bonds from this distribution, 
but keeps the low-energy tail of the distribution unchanged.
We thus expect a line of (classical) IAF fixed points, 
with properties varying smoothly with the distance from 
the criticality. 
The density of spin degrees of freedom $n_\Gamma$
in the renormalized theory is expected to decrease as
$n_\Gamma \sim \Gamma_{\delta_{\rm IAF}}^{-2}
e^{-c\Gamma/\Gamma_{\delta_{\rm IAF}}}$ 
below the crossover scale $\Omega_{\delta_{\rm IAF}}$, 
with $c$ some order one constant.  This immediately
gives us the density of states 
$\rho(\omega) \sim \omega^{-1} n_{\Gamma_{\omega}}
\Gamma_{\delta_{\rm IAF}}^{-1} 
\sim \delta_{\rm IAF}^{3\theta} \omega^{-1+1/z_{\rm IAF}}$, 
with the continuously varying dynamical exponent
$z_{\rm IAF} \sim \delta_{\rm IAF}^{-\theta}$.
The typical size of the excitations dominating the density 
of states scales as 
$l_{\rm dw}(\omega) \sim l_v \Gamma_{\delta_{\rm IAF}} 
\Gamma_\omega$ and is much smaller than their typical
separation $\sim \omega^{-1/z_{\rm IAF}}$.  This can be 
readily seen from the qualitative picture of `preformed tails':
the length $l_{\rm dw}(\omega)$ of a renormalized bond with 
$\zeta^z = 0$ in the theory with cutoff 
$\omega \ll \Omega_{\delta_{\rm IAF}}$ scales in the same way 
as the length of a bond with 
$\zeta^z \sim \ln(\Omega_{\delta_{\rm IAF}}/\omega)$ 
in the theory at the crossover scale $\Omega_{\delta_{\rm IAF}}$.
On the other hand, the distribution of the log-couplings
$\zeta^\perp \equiv \ln(\Omega/J^\perp)$ is expected to broaden
exponentially as a function of $\Gamma$: for example, 
when we combine $n$ spins that are active at the crossover scale
into a cluster, the effective transverse coupling acting
on this cluster will be of order 
$\zeta^\perp \sim n \Gamma_{\delta_{\rm IAF}}$.  This then is 
our scaling picture for the IAF phase; the important conclusion 
that emerges from this analysis is the fact that the transition 
to the RS state is preceded by a Griffiths phase---the 
{\em IAF Griffiths phase}---with a continuously varying 
power-law singularity in the low-energy density of states.

Finally, it is also possible to obtain a rather direct 
identification of the low-energy modes in terms of the rare 
regions that dominate the low-energy dynamics; we conclude by 
sketching this briefly here.  The RG picture suggests 
that in the IAF phase the typical excitations at low 
energies $\omega$ are classical domain wall excitations that 
live on the effective bonds with weak effective couplings 
$\tilde J^z \sim \omega$.  Such a weak effective $\tilde J^z$
can appear only across a long region that is locally
in the RS phase.  More quantitatively, a region of length $L$ 
locally in the RS phase effectively corresponds to a weak bond 
with $\tilde J^z \sim \Omega_0 e^{-c_z L}$.  The number density
of such regions is roughly $\sim p^L$ for some $p < 1$.
The density of such regions with $\tilde J^z \alt \omega$
is thus some power of $\omega$ that we choose to write as 
$n(\omega) \sim \omega^{1/z}$ with some exponent $z$;
the most numerous such regions will have some ``optimal'' 
(for a given $\delta_{\rm IAF}$) microscopic structure, 
but whatever this structure is, the corresponding optimal
exponent $z$ can be directly identified with the dynamical 
exponent $z(\delta_{\rm IAF})$ of this phase. 
This picture thus predicts that the typical separation 
of such regions is of order $\omega^{-1/z}$, while their 
lengths are only of order $|\ln \omega|$, in complete agreement 
with the schematic RG approach.

\subsubsection{Singlet RG description of the Random Dimer phases: 
A review}
\label{dimerphase}
While the effects of dimerization are not understood in detail 
in all regimes, it is possible~\cite{hybg} to use the singlet RG 
and follow the flows for a chain with full Heisenberg symmetry
and for a chain in the vicinity of the XX-RS point.  In these 
cases, a mapping to the off-critical flows of the RTFIM provides 
a detailed characterization of the so-called `Random Dimer' (RD) 
phases that result.  In either case, the picture that emerges 
can be summarized as follows:~\cite{hybg}  
For concreteness, assume $\delta > 0$.  If disorder is strong 
and $\delta \ll 1$, then the even and the odd bonds renormalize 
essentially as in the corresponding RS state till the log-energy 
scale $\Gamma \sim \Gamma_\delta \equiv 1/\delta$. 
Beyond this scale, the remaining odd bonds rapidly become 
much weaker relative to the remaining even bonds; 
the distribution of the even log-couplings 
$P^{\rm e}(\zeta|\Gamma) = \int dl P^{\rm e}(\zeta,l|\Gamma)$
approaches some limiting distribution with a finite 
but large width, while distribution of the odd log-couplings 
$P^{\rm o}(\zeta|\Gamma) = \int dl P^{\rm o}(\zeta,l|\Gamma)$ 
grows infinitely broad.  In the RG language, the system 
renormalizes to some point on a line of RD fixed points 
(from this point of view, the RS states at $\delta = 0$ 
represent critical points separating RD fixed points with 
opposite dimerization---see Fig.~\ref{phasesRD}). 
The corresponding joint distributions of the log-couplings
and the lengths have been worked out in Ref.~\onlinecite{fising}; 
here we only note that 
$P^{\rm e}(\zeta|\Gamma) = \tau_0(\Gamma) e^{-\zeta \tau_0(\Gamma)}$
with $\tau_0(\Gamma) \approx 2\delta$, while 
$P^{\rm o}(\zeta|\Gamma) = u_0(\Gamma) e^{-\zeta u_0(\Gamma)}$
with $u_0(\Gamma) \approx 2\delta e^{-2\delta\Gamma}$.
The ground state again consists of singlet pairs made up of
one spin on an even site $i$ and a second spin on some odd site $j$.
Note however, that while $i>j$ and $i<j$ are equally probable in
the RS state, in the RD phase with $\delta>0$ one almost always 
has $j>i$ (with the exception of a few high-energy pairs of small 
spatial extent).

%%%%%%%%%%%%%%% XXZ-1/2::STRUCTURE FACTOR %%%%%%%%%%%%%%%%%%%%%
\subsection{Dynamic structure factor}
In this section, we summarize our calculations of the dynamic
structure factor in the different regions of the phase diagram 
of spin-1/2 XXZ chains.  Our approach has already been reviewed 
in general terms in Sec.~\ref{strategy}, and our calculations
here represent one of the simplest examples of this approach 
at work.

We begin by considering only the leading term in the 
perturbative expansion for the renormalized spin operators; 
the results obtained in this manner give the correct leading 
behavior at low frequencies (some justification of this 
is given in Sec.~\ref{xxzvalidity}, where we discuss the role 
of higher-order terms).

\subsubsection{Random Singlet states}
\label{RSxxz}
The leading-order `operator renormalizations' needed are 
particularly simple: the spin operator $\vec s$ remains 
unchanged for each of the `surviving' spins and is effectively 
zero for each of the `decimated' spins (i.e., spins that are 
already locked into singlets with other spins).

Consider first $S^{zz}(k,\omega)$; in our formulation of 
the singlet RG, Sec.~\ref{reviewXXZ}, the following analysis 
applies to a general XXZ singlet state (i.e., remains valid 
so long as the ground state does not have IAF order).  
Consider two spins $L$ and $R$ connected by a strong bond 
$(\tilde{J}^\perp,\tilde{J}^z)$ in the renormalized
theory with cutoff $\Omega_{\rm final}$.  The spin operators 
$s^z_{L/R}$ connect the singlet ground state of this pair 
only to the triplet state $|t_0\rangle$ (with $m_z=0$), which 
is separated from the singlet state by a gap $\tilde{J}^\perp$.
Therefore, the energy scale $\Omega_{\rm final}$ at which 
we stop the RG is $\Omega_{\rm final} = \omega$ in this case 
(remember that the cutoff was defined as 
$\Omega = \max\{J^\perp\}$).  We thus consider 
the renormalized spectral sum
\begin{equation}
S^{zz}(k,\omega) =
\frac{1}{L} \tilde{\sum_m} 
|\langle m| \tilde{\sum_j} e^{ikx_j} \tilde s^z_j |0 \rangle|^2
\delta(\omega - \tilde E_m) \, ,
\label{dynam2}
\end{equation}
where the tildes remind us of the fact that this spectral sum
now refers to the new Hamiltonian with energy cutoff
$\Omega_{\rm final}=\omega$; this renormalized Hamiltonian 
has $n_{\Gamma_\omega}$ spins per unit length with the distribution 
of couplings and bond lengths characteristic of the fixed point 
to which the system flows in the low-energy limit.
The sum Eq.~(\ref{dynam2}) is dominated by the excitations
to the triplet state $|t_0 \rangle$ of pairs of spins connected
by the (renormalized) bonds with $\tilde J^\perp = \omega$; 
these pairs are precisely the ones that are being eliminated 
at this energy scale.  The corresponding matrix element
for each such pair is simply $(1-e^{ik \tilde l})/2$, 
where $\tilde l$ is the length of the bond connecting the pair; 
this allows us to write
\begin{equation}
S^{zz}(k,\omega) \sim n(\Gamma_\omega) \int\! dl\,d\zeta\,
|1-e^{ikl}|^2 P(\zeta,l | \Gamma_\omega)
\delta(\omega-\omega e^{-\zeta})
\label{SzzP}
\end{equation}
for $\omega \ll \Omega_0$ in any RS state.

The calculation of $S^{+-}(k,\omega)$ is more involved 
since the gap to the relevant triplet excited state $|t_1 \rangle$ 
(with $m_z = 1$) of a pair of spins connected by a strong 
bond $(\tilde J^\perp,\tilde J^z)$ is now 
$(\tilde J^\perp + \tilde J^z)/2$.  We consider
each of the three cases (XX, XXX, and XXZC) separately:
{\it (1)} In the XXX case, the Heisenberg symmetry of the problem
guarantees that $S^{xx}=S^{yy}=S^{zz}$. 
{\it (2)} When the system approaches the XX point at low energies,
we have $\tilde J^z \ll \tilde J^\perp$ implying that 
the relevant gap is approximately $\tilde J^\perp/2$. 
Thus, to calculate $S^{+-}(k,\omega)$ we now have to stop 
the RG at the scale $\Omega_{\rm final} = 2\omega$.
>From the Eq.~(\ref{SzzP}), it is clear that this leaves 
our answer unchanged except for the values of various 
non-universal scale factors.
{\it (3)} The XXZC critical point needs special attention.  
In this case $\tilde J^z/\tilde J^\perp$ can have a range of 
values.  As a result, the excited states that dominate 
the spectral sum Eq.~(\ref{struc}) are not simply obtained 
by stopping the RG at any particular $\Omega_{\rm final}$
and looking at the singlets forming only at this scale.
Instead, for any $\Omega_{\rm final} \in (0,2\omega)$
there will be some singlets formed at this scale
that will contribute to the spectral sum---namely,
the pairs coupled by strong bonds with 
$\tilde{J}^\perp = \Omega_{\rm final}$ and
$\tilde{J}^z = 2\omega - \Omega_{\rm final}$.
Note that there is no double-counting here since 
we are considering only the pairs that are being 
eliminated at each energy scale.  Thus, we have
\begin{eqnarray}
S^{+-}(k,\omega) & \sim & 
\int\! d\Gamma \, dl \, dD \; n(\Gamma) |1-e^{ikl}|^2 
P(0,D,l | \Gamma) \, \times \nonumber \\
&& \times \, \delta(\omega-\Omega_0 e^{-\Gamma}(1+e^D)/2) \, .
\end{eqnarray}
Rewriting this in terms of the scaling probability distribution
${\cal P}_2$ and using the delta function to do 
the $\Gamma$ integral gives us
\begin{eqnarray}
S^{+-}(k,\omega) & \sim & 
\frac{1}{\omega \Gamma_\omega} 
\int\! d\bar l\, d\bar D \; 
\frac{n\left(\Gamma_\omega \Upsilon_\omega(\bar D) \right)}
     {\Upsilon^{3+\psi}_\omega(\bar D)}
|1-e^{i\bar k \bar l}|^2 \, \times \nonumber \\
&& \times \, {\cal P}_2
\left(0,\frac{ \bar D }{ \Upsilon^\psi_\omega(\bar D) }, \,
        \frac{ \bar l }{ \Upsilon^2_\omega(\bar D) } \right) \, ,
\label{scalRApt}
\end{eqnarray}
where we have defined
$\bar D = D/\Gamma^\psi_\omega$, 
$\bar l = l/\Gamma^2_\omega$, $\bar k = k \Gamma^2_\omega$, 
and
\begin{equation}
\Upsilon_\omega(\bar D) = 1 + 
\frac{ \ln(1+e^{\bar D \Gamma_\omega^\psi}) - \ln 2 }
     { \Gamma_\omega } \, .
\end{equation}
Now, since $\psi < 1$, it is permissible to take the 
$\Gamma_\omega \to \infty$ limit of $\Upsilon_\omega(\bar D)$ 
before doing the $\bar D$ integral---in other words, we can
replace $\Upsilon$ by $1$ in the low-energy limit.
The $\bar D$ integral can then be done trivially, and
the final expression is identical in form to Eq.~(\ref{SzzP}).
More physically, a given bond $(J^\perp,J^z)$ is described
fairly well (on a logarithmic scale) by one of these two 
couplings---we chose the characteristic scale to be $J^\perp$.
Now, the random anisotropy leads to an uncertainty 
$|\ln(J^z/J^\perp)| \sim \Gamma^\psi$ in the corresponding 
log-energy scale.  This uncertainty is much smaller than the 
already existing typical spread in 
the log-energies or the typical log-energies themselves, 
which are both of order $\Gamma$.  The leading behavior at low 
frequencies is therefore not affected.

Thus, in the limit of low frequencies both $S^{zz}(k,\omega)$ 
and $S^{+-}(k,\omega)$ can be expressed in terms of the
scaling probability distribution ${\cal P}_1$ as
\begin{equation}
S(k,\omega) \sim \frac{n(\Gamma_\omega)}{\omega \Gamma_\omega}
\int\! d\bar l \, |1-e^{i\bar k \bar l}|^2 {\cal P}_1(0,\bar l) \, .
\label{Szzp1}
\end{equation}
Let us first focus on the regime $|q| \equiv |k-\pi/a| \ll a^{-1}$. 
Note that all {\it unscaled} lengths $l$ are {\it odd} multiples 
of the unit length $a$, and therefore $e^{ikl} = -e^{iql}$. 
The integral in Eq.~(\ref{Szzp1}) can be evaluated using 
the characterization of the function ${\cal P}_1(0,y)$ 
available in Ref.~\onlinecite{fising}; the result is the following 
rather unusual scaling form at all three RS fixed points:
\begin{equation}
S(k=\frac{\pi}{a}+q,\omega) =
\frac{\cal A}{l_v \omega \ln^3(\Omega_0/\omega)}
\Phi\left( |q l_v|^{1/2}\ln(\Omega_0/\omega) \right) \, .
\label{S}
\end{equation}
Note that we have suppressed the component labels on $S(k,\omega)$
as the two independent components obey the same scaling form,
but with different values in general of the numerical
constant ${\cal A}$ and the microscopic length scale $l_v$. 
The {\it universal} function $\Phi(x)$ can be written as
\begin{equation}
\Phi(x)=1+x\frac{\cos(x)\sinh(x) + \sin(x)\cosh(x)}
                {\cos^2(x)\sinh^2(x) + \sin^2(x)\cosh^2(x)} \, .
\label{Phi}
\end{equation}
The resulting $S(k,\omega)$ is shown on Fig.~\ref{figSXXZ}.
There is a fairly straightforward interpretation of the
main features of this lineshape:  The peak at $q=0$ 
(i.e., at $k=\pi/a$) reflects the predominantly antiferromagnetic 
character of the low-energy fluctuations; in our language, 
this is a direct consequence of the fact that the (renormalized) 
bonds all have odd lengths in units of $a$.  The strongly damped 
oscillations with the period and the decay scale both of order 
$\Gamma_\omega^{-2}$ express the properties of the distribution 
of lengths of the strong bonds: both the average and the 
RMS fluctuation of this distribution of lengths are of order 
$\Gamma_\omega^2$.

While this result is interesting, one needs to analyze
the effects of higher-order terms in the operator renormalizations
before accepting its consequences for possible neutron scattering
experiments.  We will argue in Sec.~\ref{xxzvalidity} 
that higher-order corrections do not modify the functional 
form Eq.~(\ref{Phi}) of the features in $S(k,\omega)$ 
at fixed $\omega$ but only add an ``incoherent''
background (of strength comparable to that of the features)
and suppress the amplitude of the features by a non-universal
multiplicative factor of order one.

A similar scaling function can be derived for the regime 
$|k| \ll a^{-1}$.  Repeating the above analysis gives
\begin{equation}
S(k,\omega) = \frac{{\cal A}'}{l_v \omega \ln^3(\Omega_0/\omega)}
\tilde\Phi \left( |kl_v|^{1/2}\ln(\Omega_0/\omega) \right) \, ,
\label{Sknear0}
\end{equation}
with $\tilde\Phi(x)=2-\Phi(x)$, and ${\cal A}'$ an order one 
numerical constant.  This scaling function {\it vanishes} for 
$k \to 0$; for small $k$ we have 
$S(k,\omega) \sim l_v k^2 \ln(\Omega_0/\omega)/\omega$.
We must therefore consider the possibility that higher-order 
corrections may overwhelm this scaling result and render
it irrelevant.  This is indeed expected to happen for 
$S^{+-}(k,\omega)$ away from the XXX point.  However, we expect 
the scaling result to be valid quite generally for 
$S^{zz}(k,\omega)$---spin conservation guarantees that 
the higher-order corrections to $S^{zz}(k,\omega)$ must also 
vanish as $k \to 0$ (see Sec.~\ref{xxzvalidity} for a detailed
discussion of this point).

\subsubsection{Dynamic structure factor in Random Dimer phases:
`sharpness' of the Griffiths regions}
\label{SdynRD}
Next, we consider spin dynamic structure factor in the XX and XXX 
Random Dimer phases introduced in Sec.~\ref{dimerphase}.  
The same approach as for the RS states goes over unchanged, 
and we write
\begin{eqnarray}
S(k,\omega) & \sim & n(\Gamma_\omega) \int\! dl\,d\zeta\,
|1-e^{ikl}|^2 \times \nonumber \\
&& \times [P^{\rm o}(\zeta,l | \Gamma_\omega) +
           P^{\rm e}(\zeta,l | \Gamma_\omega)] 
\delta(\omega-\omega e^{-\zeta})
\label{RDdyn1}
\end{eqnarray}
for both $S^{zz}(k,\omega)$ and $S^{+-}(k,\omega)$ in both 
the XX and XXX RD phases (we are again being sloppy about 
the distinction between $\Gamma_\omega$ and $\Gamma_{\omega/2}$, 
as this can be absorbed in the definition of the non-universal 
scale factors that enter our expressions).

Using the results of Ref.~\onlinecite{fising}, it is 
a simple matter to obtain the full crossover from the RS-like 
behavior of the structure factor in the regime 
$1 \ll \Gamma_\omega \alt \Gamma_\delta$ to the behavior 
characteristic of the RD phase in the regime 
$\Gamma_\omega \gg \Gamma_\delta$.  Here, we focus on 
the behavior in the regime $\Gamma_\omega \gg \Gamma_\delta$, 
as this exhibits some rather unusual features.  At these low 
energies, the even bonds dominate over the odd bonds, 
and the contribution of the odd bonds to the sum 
Eq.~(\ref{RDdyn1}) is negligible 
(we are assuming $\delta > 0$ for concreteness). 
For wavevectors in the vicinity of $k=\pi/a$ 
with $|q| \equiv |k-\pi/a| \ll \delta^2/l_v$ 
(i.e., probing lengths larger than the correlation length 
$\xi_{\rm av} \sim l_v/\delta^2$) we obtain
\begin{equation}
S(k, \omega) = \frac{ {\cal C}|\delta|^3 \Omega_0^{-1/z_{\rm RD}} }
                    { l_v \omega^{1-1/z_{\rm RD}} }
\left[ 1 + \cos(l_v q\Gamma_\omega/|\delta|)
      e^{-c l_v^2 q^2 \Gamma_\omega/|\delta|^3} \right]\, ,
\label{SkoRD}
\end{equation}
where ${\cal C}$ and $c$ are some order one constants, and
we have chosen to write the power-law prefactor in terms of
the dynamical exponent $z_{\rm RD}$ (as far as our RG calculations 
are concerned $z_{\rm RD}^{-1} = 2|\delta|$ for small 
$|\delta|$---however, the effective value of $\delta$ that enters 
this expression is expected to acquire a non-universal 
multiplicative renormalization from the high-energy physics, 
and the only reliable statement we can make is that 
$z_{\rm RD}^{-1} \sim |\delta|$ for small enough $|\delta|$).
This result has a striking oscillatory structure 
(see~Fig.~\ref{figSRD}) that is {\it not} suppressed 
significantly by the exponential factor, since
$\sqrt{\Gamma_\omega/|\delta|^3} \ll \Gamma_\omega/|\delta|$ 
in the regime under consideration.  This is best understood 
as a novel signature of the sharply defined geometry of 
the rare Griffiths regions that contribute to the scattering 
at a given low energy (i.e., that are {\it filtered out} by their 
energy).  More precisely, the average length of such regions is 
of order 
$l_v\Gamma_\delta^2(\Gamma_\omega/\Gamma_\delta) 
 = l_v\Gamma_\omega/|\delta|$,
while the RMS fluctuations in the length are only of order 
$l_v\Gamma_\delta^2\sqrt{\Gamma_\omega/\Gamma_\delta}
 = l_v \sqrt{\Gamma_\omega/|\delta|^3}$.
Our results thus suggest that low-energy INS experiments
would be able to pick up the sharply defined geometry of
such Griffiths regions in the RD phases in one dimension.

This feature of the Griffiths regions in one dimension 
was noted in Ref.~\onlinecite{fising}~Sec.~IVB in the context 
of the RTFIM, where it was conjectured also that other 
properties of such low-energy regions are likewise sharply 
defined: for example, in the disordered phase of the RTFIM, 
the magnetic moment of the Griffiths regions with a {\it given} 
characteristic energy is sharply defined~\cite{saddle} 
and proportional to the (sharply defined) length of such regions.
%---this implies, in particular, that {\it all} 
%(i.e., independent of the energy scale) such Griffiths regions 
%have a sharply-defined magnetization density, and can be thought 
%of as rare regions which are locally the same distance into the
%ordered phase.
[In fact, similar ``sharpness'' is expected to hold for any 
bond ``property'' that ``rides'' on top of the singlet RG 
via recursion relation $\tilde x=x_1+x_3+\Upsilon x_2$ 
when bond $J_2$ is eliminated.]
We expect to see a signature of this sharpness of the
Griffiths regions also in the dynamic structure factor 
$S^{zz}$ in the IAF Griffiths phase (see below) and also
in the Griffiths phases of the one-dimensional RTFIM 
(Sec.~\ref{SdynRTFIM}).  Finally, an interesting question, 
which we leave unanswered for now, is whether 
similar sharpness in the properties of the Griffiths regions 
at a given energy occurs and has observable consequences in 
higher dimensions as well, e.g., in the disordered phase of 
the $d>1$ RTFIM.

\subsubsection{IAF Griffiths phase}
\label{SdynIAF}
Let us first consider $S^{zz}(k,\omega)$ in the IAF Griffiths 
phase.  As discussed in Sec.~\ref{scalingIAF}, the dominant 
low-energy excitations in this phase are classical domain walls.
However, it is clear that such excitations do not contribute 
at all to $S^{zz}(k,\omega)$, since they cannot be excited 
from the ground state by the action of the operators
like $\hat s^z_k$, which conserve the total $s^z_{\rm tot}$.
The leading excitations that do contribute to $S^{zz}$ can clearly
be identified in the RG picture with the $m_z=0$ excited
states of pairs of super-spins, with each pair connected by a bond 
with $\tilde J^\perp \sim \omega$ and forming a singlet
(note that this is true {\em regardless} of the value of the 
corresponding $\tilde J^z$).  Now, it is easy 
to generate a weak $\tilde J^\perp$ coupling of order $\omega$ 
in the IAF phase, since any typical region of length $L$ will have 
an effective $\tilde J^\perp$ of order $\Omega_0 e^{-c_x L}$ 
(and an effective $\tilde J^z$ typically much stronger).  
What is more difficult is to {\it isolate} such a region from 
becoming a part of a larger cluster---otherwise this region can 
not support spin fluctuations at frequency $\omega$.  
For this, we need two rare RS-like segments (domain walls) with 
$\tilde J^z \alt \omega$, one on each side of our (typical) 
region.  Thus, we need two domain walls, which are usually 
separated by a large distance of order $\omega^{-1/z}$, 
to occur close to each other---the ``density'' of such 
occurrences is $\sim \omega^{2/z_{\rm IAF}}$. 
The separation of the two domain walls---the length of 
the IAF-ordered cluster that they isolate---must be of order 
$|\ln \omega|$.  More precisely, if the IAF-ordered 
cluster has length $L$, it can be thought of as consisting of the 
$n \sim L/\Gamma_{\delta_{\rm IAF}}^2$ strongly Ising 
coupled spins that are active at the crossover scale; 
the effective bonds connecting these spins at the crossover 
scale typically satisfy 
$\ln(\tilde J^\perp/\tilde J^z) \sim -\Gamma_{\delta_{\rm IAF}}$.
The requirement that the spin-flip coupling for this cluster 
is $\omega$ fixes the length of this cluster to be 
$L = l_v \Gamma_\omega \Gamma_{\delta_{\rm IAF}}$, 
while the uncertainty in this length can only be of order 
$l_v \sqrt{\Gamma_\omega \Gamma_{\delta_{\rm IAF}}^3} \ll
l_v \Gamma_\omega \Gamma_{\delta_{\rm IAF}} $.

We are now ready to calculate $S^{zz}(k=\pi/a+q,\omega)$
in the regime $|q|^{-1} \gg l_v \Gamma_{\delta_{\rm IAF}}^2$, 
in addition to $\omega \ll \Omega_{\delta_{\rm IAF}}$.
The leading-order renormalization of the $s^z_j$ in the cluster
RG is simple: $s^z_j$ is renormalized to $(-1)^j s^z_c$ 
for each spin $j$ that is active in some cluster $c$, and 
renormalizes to zero for every spin that forms a singlet.
Assuming that such clusters ``look'' fairly uniform on the length
scales larger than $l_v \Gamma_{\delta_{\rm IAF}}^2$, and 
adding up the contributions from all such isolated clusters 
with effective spin fluctuation frequency $\omega$, we obtain 
\begin{eqnarray}
S^{zz}(k & = & \frac{\pi}{a}+q,\omega) =
\frac{ {\cal C}' |\delta_{\rm IAF}|^{7\theta}
                \Omega_0^{-2/z_{\rm IAF}} }
     { q^2 l_v^3 \omega^{1-2/z_{\rm IAF}} } \nonumber \\
& \times &
\left[ 1 - \cos(q l_v \Gamma_\omega/|\delta_{\rm IAF}|^\theta)
  e^{-c q^2 l_v^2 \Gamma_\omega/ |\delta_{\rm IAF}|^{3\theta}} 
\right] \, ,
\label{SkzzIAF}
\end{eqnarray}
where ${\cal C}^{\prime}$ and $c$ are some order one constants
and the power of the $\delta_{\rm IAF}$ that appears in the 
prefactor has been fixed by demanding consistency with 
the off-critical scaling form
\begin{equation}
S^{zz}(k=\frac{\pi}{a}+q,\omega) =
\frac{\cal A}{l_v \omega \Gamma_\omega^3}
\Psi\left( \frac{\Gamma_\omega}{\Gamma_{\delta_{\rm IAF}}},
|q l_v|^{1/2} \Gamma_\omega \right) \, ,
\end{equation}
with $\Psi(0,y) = \Phi(y)$.  Note also that the overall $1/q^2$ 
dependence is a consequence of the fact that the spins contributing 
to the scattering have been taken to be distributed uniformly 
over a sharply defined region (the cluster); we expect this to 
cross over to a much faster decay at large momenta (such that 
$|q|^{-1} \sim l_v \Gamma_{\delta_{\rm IAF}}^2$) well outside 
the range of validity of our scaling picture.

The situation is quite different for $S^{+-}(k, \omega)$. As
we shall see in Section~\ref{autocorrxxz}, the renormalization
of the $s^{\pm}_j$ spin operators is quite non-trivial, and we 
are unable to make an equally detailed prediction for $S^{+-}$. 
However, we expect that the matrix element for producing domain
wall excitations with energies of order $\omega$ by the action 
of $s^{\pm}$ on the ground state is strongly suppressed as some 
power of $\omega$, giving rise to a correspondingly small value 
for $S^{+-}(k,\omega)$ at small $\omega$.

%%%%%%%%%%%%%%% XXZ-1/2::AUTOCORRELATIONS %%%%%%%%%%%%%%%%%%%%%%
\subsection{Average local autocorrelations}
\label{autocorrxxz}
The same approach can be used to calculate average autocorrelation 
functions, and this section is devoted to a brief account of our 
results.

We consider the local dynamical susceptibilities
\begin{equation}
\chi^{\alpha\alpha}_{jj}(\omega)=\sum_m 
|\langle m| s^\alpha_j |0 \rangle|^2 \delta(\omega-E_m) \, ,
\end{equation}
where $\alpha=z$ or $\alpha=x$.  A knowledge of the low-frequency 
behavior of these susceptibilities can immediately be translated 
into information about the long-time limit of the corresponding 
imaginary-time autocorrelation functions
\begin{equation}
C^{\alpha\alpha}_{jj}(\tau)=
\langle s^\alpha_j(\tau) s^\alpha_j(0) \rangle \, .
\end{equation}

\subsubsection{RS states and RD phases}
As long as one is only interested in averages of such local
quantities (over different realizations of disorder), it again
suffices to consider only the leading-order spin operator
renormalizations.  We thus already have all the ingredients 
needed to calculate these average dynamical susceptibilities: 
our basic approach is familiar enough by now, and the relevant 
results of Ref.~\onlinecite{fraf} for the renormalized bond 
distributions have already been reviewed in Sec.~\ref{reviewXXZ}.  
Below, we will be correspondingly brief.  We first give 
our results for the average local dynamical susceptibilities 
and then translate these to results for the long-time behavior 
of the corresponding average autocorrelation 
functions.~\cite{acknum}  The leading behavior is the same for 
both $\alpha=z$ and $\alpha=x$, so we drop all superscripts.

For a {\it bulk-spin}, we obtain
\begin{equation}
\left[ \chi_{\rm loc} \right]_{\rm av}(\omega) \sim
\frac{n(\Gamma_\omega)}{\omega}
(P^{\rm e}_0(\Gamma_\omega)+P^{\rm o}_0(\Gamma_\omega)) \, .
\end{equation}
For the {\it critical} RS states ($P^{\rm e}=P^{\rm o}$) we find
\begin{eqnarray}
\left[ \chi_{\rm loc} \right]_{\rm av}(\omega)
& \sim & \frac{1}{\omega |\ln \omega|^3} \, , \nonumber \\
\left[ C_{\rm loc} \right]_{\rm av}(\tau)
& \sim & \frac{1}{|\ln \tau|^2} \, ,
\end{eqnarray}
while {\it off-critical}---in the RD phases---we find
\begin{eqnarray}
\left[ \chi_{\rm loc} \right]_{\rm av}(\omega)
& \sim & \frac{|\delta|^3}{\omega^{1-1/z_{\rm RD}}} \, , 
\nonumber \\
\left[ C_{\rm loc} \right]_{\rm av}(\tau)
& \sim & \frac{|\delta|^3}{\tau^{1/z_{\rm RD}}} \, .
\label{chiRD}
\end{eqnarray}

Similarly, for an {\it end-spin} $s_1$ of a semi-infinite 
chain (with $j \geq 1$) we obtain
\begin{equation}
\left[ \chi_1 \right]_{\rm av}(\omega) \sim
\frac{ P^{\rm e}_0(\Gamma_\omega) P^{\rm o}_0(\Gamma_\omega) }
     { \omega } \, .
\end{equation}
For the RS states we find
\begin{eqnarray}
\left[ \chi_1 \right]_{\rm av}(\omega)
& \sim & \frac{1}{\omega |\ln \omega|^2} \, , \nonumber \\
\left[ C_1 \right]_{\rm av}(\tau)
& \sim & \frac{1}{|\ln \tau|} \, ,
\end{eqnarray}
and in the RD phases
\begin{eqnarray}
\left[ \chi_1 \right]_{\rm av}(\omega)
& \sim & \frac{\delta^2}{\omega^{1-1/z_{\rm RD}}} \, , \nonumber \\
\left[ C_1 \right]_{\rm av}(\tau)
& \sim & \frac{\delta^2}{\tau^{1/z_{\rm RD}}} \, .
\end{eqnarray}

\subsubsection{IAF Griffiths phase}
In the IAF phase, unlike in the singlet states, we need to make a
distinction between $\chi^{zz}$ and $\chi^{xx}$.  Consider first 
$\left[ \chi_{\rm loc}^{zz} \right]_{\rm av}(\omega)$.
>From our previous discussion of the IAF phase, it is clear
that, in the regime $\omega \ll \Omega_{\delta_{\rm IAF}}$, 
the dominant contributions come from IAF-ordered clusters
of lengths $\sim \Gamma_\omega \Gamma_{\delta_{\rm IAF}}$
(i.e., with effective spin-flip couplings of order $\omega$)
that are isolated from the rest of the system by domain 
walls with $\tilde J^z \alt \omega$ on either side.  
>From the scaling picture of the phase, we get
\begin{eqnarray}
\left[ \chi^{zz}_{\rm loc} \right]_{\rm av}(\omega) 
& \sim & \delta_{\rm IAF}^{4\theta} 
\frac{ \Omega_0^{-2/z_{\rm IAF}} \ln(\Omega_0/\omega) }
     { \omega^{1-2/z_{\rm IAF}} }  
\, , \nonumber \\
\left[ C^{zz}_{\rm loc} \right]_{\rm av}(\tau) 
& \sim & \delta_{\rm IAF}^{4\theta} 
\frac{ \ln(\Omega_0\tau) }{ (\Omega_0 \tau)^{2/z_{\rm IAF}} } \, .
\label{IAFautozz}
\end{eqnarray}

The analysis is more complicated for
$\left[ \chi^{xx}_{\rm loc} \right ]_{\rm av}(\omega)$, and
we can only make a plausible estimate for this quantity.
This is because the $x$ and $y$ components of the spin operators 
renormalize in a non-trivial way under the cluster RG.
The origin of this difficulty may be seen as follows:
Consider, for example, combining three spins $s_2$, $s_3$, 
and $s_4$, connected by strong $J^z_2$ and $J^z_3$, into 
a super-spin $\tilde s_{(234)}$.  To zeroth order, all 
three operators $s^+_2$, $s^+_3$, and $s^+_4$, renormalize 
to zero.  To first order, $s^+_2$ and $s^+_4$ renormalize to 
$(s^+_2)^{\rm eff}= - s^+_1 J^\perp_1/J^z_2 
                    - \tilde{s}^+_{(234)} 2J^\perp_3/J^z_2$,
$(s^+_4)^{\rm eff}= - s^+_5 J^\perp_4/J^z_3 
                    - \tilde{s}^+_{(234)} 2J^\perp_2/J^z_4$,
while $s^+_3$ {\em renormalizes to zero to this order}.
Roughly speaking, the original spin-flip operators of the
(active) spins have projections onto the remaining effective 
cluster spin-flip operators with components given by the ratio 
of the corresponding effective spin-flip couplings to 
the original spin-flip couplings.

Now, the dominant contributions to
$\left[ \chi^{xx}_{\rm loc} \right]_{\rm av}(\omega)$ come from
the low-energy (of order $\omega$) domain wall excitations, 
which are represented in the RG picture by the bonds with 
$\tilde{J}^z \sim \omega$ connecting the effective spins 
(clusters) in the effective theory with the renormalized 
cutoff $\omega$.  The matrix element for producing such 
an excitation by a bare spin-flip operator of a spin active 
in one of these clusters will be of order the corresponding 
$\tilde{J}^\perp$, while the number of such spins contributing 
will be of order some effective ``moment'' $\mu_x$ of this 
cluster.  Because of the matrix element proportional to 
$\tilde{J}^\perp$, there will be a significant contribution 
only if this $\tilde{J}^\perp$ is also of order $\omega$.  
As we have already seen, this can happen only if such an 
IAF-ordered cluster has length of order $\Gamma_\omega$ and is 
isolated from the rest by RS-like regions (domain walls) with
$J^z \alt \omega$ on either side.  We already know 
how to estimate the number density of such Griffiths regions.  
As far as the effective moment $\mu_x$ of such an IAF cluster
is concerned, we can only make a crude estimate that bounds it 
from above by the number of spins that are active in this cluster: 
$\mu_x(\omega) \alt \Gamma_\omega$; however, we are unable 
to obtain the precise power of the logarithm that enters
the energy dependence of the effective moment.  We therefore 
leave out the logarithmic correction, and only write 
the dominant power-law part of our estimate:
\begin{eqnarray}
\left[ \chi^{xx}_{\rm loc} \right]_{\rm av}(\omega) 
& \sim & \omega^{1+2/z_{\rm IAF}} \, , \nonumber \\
\left[ C^{xx}_{\rm loc} \right]_{\rm av}(\tau) 
& \sim & \frac{1}{\tau^{2+2/z}} \, .
\label{IAFautoxx}
\end{eqnarray}

%%%%%%%%%%%%%%%% XXZ-1/2::TRANSPORT %%%%%%%%%%%%%%%%%%%%%%%%%%%%
\subsection{Spin transport}
\label{SigmaXXZ}
This section is devoted to a discussion of the dynamical spin 
conductivity $\sigma'(\omega)$ in the spin-1/2 XXZ chains.
Our task here is to evaluate the Kubo formula Eq.~(\ref{sigmakubo}) 
in the low-frequency limit.  For the RS and RD states, 
we will use information available from the scaling solutions to 
the singlet RG recursion relations to achieve this, while in 
the IAF phase, we will use the scaling picture of the Griffiths 
phase we have developed earlier.  Our results
for the dynamical conductivity are summarized in 
Figs.~\ref{phasesXXZ}~and~\ref{phasesRD}.

\subsubsection{Random singlet states}
We first need to work out the rules that govern 
the renormalizations of the current operators.  Assume
once again that $J^\perp_{23}$ is the strongest bond.
We wish to work out perturbatively the renormalized operators 
$\tilde\tau_{1/2/3}$ that we trade in $\tau_{1/2/3}$ for, 
when we freeze spins $2$ and $3$ in their singlet ground state
(the other current operators to the left and right of this segment 
are left unchanged to leading order by the renormalization).  
Now, note that these other operators have overall scale factors in 
them that are nothing but the corresponding $J^\perp$ couplings.
In order to be consistent, we clearly need to work out 
$\tilde\tau_{1/2/3}$ correct to $O(\tilde J^\perp_{14})$
(where $\tilde J_{14}$ is the effective bond connecting 
spins $1$ and $4$ after we freeze out spins $2$ and $3$)
by adding the effects of virtual fluctuations to the 
projections of $\tau_{1/2/3}$ into the singlet subspace.
An explicit calculation gives the simple result
that all three operators renormalize to the same operator
$\tilde\tau_{1/3} = \tilde\tau_2 = 
i \tilde J^\perp_{14} (s^+_1 s^-_4 - s^+_4 s^-_1)/2$,
which we will denote henceforth by $\tilde\tau_1$ for 
consistency of notation.

As we carry out the RG, the above result implies that the total 
current operator $\sum_{j=1}^L \tau_j$ entering 
Eq.~(\ref{sigmakubo}) renormalizes to 
$\tilde{\sum}_j \tilde l_j \tilde \tau_j$, where $j$ now labels 
the remaining sites of the renormalized system, and the 
$\tilde l_j$ are the lengths of the corresponding renormalized 
bonds.  [Note that this result makes sense physically and is a
consequence of spin conservation: when a magnetic field with
a uniform gradient is applied along the length of the chain,
the effective lengths $\tilde l_j$ measure the ``phase'' 
along the chain of this ``driving potential''.]
Consider two spins connected by a strong bond 
$(\tilde J^\perp,\tilde J^z)$ in the renormalized theory with 
cutoff $\Omega_{\rm final}$.  Since the current operator 
living on this bond connects the singlet ground state of 
the pair only to the triplet state $|t_0\rangle$ separated 
from the singlet by a gap $\tilde J^\perp$, we choose 
$\Omega_{\rm final}=\omega$ and consider the renormalized 
spectral sum
\begin{equation}
\sigma'(\omega) = \frac{1}{\omega L} \tilde{\sum_m} 
|\langle m| \tilde{\sum_j} \tilde l_j \tilde\tau_j |0 \rangle|^2
\delta(\omega - \tilde E_m) \, .
\end{equation}
This spectral sum is dominated by precisely the $|t_0\rangle$ 
triplet excitations of pairs of spins that are connected by the 
(effective) bonds with $\tilde J^\perp=\omega$ and are being 
eliminated at this energy scale; the corresponding matrix element
is just $\tilde l\omega/2$, where $\tilde l$ is the length of 
the bond connecting the pair.  In the thermodynamic limit, 
we thus have
\begin{eqnarray}
\sigma'(\omega) & \sim & \frac{n(\Gamma_\omega)}{\omega}
\int\! dl\, d\zeta\, \omega^2 l^2 \, P(\zeta,l | \Gamma_\omega)
\delta(\omega - \omega e^{-\zeta})\, . 
\label{RSsigma}
\end{eqnarray}
This immediately yields our central result
\begin{equation}
\sigma'(\omega) = {\cal K}_{\rm RS}l_{v}\ln(\Omega_0/\omega) \, ,
\label{RSsigma1}
\end{equation}
valid for $\omega \ll \Omega_0$.  Here, ${\cal K}_{\rm RS}$ 
is an order one numerical constant, $l_v$ is the microscopic 
length scale defined earlier, and $\Omega_0$ is the microscopic 
energy cutoff.  Notice that this analysis holds equally well 
at all three RS fixed points, which differ only in the 
corresponding values of the non-universal scale factors.

A brief digression is in order, before we go on to discuss 
this result:  The real part of the dynamical conductivity 
can be related (on general grounds) to the behavior of 
the dynamic structure factor $S^{zz}(k,\omega)$ near $k=0$
\begin{equation}
\sigma'(\omega) =  \omega \frac{1}{2} \frac{d^2}{d k^2} 
S^{zz}(k,\omega) \, ;
\end{equation}
this can be checked by comparing directly the corresponding
spectral sums and noticing that the action of the two operators 
${\cal T}=\sum_j \tau_j$ and ${\cal V}=\sum_j j\sigma^z_j$ 
on the eigenstates of the Hamiltonian ${\cal H}$ are related
through ${\cal T} = i[{\cal H}, {\cal V}]$.  It is easy to check, 
using the scaling form Eq.~(\ref{Sknear0}), that our result 
for the conductivity is consistent, as it must be, with our 
previously derived result for the dynamic structure factor.

Going back to Eq.~(\ref{RSsigma}), we see that $\sigma'(\omega)$ 
{\it diverges logarithmically} for small $\omega$ in the unusual 
`spin-metal' phase controlled by the XX fixed point 
{\it as well as at the critical points} (XXX and XXZC) separating 
this phase from the `insulating' phase with Ising antiferromagnetic 
order in the ground state.  Note that this `metal-insulator'
transition has the curious feature that the quantum critical
points separating the conducting phase from the insulating phase
have the same $T=0$ transport properties as the conducting phase.

\subsubsection{IAF Griffiths phase}
\label{IAF}
On the insulating side, we expect $\sigma'(\omega)$ to be 
suppressed below the crossover scale $\Omega_{\delta_{\rm IAF}}$; 
the dominant contributions for 
$\omega \ll \Omega_{\delta_{\rm IAF}}$ 
come from some rare regions that contain long finite segments 
locally in the `metallic' phase.

We begin by providing a rough estimate of these contributions
to $\sigma'(\omega)$:  In our sample, consider a (large) region 
of length $L$ locally in the RS phase; the number density of 
such regions is roughly $p^L$, with some $p<1$ (which depends 
on the distance from the transition).  If these regions are 
effectively isolated from the rest of the system, the average 
power absorption per spin in each such region is proportional 
to the finite-size conductivity calculated in 
Appendix~\ref{finite_size}:
\begin{equation}
W = L \sigma'_{\rm RS}(\omega,L) 
\sim L^{3/2} \exp(-c |\ln\omega|^2/L) \, ,
\label{Lsigma}
\end{equation}
where we have assumed that $L$, although large, satisfies 
$L \ll |\ln\omega|^2$ (this assumption will turn out to be 
self-consistent).  The total power absorbed in the sample is 
then obtained by summing over all such regions:
\begin{equation}
\sigma'(\omega) \sim 
\int\! dL\, p^L \, L^{3/2} \exp(-c |\ln\omega|^2/L) \, .
\end{equation}
Evaluating this integral by a saddle-point method, we find that 
the lengths that dominate are of order $|\ln\omega|$ 
(our assumption about the lengths is thus valid), and 
arrive at the following estimate
\begin{equation} 
\sigma'(\omega) \sim \omega^\alpha |\ln \omega|^2 \, ,
\label{IAFsigma}
\end{equation}
where $\alpha = \alpha(\delta_{\rm IAF}) > 0$ is a continuously 
varying exponent vanishing at the transition.  While this argument 
is suggestive, we find it more convincing~\cite{worry1} to take 
an alternative route based on the scaling picture we have 
developed earlier for the IAF phase---this has the added advantage 
that it allows us to relate the exponent $\alpha$ to the dynamical 
exponent $z(\delta_{\rm IAF})$.  This is what we turn to next.

We have already seen that the most numerous low-energy excitations 
in the IAF Griffiths phase are domain walls, with the integrated
density of states $n_\omega \sim \omega^{1/z_{\rm IAF}}$.  
Such {\it classical Ising} excitations, however, do not contribute 
to the dynamical conductivity.  The dominant contributions come
from IAF-ordered clusters of lengths 
$L \sim \Gamma_\omega \Gamma_{\delta_{\rm IAF}}$ (i.e., with
effective spin-flip couplings of order $\omega$) that are isolated
from the rest of the system by domain walls with $J^z \alt \omega$. 
Remembering that the number density of such Griffiths regions 
is $\sim \omega^{2/z_{\rm IAF}}$, and noting that the corresponding
``phase lengths'' are of order $L \sim |\ln \omega|$,
we immediately obtain Eq.~(\ref{IAFsigma}) with 
$\alpha = 2/z_{\rm IAF}$.
More formally, we sum over the possible separations of 
two such domain walls, with the constraint that the typical 
IAF-ordered region isolated by the two has significant spin 
fluctuations at the characteristic frequency $\omega$:
\begin{equation}
\sigma'(\omega) \sim \frac{n_\omega^2}{\omega} \int\! dL\, 
\omega^2 L^2 \delta(\omega-\Omega_0 e^{-c_x L}) \, .
\end{equation}
We thus obtain for the dynamical conductivity in the IAF phase
\begin{equation} 
\sigma'(\omega) = {\cal K}_{\rm IAF}l_{v}
(\omega/\Omega_0)^{2/z_{\rm IAF}} \ln^2 (\Omega_0/\omega) \,,
\end{equation}
where ${\cal K}_{\rm IAF}$ is a numerical prefactor
that depends continuously on $\delta_{\rm IAF}$.
The scaling of ${\cal K}_{\rm IAF}$ with $\delta_{\rm IAF}$
for small $\delta_{\rm IAF}$ can be obtained
by demanding consistency with the off-critical
scaling form for the conductivity
\begin{equation}
\sigma'(\omega) = {\cal K}_{\rm RS}l_v\ln(\Omega_0/\omega)
\Sigma_{\rm IAF}(\Gamma_{\omega}/\Gamma_{\delta_{\rm IAF}}) \, ,
\end{equation}
which immediately implies that ${\cal K}_{\rm IAF} \sim
\delta_{\rm IAF}^{(2-\psi)/\lambda} \sim z_{\rm IAF}^{-1}$.

\subsubsection{Dynamical conductivity in RD phases}
We now calculate the dynamical spin conductivity in the XX and XXX
Random Dimer phases.  Here, the same singlet RG can be employed 
all the way across the crossover scale 
$\Gamma_\delta \equiv 1/|\delta|$, and into the energy regime 
of a well-developed RD phase.  The dynamical conductivity is 
given by the same expression Eq.~(\ref{RSsigma}) as for the
RS states: we simply add contributions from the even ($P^{\rm e}$) 
and the odd ($P^{\rm o}$) bonds in complete analogy with the 
calculation of the dynamic structure factor.  Using the scaling 
solutions of Ref.~\onlinecite{fising}, it is quite simple to 
calculate the full scaling function for the dynamical conductivity
\begin{equation}
\sigma'(\omega,\delta)={\cal K}_{\rm RS} l_v \ln(\Omega_0/\omega) \,
\Sigma_{\rm RD} \left( |\delta| \ln(\Omega_0/\omega) \right) \, .
\end{equation}
Here, we restrict ourselves to noting that
$\Sigma_{\rm RD}(x) \sim {\rm const} \;$ 
for $x \ll 1$, while for $x \gg 1$, $\Sigma_{\rm RD}(x)$ scales
as $ \Sigma_{\rm RD}(x) \sim x e^{-2x} \;$.  Thus, at frequencies 
$\omega$ well below the crossover scale $\Omega_\delta$, we have
\begin{eqnarray}
\sigma'(\omega) & = &
{\cal K}_{\rm RD} l_v (\omega/\Omega_0)^{1/z_{\rm RD}} 
\ln^2(\Omega_0/\omega) \, ,
\label{RDsigma}
\end{eqnarray}
with the numerical prefactor ${\cal K}_{\rm RD} \sim |\delta|$
and the dynamical exponent $z_{\rm RD} \sim |\delta|^{-1}$
for small $|\delta|$.  We can now interpret this form
directly in terms of the rare regions that dominate the 
conductivity:  Assume, for concreteness, that $\delta > 0$, i.e., 
that the even bonds are dominating; the main contribution to 
the dynamical conductivity at frequency $\omega \ll \Omega_\delta$ 
then comes from the even bonds with effective 
$\tilde J_{\rm e} = \omega$.  Such weak even bonds are generated 
only across rare long regions that are locally in the opposite 
dimerized phase, and these are precisely the regions that dominate 
the low-energy density of states and thus determine the dynamical 
exponent $z_{\rm RD}(\delta)$---this explains the factor 
$\omega^{1/z_{\rm RD}}$ in Eq.~(\ref{RDsigma}).  Moreover, all 
such bonds have a well-defined length proportional to 
$\ln(\Omega_0/\omega)$, which explains the $\ln^2(\Omega_0/\omega)$ 
in Eq.~(\ref{RDsigma}). 
%[Note that the subdominant contribution from the odd bonds,
%which is of order $\omega^{2/z_{\rm RD}} \ln^2(\Omega_0/\omega)$,
%can be also easily interpreted through the appropriate 
%construction of the much more rare regions that produce 
%anomalously strong odd bonds with $\tilde J_{\rm o} = \omega$.]

%%%%%%%%%%%%%%%%%% XXZ-1/2::FERMIONS %%%%%%%%%%%%%%%%%%%%%%%%%%%%%%
\subsubsection{Perspective: spinless interacting fermions
with particle-hole symmetric disorder}
To put these transport properties in perspective, we recall
that the spin-1/2 XXZ chain is equivalent, via the usual 
Jordan-Wigner transformation, to a system of spinless 
interacting fermions with particle-hole symmetric disorder.
More specifically, we write the spin operators
$s^\pm_j \equiv s^x_j \pm i s^y_j$ in terms of fermion
creation~(annihilation) operators $c^\dagger_j$~($c_j$) as
\begin{eqnarray}
s^{+}_j & = & \prod_{j'<j} (1-2n_{j'})\; c^\dagger_j \,,\nonumber\\
s^{-}_j & = & \prod_{j'<j} (1-2n_{j'})\; c_j \, ,
\end{eqnarray}
while $s^z_j = n_j-1/2$ (here $n_j \equiv c^\dagger_j c_j$ 
is the fermion number operator at site $j$).
In this language, ${\cal H}_{\rm XXZ}$ can be written as
\begin{equation}
{\cal H} = \sum_{j=1}^{L-1} \left[ t_j (c^\dagger_{j+1} c_j + 
c^\dagger_j c_{j+1}) + V_j(n_j-1/2)(n_{j+1}-1/2) \right] \, ,
\label{Hferms}
\end{equation}
with $t_j = J^\perp_j/2$ and $V_j = J^z_j$.  The coupling $J^z$ 
thus controls the strength of the nearest-neighbor 
particle-hole--symmetric repulsive interaction between the 
fermions.  The IAF phase that obtains for large $J^z$ corresponds 
to a charge density wave state stabilized by interactions.  
In the absence of interactions (XX chain) we obtain a free-fermion 
random-hopping problem at zero chemical potential.
  
This free-fermion problem has been extensively studied in the past,
and is known to have rather unusual localization properties due to 
the additional particle-hole symmetry present.~\cite{er,lbmpaf} 
For instance, an elementary calculation immediately reveals that 
the zero-temperature average Landauer conductance $[g_L]_{\rm av}$ 
of a finite segment of length $L$ connected to perfect leads 
scales as $[g_L]_{\rm av} \sim 1/\sqrt{L}$, in sharp contrast to 
the usual exponentially-localized behavior in one dimension; 
the corresponding conductivity, of course, scales as $\sqrt{L}$. 
Now, the strong-disorder RG predicts that lengths scale as 
the square of the logarithm of the energy scale in the low-energy 
effective theory describing the XX-RS state---our result for 
the dynamical conductivity is thus consistent with the elementary 
Landauer calculation (see also our explicit finite-size scaling 
calculations of Appendix~\ref{finite_size}).~\cite{wrongsigma}  
Notice, however, that our approach is not limited to 
the non-interacting case.  It allows us to reliably treat 
the effects of interactions, and follow the dynamical 
conductivity through a `metal-insulator' transition that is 
driven by strong interactions in the presence of strong disorder.

%%%%%%%%%%%%%%%% XXZ-1/2::NUMERIC_SIGMA %%%%%%%%%%%%%%%%%%%%%%%%%%
\subsubsection{Numerical study of the dynamical conductivity
at the XX fixed point}
\label{numeric_sigma}
At the XX point, the Hamiltonian Eq.~(\ref{Hferms}) describes 
non-interacting fermions with random hopping amplitudes,
and we are essentially faced with the problem of finding
the low-energy eigenvalues and eigenstates of the corresponding
single-particle Hamiltonian (an \mbox{$L \times L$} matrix operator)
\mbox{${\mathbf H}=\sum_{j=1}^{L-1} 
t_j(|j+1\rangle \langle j|+|j\rangle \langle j+1|)$},
which defines the Schroedinger equation for this problem.
Any fermionic state can then be represented as a Slater 
determinant of the corresponding (normalized) single-particle 
eigenstates $|\phi_\mu \rangle$ with eigen-energies $\epsilon_\mu$.
In the single-particle language, the Kubo formula for 
the conductivity $\sigma'(\omega)$ at zero chemical potential
and at a finite temperature $T$ reads
\begin{eqnarray}
\sigma'(\omega) = \frac{1}{\omega L} &&  \sum_{\mu_1,\mu_2} 
|\langle \phi_{\mu_2}|\sum_j {\mathbf T}(j)|\phi_{\mu_1}\rangle|^2
\, \times \nonumber\\
&& \times \,
[f(\epsilon_{\mu_1}) - f(\epsilon_{\mu_2})]
\delta(\omega - \epsilon_{\mu_2} + \epsilon_{\mu_1}) \, ,
\label{freekubo}
\end{eqnarray}
where \mbox{${\mathbf T}(j) \equiv 
i t_j ( |j\rangle \langle j+1| - |j+1\rangle \langle j| )$}
is the current operator on the link $j$ and
$f(\epsilon) \equiv 1/(e^{\epsilon/T}+1)$.  [This version of
the Kubo formula will also prove useful when we analyze the full 
temperature dependence of the dynamical conductivity in 
Sec.~\ref{FiniteT}.] 

\narrowtext
\begin{figure}[!t]
\epsfxsize=\columnwidth
\centerline{\epsffile{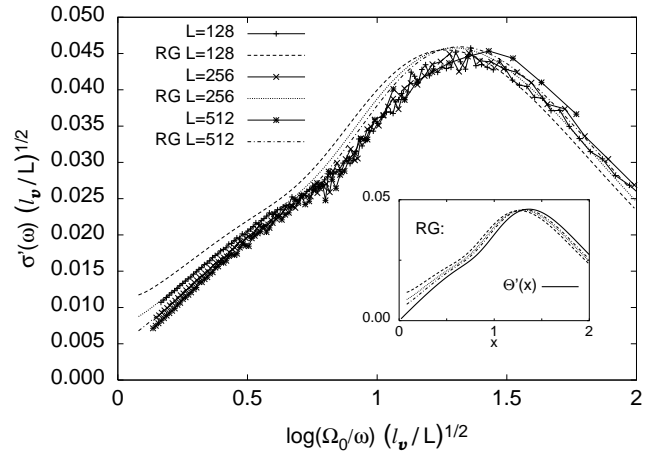}}
\vspace{0.15in}
\caption{Scaling plot of the dynamical conductivity 
$\sigma'(\omega,L)$ at the XX point for finite systems 
of sizes $L=128$, $256$, and $512$, with free boundary 
conditions, calculated by exact numerical diagonalization 
and---for {\it exactly the same} systems---by the finite-size RG 
analysis of Appendix~\ref{finite_size}.  Free-fermion hopping 
amplitudes $t_j$ are drawn independently from a uniform distribution
over $[0,1]$, and the (bare) phase lengths are set to $l_j\!=\!1$.  
We used $\Omega_0 \!=\! 2$ corresponding to the 
initial energy cutoff in the equivalent spin system; also, 
we used $l_v \!=\! 1$ corresponding to the microscopic length 
scale in the problem.  
The agreement of the RG predictions with the results of the exact 
diagonalization is fairly good (given the not so strong initial 
disorder), and the dynamical conductivity is indeed increasing 
all the way to the (finite-size) cross-over scale
$\ln(\Omega_0/\omega) \sim \sqrt{L/l_v}$.  Note that with 
the sizes studied we can only partially access the bulk scaling 
regime $\ln(\Omega_0/\omega) \ll \sqrt{L/l_v}$, in which we expect 
$\sigma'(\omega, L) \sim \ln(\Omega_0/\omega)$ and which 
on the plot is towards the left of the horizontal axis.
Also note that in the opposite regime 
$\ln(\Omega_0/\omega) > \sqrt{L/l_v}$, unlike in the bulk
regime, $\sigma'(\omega, L)$ is not self-averaging; in this 
regime, the plotted average over different samples represents 
roughly the distribution of the lowest gap in the system.  
Inset shows how the $L \!\to\! \infty$ scaling form (thick line)
is approached by the finite-size $\sigma'(\omega,L)$ calculated 
from the RG for the given initial conditions (the lines plotted
here are the same as in the main panel).
Note that the vertical scale is set by the (numerical) prefactor
of the scaling function in the bulk scaling regime (where 
$\sigma'(\omega) \approx \frac{7}{180} \ln(\Omega_0/\omega)$)
}
\label{sigmafig}
\end{figure}

Here, we test the $T=0$ predictions by evaluating 
$\sigma'(\omega)$ using exact numerical diagonalization of 
finite systems.  The results of such calculations for system 
sizes $L=128$, $256$ and $512$ with the hopping amplitudes $t_j$ 
drawn independently from a uniform distribution over $[0,1]$ are 
shown on Fig.~\ref{sigmafig}, where we have averaged over $100,000$ 
samples for each $L$.  In an infinite sample we expect 
the conductivity to diverge logarithmically, but with the system 
sizes studied here, we cannot quite probe this infinite-sample 
regime $1 \ll \ln(\Omega_0/\omega) \ll \sqrt{L}$---rather, we are 
in the regime $1 \ll \ln(\Omega_0/\omega) \alt \sqrt{L}$.
Nevertheless, the numerical results of Fig.~\ref{sigmafig}
clearly show that the dynamical conductivity is increasing 
as the frequency is lowered all the way to the crossover scale 
$\ln(\Omega_0/\omega) \sim \sqrt{L}$, thus supporting our claim 
that $\sigma'(\omega)$ diverges logarithmically at low frequencies.

For a more detailed test of our theoretical results, we need to 
quantitatively analyze the effects of a finite system size on our 
predictions for the dynamical conductivity.  The calculation is 
summarized in Appendix~\ref{finite_size}.  Here, we only note that 
this analysis allows us to write the following scaling form for 
the conductivity
\begin{equation}
\sigma'(\omega, L) = l_v \ln(\Omega_0/\omega)
\Theta\left(l_v \ln^2(\Omega_0/\omega)/L \right) \, ;
\label{Lsigma_scale}  
\end{equation}
the scaling function $\Theta$ is characterized in 
Appendix~\ref{finite_size}, and the above result is expected 
to hold for large enough $L$ and $\ln(\Omega_0/\omega)$ (with no
restrictions on the ratio $\ln^2(\Omega_0/\omega)/L$).
However, the numerical results cannot be compared directly 
with this scaling result since it assumes that the distribution 
of bond lengths has reached the form characteristic of the XX 
fixed point, which is not the case for the sizes that we can 
diagonalize numerically: the ``length part'' of the distribution 
$P(\zeta,l|\Gamma)$ is still evolving towards the corresponding 
scaling form from the initial condition 
$P(\zeta,l|\Gamma_I) = e^{-\zeta}\delta(l-1)$.
Nevertheless, we can compare the results of the exact numerical 
diagonalization with the (formal) predictions of the RG for the 
{\it same} systems.  This can be done by either running the RG 
on the same samples or by evaluating the analytical (within the RG) 
expression, given by the inverse Laplace transform 
Eq.~(\ref{InvLaplace}) for these initial conditions.  
In Fig.~\ref{sigmafig}, we compare the RG result obtained in 
this manner with the numerically evaluated conductivity.
Given that the initial disorder is not very strong, the agreement 
of the RG predictions with the $\sigma'(\omega,L)$ from the exact 
diagonalization is fairly good.

%%%%%%%%%%%%%%%%%%%% XXZ-1/2::VALIDITY %%%%%%%%%%%%%%%%%%%%%%%%%
\subsection{On validity of results}
\label{xxzvalidity}
So far, our calculations have relied on the leading-order
renormalizations of the spin operators; in this subsection we 
will try to justify validity of this approximation.  
We will not address the corresponding question for the RG
itself---this has been analyzed with great care in 
Refs.~\onlinecite{fising}~and~\onlinecite{fraf},
and we have nothing to add here.  Instead, we focus on issues 
specific to our calculation of dynamical quantities.  Here, 
we provide a (partial) justification of our leading-order 
results by analyzing the effects of the first corrections to 
the leading-order expressions for the renormalized 
operators---this can be done consistently within the framework 
of the RG approach.  Any consistent analysis of further 
corrections would require that we also consider higher-order 
corrections to the RG rules themselves, and we stop well short 
of doing that.~\cite{hfs}

As an illustrative example, we consider the dynamic structure factor 
in the RS states.  
Our leading-order calculations used only the zeroth-order 
result for the renormalized spin operators.  The renormalized 
operators can also be easily worked out to first order---these 
were considered in Ref.~\onlinecite{fraf} in the discussion of 
typical correlations.  When a pair of spins $2$ and $3$ 
connected by a strong bond is frozen into a singlet state, 
the neighboring spin operators $\vec s_1$ and $\vec s_4$ 
do not change even to first order, while the spin operators 
$\vec s_2$ and $\vec s_3$ renormalize to
\begin{eqnarray}
(s^z_2)^{\rm eff} = -(s^z_3)^{\rm eff} & = &
      -\frac{J^z_1}{2 J^\perp_2} s^z_1
      +\frac{J^z_3}{2 J^\perp_2} s^z_4 \, , \nonumber \\
(s^+_2)^{\rm eff} = -(s^+_3)^{\rm eff} & = &
      -\frac{J^\perp_1}{J^\perp_2+J^z_2} s^+_1
      +\frac{J^\perp_3}{J^\perp_2+J^z_2} s^+_4 \, .
\label{1orderren}
\end{eqnarray}
Thus, the decimated spins $\vec s_2$ and $\vec s_3$ obtain 
small ``components'' of order $J_1/J_2$, $J_3/J_2$, onto 
the neighboring spins $\vec s_1$ and $\vec s_4$.  As we 
run the RG and renormalize down to scale $\Gamma$, the system  
consists of $n_\Gamma$ active spins per unit length, separated 
from each other by ``dead'' regions (with lengths of order 
$\Gamma^2$) of decimated spins.  Each decimated spin $r$ in 
the dead region between two remaining active spins $j$ and $k$
(where $j$ and $k$ are nearest neighbors in the effective 
Hamiltonian at scale $\Gamma$) will have some components 
$C_{rj}$ and $C_{rk}$ on spins $\vec s_j$ and $\vec s_k$.  
>From the point of view of calculating the spectral 
sum~Eq.~(\ref{struc}), each active spin $j$ acquires some 
($k$-dependent) effective moment $\tilde \mu_j(k)$ coming 
from all decimated spins with non-zero components on $\vec s_j$:
\begin{equation}
\tilde\mu_j(k)=1+{\sum_{i_1<r<i_2}\!\!\!\!}' \; 
C_{rj} e^{ik(r-j)} \, ,
\label{mueff}
\end{equation}
where the sum is over all previously decimated spins $r$ between 
the effective neighbors $i_1$ and $i_2$ of the spin $j$, 
$i_1<j<i_2$.  The components $C_{rj}$ are simply the ground-state 
correlations $\langle s_r s_j \rangle$; such {\it typical} 
correlators decay as a stretched exponential 
$-[\ln C_{rj}]_{\rm av} \sim |r-j|^{1/2}$.  Note that the 
characteristic length scale for this decay is the {\it microscopic} 
length scale $l_v$.
%[Here we specialized our discussion to the critical RS states; 
%in a dimerized phase we would have
%$-[\ln C_{rj}]_{\rm av} \sim |r-j|/\xi_{\rm typ}$ with a typical
%correlation length $\xi_{\rm typ}(\delta) \sim l_v/|\delta|$ 
%that is much smaller than an average correlation length
%$\xi_{\rm av}(\delta) \sim l_v/\delta^2$---this would 
%require only minor changes in our analysis.]
It is thus clear that the sum over $r$ in Eq.~(\ref{mueff}) 
converges quickly, and the renormalization of the moment 
$\tilde \mu_j$ away from its bare value of $1$ comes mainly 
from the nearby spins that were decimated early in the RG.  
This renormalization is of order one, but only weakly 
$k$-dependent.

We now analyze the consequences of this renormalization
of the moments for the two scaling forms of the dynamic 
structure factor derived earlier in the limit of low frequencies, 
one in the vicinity of $k=\pi/a$, and the other in the vicinity 
of $k=0$.  First, consider $k=\pi/a + q$, with $|q| \ll l_v^{-1}$.
For such small values of $q$, we can neglect the $q$-dependence
of the moments and evaluate them at $k=\pi/a$.  To evaluate 
the spectral sum~Eq.~(\ref{struc}), we need to add up 
the contributions coming from the strong bonds at scale 
$\Omega_{\rm final}$.  Each strong bond contributes  
$|\tilde\mu_L+\tilde\mu_R e^{iql}|^2$, where $\mu_L$
and $\mu_R$ are the moments (evaluated at $k = \pi/a$)
of the two spins connected by this strong bond. 
We can now proceed in two steps:  First, we fix $l$ and 
average over the moments of all strong bonds with a given length. 
This gives us a quantity $c_1+c_2|1+e^{iql}|^2$ which we now
need to average over the length distribution of the strong bonds;
here $c_1$ and $c_2$ are now some fixed numbers of order one,
since we expect that the main renormalization of each moment
comes from few nearby spins and is roughly independent of 
the lengths of the adjoining bonds.  Thus, we see that 
Eq.~(\ref{S}), with $\Phi$ given by Eq.~(\ref{Phi}), indeed 
describes the dynamic structure factor for $k$ close to $\pi/a$ 
and fixed low $\omega$; the higher-order corrections renormalize 
the overall amplitude by a factor of order one, and also produce 
an ``incoherent'' background of a comparable strength that depends 
only weakly on $k$ (i.e., that changes significantly only when 
$k$ is changed by an amount of order $l_v^{-1}$).

For $k \ll l_v^{-1}$ (i.e in the scaling regime near $k=0$), 
the discussion is very similar; each strong bond contributes 
$|\tilde\mu_L-\tilde\mu_R e^{ikl}|^2$, where the moments are now 
evaluated at $k=0$.  This again gives us a quantity 
$c_1+c_2|1+e^{iql}|^2$ to be averaged over the length distribution 
of the strong bonds.  In general (away from the XXX point), we now 
have to consider the $S^{zz}(k,\omega)$ component separately from 
the $S^{+-}(k,\omega)$ component, since the total $s^z_{\rm tot}$ 
conservation constrains the constant $c_1$ to be 
{\it identically zero} for the case of $S^{zz}(k,\omega)$.  
Thus, in the case of $S^{zz}(k,\omega)$, higher-order corrections 
only produce an order one renormalization to the overall scale 
of our scaling result Eq.~(\ref{Sknear0}); of course, there will 
be additional corrections, but these will vanish faster than the 
scaling result in the low-frequency limit.
In the case of $S^{+-}(k,\omega)$, an inspection of the 
renormalization rules~Eq.~(\ref{1orderren}) shows that 
to this order $c_1$ will be zero for $S^{+-}(k,\omega)$ 
as well, even in the absence of full Heisenberg symmetry;
however, this is not expected to be true in general 
(to all orders), and we expect a small but non-zero background 
to be present in the general case.  Thus, $S^{+-}(k,\omega)$ 
near $k=0$ will in general consist of two parts: the scaling part 
given by Eq.~(\ref{1orderren}) with an order one non-universal
overall scale (this part vanishes as $\sim k^2$ for small $k$),
and a non-scaling weakly $k$-dependent additive background 
of the same order as the scaling part.

The above arguments typify the general logic behind our 
justification of the leading-order results for all of our 
calculations; in some cases such a programme can be carried 
out analytically (e.g., for the average boundary spin 
autocorrelations---see also Sec.~\ref{RTFIM_ERRORS}), 
while in other cases we have to be satisfied with arguments 
like the ones presented above.  Such arguments can also be 
bolstered by numerically implementing the higher-order operator 
renormalizations to calculate corrections within the RG to our 
leading order results (indeed, we have confirmed that such 
a numerical check for $S(k,\omega)$ in the Heisenberg model 
is in qualitative agreement with the arguments presented above).
%%%%%%%%%%%%%%%%%%%%%%%%%%%%%%%%%%%%%%%%%%%%%%%%%%%%%%%%%%%%%%%

%%%%%%%%%%%%%%%%%%%%%%%%%%%%%%%%%%%%%%%%%%%%%%%%%%%%%%%%%%%%%%%
%%%%%%%%%%%%%%%%%%%%%% SPIN-1 TRANSPORT %%%%%%%%%%%%%%%%%%%%%%%
\section{Transport in strongly random spin-1 chains}
\label{HS1}
\subsection{Singlet RG description of the phases: A review}
\label{reviewS1}
The strong-randomness quantum critical point, which controls 
the transition from the Haldane state to the Random Singlet state 
in the spin-1 chains, and the immediate vicinity of this critical 
point, can be analyzed by a somewhat extended RG procedure 
introduced in Refs.~\onlinecite{hPhD}~and~\onlinecite{hy}, 
or by a variant of the same used in Ref.~\onlinecite{mgj}.

The basic idea is to replace the original spin-1 chain by 
an {\it effective model} that is argued to describe the
low-energy physics of the original system---as we shall see later,
this effective model can be made plausible by thinking
in terms of a bond-diluted chain (it is also possible to arrive 
at essentially the same model by starting with a random 
antiferromagnetic spin-1 chain and using the 
{\it approximate}~\cite{cheat} RG procedure of 
Ref.~\onlinecite{mgj}).  This effective model is written 
entirely in terms of spin-1/2 degrees of freedom coupled by 
nearest-neighbor Heisenberg exchange couplings.  
All {\it even} bonds are always {\it antiferromagnetic} and 
are drawn from an appropriate distribution of positive bonds, 
while {\it odd} bonds can be of {\it either} sign and are drawn 
from a different distribution.

This effective model can be analyzed using 
the extension of the singlet RG introduced in
Refs.~\onlinecite{hPhD}~and~\onlinecite{hy}.  One begins by
looking for the largest {\it antiferromagnetic} bond in 
the system, say $J_2$ connecting spins 2 and 3; 
this defines our bare energy cutoff $\Omega_0$.  
Further analysis can be split into three cases: 
{\it (i)} If the bonds adjacent to the largest AF bond 
are smaller in magnitude, the two spins are frozen into a singlet
state and an effective coupling $\tilde{J}_{14}$ is generated 
between spins 1 and 4 exactly as in the singlet RG for 
the spin-1/2 chain.
{\it (ii)} If both adjacent bonds are larger in magnitude 
than $J_2$, then spins 1 and 2 and spins 3 and 4 are first 
combined to make effective spin-1 objects (since in this case 
$J_1$ and $J_3$ are necessarily ferromagnetic), 
and these effective spin-1 degrees of freedom are then 
frozen into a singlet state, generating an effective coupling 
$\tilde{J}_{05} = 4J_0 J_4/3J_2$ between spins 0 and 5. 
{\it (iii)} If only one of the adjacent bonds, say $J_3$, 
is larger in magnitude than $J_2$, then spins 3 and 4 are first 
combined into an effective spin-1 object.  The system is then 
frozen into the subspace in which spin 2 and this effective 
spin-1 object are coupled together to form an effective 
spin-1/2 object which we label $s_2$ for consistency of notation.
The corresponding renormalized couplings are given as
$\tilde{J}_{12} = -J_1/3$ and $\tilde{J}_{25} = 2J_4/3$. 
This procedure is now iterated with the energy cutoff 
$\Omega$ being gradually reduced.  It is important to note 
that there is no inconsistency in leaving untouched ferromagnetic
bonds $J < -\Omega$ that are not adjacent to any 
antiferromagnetic bonds at the cutoff scale---we could 
equally well have combined {\it all} pairs of spins connected 
by such strong ferromagnetic bonds into effective spin-1 objects 
at the cost of cluttering up our notation.

A detailed analysis of this iterative procedure can be
summarized as follows:~\cite{hPhD,hy,mgj}
Let $\Gamma \equiv \ln(\Omega_0/\Omega)$ and let $n_\Gamma$ 
be the fraction of active spins at log-cutoff $\Gamma$.
For the even bonds, we introduce the distribution $P(\zeta|\Gamma)$ 
of the corresponding logarithmic couplings
$\zeta \equiv \ln(\Omega/J)$.  For the odd bonds,
let $N(\Gamma)$ be the fraction of odd bonds at scale
$\Gamma$ that are {\it strongly} ferromagnetic with 
$J < -\Omega$; for large $\Gamma$, the remainder of 
the odd bonds are symmetrically distributed about zero
and are therefore described by a distribution for $|J|$ 
that we characterize by the distribution $Q(\zeta|\Gamma)$ 
of the corresponding logarithmic couplings 
$\zeta \equiv \ln(\Omega/|J|)$. 
When $W$, the width of the distribution of the log-exchanges
in the original spin-1 Hamiltonian~Eq.~(\ref{Hs1}), 
exceeds a critical value $W_c$, the system is in 
a spin-1 Random Singlet phase.  In the language of the spin-1/2 
effective model, this RS phase is described by a fixed point with
$P(\zeta|\Gamma)=\Gamma^{-1} e^{-\zeta/\Gamma}$, $N(\Gamma)=1$, 
$n_\Gamma \sim 1/\Gamma^2$, and 
$Q(\zeta|\Gamma) = Q_0e^{-Q_0\zeta}$ for large $\Gamma$ 
($Q_0$ is some non-universal O(1) number).
As $W$ is decreased, the system undergoes a quantum 
phase transition to the so-called Gapless Haldane (GH)
phase; both the quantum critical point and the GH phase
in the vicinity of it are still controlled by strong-disorder 
fixed points.  At the critical fixed point (which is an
infinite-disorder fixed point) we have 
$P(\zeta|\Gamma)=Q(\zeta|\Gamma)=2\Gamma^{-1}e^{-2\zeta/\Gamma}$, 
$n_{\Gamma} \sim 1/\Gamma^3$, and $N(\Gamma) = 1/2$.  
The GH phase in the vicinity of the quantum critical point 
is controlled by a line of fixed points; each point on this line 
is characterized by some constant $P_0$ (which depends on the 
strength of disorder $W$).  At a point labeled by $P_0$, we have
$P(\zeta|\Gamma) = P_0 e^{-P_0\zeta}$,
$Q(\zeta|\Gamma) = Q_0(\Gamma) e^{-Q_0(\Gamma)\zeta}$
where $Q_0(\Gamma) \sim e^{-P_0 \Gamma}$, $N(\Gamma) \to 0$,
and $n_\Gamma \sim P_{0}^3 e^{-P_0 \Gamma}$.
The continuously varying $P_0(W)$ vanishes at the transition as 
$P_0 \sim (W_c - W)^{\nu/3}$, where $\nu$ is the correlation 
length exponent obtained in 
Refs.~\onlinecite{hPhD}~and~\onlinecite{hy} --- the GH phase 
is thus similar to the dimerized phases of the spin-1/2 chains.

\subsection{Spin transport}
\subsubsection{Doing calculations in the effective model}
Before we calculate anything, we need to describe how we think 
about the spin transport in this case.  This is somewhat
non-trivial, for we are working in an effective model of 
spin-1/2 degrees of freedom, and some thought is required to
decide what is the correct quantity to calculate.

For this, we go back for a moment to the original random 
spin-1 chain and review an intuitive construction that leads 
to the effective model in terms of spin-1/2 variables only.  
Consider the case of dilute randomness,~\cite{hPhD} that is, 
consider a uniform spin-1 chain with a small fraction of 
very weak bonds that effectively break the chain into pure 
finite segments weakly coupled with each other.
The low-energy effective degrees of freedom of such a segment 
are two half-spins localized near the two edges of 
the segment---these are the spin-1/2's of the effective model.  
The coupling of the edge spins on neighboring segments is given 
roughly by the original coupling of the two segments, 
and is always antiferromagnetic---these are the even bonds 
of the effective model.  On the other hand, the coupling of 
the two edge spins of the same segment can be either 
antiferromagnetic or ferromagnetic depending on whether 
the length of the segment is even or odd---these coupling 
are represented by the odd bonds in the effective model.

We now need to express dynamical properties of the system 
in terms of these effective spin-1/2 degrees of freedom. 
In particular, we want to analyze the low-frequency power 
absorption when an oscillating magnetic field with a uniform 
gradient is applied to the system; this will give us 
the dynamical conductivity $\sigma'(\omega)$.
Since the magnetic field couples to the conserved `charge'
in the system, the corresponding current operators that we
need to use when working out the Kubo formula for the
effective model are uniquely determined by spin conservation:
The current operator on the odd bonds connecting 
the edge half-spins $\vec s_1$ and $\vec s_2$ of 
the same segment (which represents the
total spin current operator of this segment) is 
$\vec \tau = J_{12} l_{12} \vec s_1 \times \vec s_2$;
here $J_{12}$ is the corresponding effective coupling
and $l_{12}$ is some effective {\it phase length}
that we expect to be given roughly by the length of 
the segment.  Naturally, the current operators on 
the even bonds connecting the edge half-spins of the 
neighboring segments have a similar form (the argument 
in this case is even simpler: one only needs to know that 
the true edge spin-1 operator of a segment ``projects'' 
onto the corresponding effective edge spin-1/2 operator with 
an amplitude of order one).  Note that the precise values 
of the phase lengths in the initial effective model 
(for the dilute spin-1 chain) are not important, since at 
still lower energies we expect the distributions of 
couplings and the corresponding bond lengths to approach
some universal distributions characteristic of the 
appropriate fixed point.

\subsubsection{Dynamical conductivity}
Having identified the appropriate
current operators in the effective problem, we now work out 
the rules that govern their renormalization in the RG scheme 
used to analyze this effective model.  As in the spin-1/2 case, 
and as discussed above, we write the part of the total current 
operator (in the spectral sum Eq.~(\ref{sigmakubo})) that is 
associated with a given bond $(j,j+1)$ in the form 
$l_j \vec\tau_j$ where $\vec\tau_j$ is the usual bond operator
$\vec\tau_j = J_j \vec s_j \times \vec s_{j+1}$ and $l_j$
is the appropriate phase length.  We can then follow 
renormalizations of the needed operators by keeping track 
of the phase lengths, in addition to the various bond-strengths.
Unlike the spin-1/2 chains, these phase lengths need not equal 
the physical distances between the corresponding spins; in fact, 
even the physical position of an effective half-spin often cannot 
be specified unambiguously, as, for example, when this half-spin 
appears as an effective doublet formed by combining 
(via a strong AF bond) an effective spin-1 (which is an 
intermediate construction in the Hyman-Yang RG rules) 
and a neighboring spin-1/2.  In such cases, our rules can 
actually be used to assign some meaning to the physical position 
of such an effective half-spin.

The rules for the phase lengths can be easily stated:
In the cases {\it (i)} and {\it (ii)}, when in the final step 
we form a singlet from either two spin-1/2 objects or two spin-1 
objects, the phase length of the new effective bond is simply 
the sum of the phase lengths of all the bonds that are eliminated.
In the case {\it (iii)} the phase lengths associated with 
effective bonds $\tilde J_{12}$ and $\tilde J_{25}$ are 
\mbox{$\tilde l_{12} = l_1 + (4/3)l_2 + (2/3)l_3$} and
\mbox{$\tilde l_{25} = l_4 - (1/3)l_2 + (1/3)l_3$} respectively.

The rules for the phase lengths in the case {\it (iii)}
are somewhat unusual---for example, negative phase lengths
can be produced.  Note, however, that there are many factors
that prevent this from happening too often, and the phase
lengths will in many instances coincide with the corresponding 
geometrical lengths: decimations in the cases {\it (i)} 
and {\it (ii)} tend to ``correct'' deviations of the phase
lengths from the geometrical lengths, and in both the RS and GH 
phases there are simply no decimations of type {\it (iii)} at 
low enough energies.  Also, the lengths $l_2$ and $l_3$ in the above
rule for the case {\it (iii)} are the lengths of the strong bonds
that are eliminated and are therefore usually smaller than 
the lengths $l_1$ and $l_4$ of the more typical bonds. 
Finally, one can argue generally that the {\it phase positions} 
of the spins as dictated by the phase lengths 
have to agree---at least roughly---with their 
{\it geometrical positions} as inferred from the order of the 
(remaining) spins in the chain (i.e., from the spin 
labels).~\cite{phase_lengths}  All of this implies that the phase 
lengths are roughly given by the geometrical distances 
between the spins; in particular their scaling with $\Gamma$ 
is given by the inverse of the density of the remaining spins,
$l \sim n(\Gamma)^{-1}$.  

We can now immediately deduce behavior of the dynamical
conductivity in the different phases exactly as in our previous 
calculations for the spin-1/2 model; as the method remains 
the same, and the relevant details about the statistics of 
the fixed point Hamiltonians have already been summarized, 
we merely state our results.

In the RS phase the same result Eq.~(\ref{RSsigma}) applies,
as is true for an RS state of an arbitrary-$S$ spin chain
at strong enough randomness.  

At the critical point separating the RS phase from the GH phase, 
we find
\begin{equation}
\sigma'(\omega) = {\cal K}_{\rm HY} l_v \ln^2(\Omega_0/\omega) \, ,
\end{equation}
which is a {\em stronger divergence} than in the RS phase
(note that this difference from the result in the RS states 
can be traced to the fact that the density of remaining spins
behaves as $n_\Gamma \sim 1/\Gamma^3$ at the critical point, 
in contrast to the $1/\Gamma^2$ decay of the corresponding 
quantity in the RS states).

Finally, In the GH phases parametrized by $P_0$ we find
\begin{equation}
\sigma'(\omega) = {\cal K}_{\rm GH} l_v 
(\omega/\Omega_0)^{1/z_{\rm GH}} \ln^2(\Omega_0/\omega) \, ,
\end{equation}
where we have introduced the continuously varying 
dynamical exponent $z_{\rm GH} \equiv P_0^{-1}$, and 
${\cal K}_{\rm GH}$ is an order one numerical prefactor 
which goes to a constant as $W \to W_c$
(note that the factor $\ln^2(\Omega_0/\omega)$ appears for exactly 
the same reasons as in the RD phases of the spin-1/2 chains: 
the lengths of the singlets that are decimated 
at scale $\omega$ are roughly $\sim \ln(\Omega_0/\omega)$).
%%%%%%%%%%%%%%%%%%%%%% END SPIN-1 %%%%%%%%%%%%%%%%%%%%%%%%%%%%

%%%%%%%%%%%%%%%%%%%%%%%%%%%%%%%%%%%%%%%%%%%%%%%%%%%%%%%%%%%%%%
%%%%%%%%%%%%%%%%%%%%%  QUANTUM ISING  %%%%%%%%%%%%%%%%%%%%%%%%
\section{Dynamics in the random transverse field Ising chain}
\label{RTFIM}
\subsection{Strong-disorder RG description of the phases: A review}
The strong-randomness cluster RG of Ref.~\onlinecite{fising},
from which the low-energy long-distance behavior of a system
near the critical point ($|\delta| \ll 1$) can be obtained, 
proceeds as follows:  One finds the largest coupling in the system,
with energy $\Omega_0 \equiv \max\{h_j,J_j\}$.
If the largest coupling is a field, say $h_2$ on spin $2$, 
this spin is frozen into the $\sigma^x_2 = +1$ ground state 
of the local field term and is eliminated from the system leaving 
an effective coupling $\tilde J_{13} = J_{12} J_{23}/h_2$ 
between the neighboring spins $1$ and $3$.
If the largest coupling is an interaction, say $J_{12}$ between 
spins $1$ and $2$, the two spins are combined into 
one new spin---a cluster---with an effective spin variable 
$\tilde \sigma_{(12)}$ [representing the two classical minimum 
energy states $\sigma^z_1 = \sigma^z_2 = \pm 1$] and an 
effective transverse field $\tilde h_{(12)} = h_1 h_2/J_{12}$; 
the couplings of this new spin to the neighbors remain unchanged 
to leading order.  Each such cluster $c$ has a moment
$\tilde \mu_c$ given by the number of initial spins in the cluster;
when two clusters are combined to form a bigger cluster,
their moments add: $\tilde \mu_{(12)} = \mu_1+\mu_2$.
This procedure is now iterated with the energy cutoff
$\Omega \equiv \max\{\tilde h_j,\tilde J_j\}$ of the new
effective Hamiltonian being gradually lowered.

A detailed analysis of this procedure was given
in Ref.~\onlinecite{fising}, of which a summary follows:
Define the log-couplings 
$\beta_i \equiv \ln(\Omega/h_i)$, 
$\zeta_i \equiv \ln(\Omega/J_i)$, and also the log-cutoff
$\Gamma \equiv \ln(\Omega_0/\Omega)$; also, let $n_\Gamma$ 
be the number density per unit length of the (remaining)
clusters at scale $\Gamma$. 
The essential feature of the RG near the critical point
is that the distributions of the log-couplings 
$R(\beta|\Gamma)$ and $P(\zeta|\Gamma)$ become broader 
and broader as the energy cutoff is lowered;
the RG flows are characterized by a special family of 
scaling solutions with 
$R(\beta|\Gamma) = R_0(\Gamma) e^{-R_0(\Gamma)\beta}$ and 
$P(\zeta|\Gamma) = P_0(\Gamma) e^{-P_0(\Gamma)\beta}$.
At the critical point, $\delta = 0$, we have 
$R_0(\Gamma) = P_0(\Gamma) = 1/\Gamma$; thus 
the widths of the two distributions grow without limit, 
and the number density decreases as $n_\Gamma \sim 1/\Gamma^2$.
Also, magnetic moments of the clusters scale as 
$\mu \sim \Gamma^\phi$, with $\phi = (1+\sqrt{5})/2$. 
In the disordered phase, $\delta > 0$, beyond 
the crossover scale $\Gamma_\delta \equiv 1/|\delta|$, 
the width of the field distribution saturates, with
$R_0(\Gamma) \approx 2\delta$ for $\Gamma \gg \Gamma_\delta$,
while the width of the bond distribution grows without limit,
with $P_0(\Gamma) \approx 2\delta e^{-2\delta\Gamma}$.
In the ordered phase, $\delta < 0$, the situation is reversed:
$R_0(\Gamma) \approx 2|\delta| e^{-2|\delta|\Gamma}$ and
$P_0(\Gamma) \approx 2|\delta|$ for $\Gamma \gg \Gamma_\delta$.  
In both phases, we have
$n_\Gamma \sim |\delta|^2 e^{-2|\delta|\Gamma}$.
Note also that the clusters that are being eliminated at 
scale $\Gamma \gg \Gamma_\delta$ all have a fairly well-defined 
length of order $|\delta|^{-1} \Gamma$ and magnetic moment
of order $|\delta|^{1-\phi} \Gamma$.

\subsection{Average autocorrelations}
In this section, we obtain the long-time asymptotics of average 
imaginary-time autocorrelations in the critical region of 
the RTFIM---we will be using heavily results of 
Ref.~\onlinecite{fising} refering to sections in that paper by, 
e.g., F~Sec.~IVB.  Our predictions can be compared with the 
extensive numerical results~\cite{numerics} available in the 
literature, and this serves as a useful check on the validity of 
our basic approach---the results of such a comparison have already 
been summarized in Sec.~\ref{reviewRTFIM}.

We consider the local dynamical susceptibilities
\begin{equation}
\chi^{\alpha\alpha}_{jj} = \sum_m 
|\langle m| \sigma^\alpha_j |0 \rangle|^2 \delta(\omega-E_m) \, ,
\end{equation}
where the sum is over all excited states $|m\rangle$ with 
excitation energies $E_m$, and $\alpha = x$ or $\alpha = z$.
The low-frequency behavior of these susceptibilities 
determines the long-time asymptotics of the corresponding
imaginary-time autocorrelation functions
\begin{equation}
C^{\alpha\alpha}_{jj}(\tau)=
\langle \sigma^\alpha_j(\tau) \sigma^\alpha_j(0) \rangle \, ,
\end{equation}
with $\alpha = z$ (local magnetic moment autocorrelation) 
or $\alpha = x$ (local energy autocorrelation);
we are considering here only the fluctuating (time-dependent)
parts of autocorrelations and will ignore any constant 
(time-independent) parts (such a constant part in, for example,
spin autocorrelation represents a non-zero magnetization
density in the system and is a static property).
In the following, we simply write $C_{\rm loc}(\tau)$ for 
the local magnetic moment autocorrelations $C^{zz}_{jj}(\tau)$ 
and $C^e_{\rm loc}(\tau)$ for the local energy autocorrelations
$C^{xx}_{jj}(\tau)$ (and similarly for susceptibilities).
We first obtain (using our basic strategy) results for average 
susceptibilities, which can be conveniently defined as
\begin{equation}
[ \chi^{\alpha\alpha}_{\rm loc} ]_{\rm av}(\omega)=
\frac{1}{L} \sum_j \chi^{\alpha\alpha}_{jj}(\omega) \, ,
\end{equation}
where $L$ is the size of the system (in the thermodynamic
limit of $L \to \infty$ this definition coincides
with an ensemble average over disorder realizations). 
We also consider a semi-infinite chain, $j \geq 1$, 
and calculate average dynamical susceptibilities 
$\chi_1^{\alpha \alpha}$ of the boundary spin $\sigma_1$
(in this case, the average is over disorder realizations). 
These results are then immediately translated to the
corresponding statements about the long-time behavior 
of average autocorrelations.  As long as we are only
interested in the asymptotic behavior of the average 
dynamical susceptibilities and autocorrelations,
it suffices to use the leading-order results for the 
renormalization of the corresponding operators---this is 
discussed further in Sec.~\ref{RTFIM_ERRORS}.

We emphasize from the outset that our calculations in this
section are closely related to the discussion in
Ref.~\onlinecite{fising} of static response functions at 
finite temperature $T$:  Such static response properties 
are calculated by assuming that {\it all} effective degrees 
of freedom that are present (in the sense of the RG) at 
energy scale $T$ contribute freely to the response at this
temperature, while in our calculations of the dynamical
properties at frequency $\omega$ only degrees of freedom 
that are being decimated at scale $\omega$ contribute 
to the dynamical response at this frequency; even this difference 
disappears when the dynamical susceptibilities are translated 
to the imaginary time autocorrelations, since average 
autocorrelations at time $\tau$ acquire contributions
from all frequencies smaller than $1/\tau$.
The intent here is merely to present a unified approach, 
{\it within the RG} of Ref.~\onlinecite{fising}, to the 
calculation of such average dynamical properties.
Also, these calculations, together with a detailed physical 
picture developed in Ref.~\onlinecite{fising} of the phases 
of the system near the critical point, serve as a valuable 
guide to our intuition in identifying the relevant Griffiths 
regions that dominate a particular response; on some occasions  
in the previous sections (particularly in the IAF phase of spin-1/2 
chains), such Griffiths arguments were our only 
source of information about the behavior of dynamical quantities, 
and the opportunity to compare such suggestive arguments against the 
results of controlled calculations is most welcome.

\subsubsection{Average local spin autocorrelation 
$[C_{\rm loc}]_{\rm av}(\tau)$}
The leading-order renormalizations of the $\sigma^z$ spin operators 
are particularly simple:  As long as a given spin $j$ is active,
the operator $\sigma^z_j$ is renormalized to the ``spin'' operator 
$\tilde \sigma^z_c$ of the cluster $c$ that the spin $j$ 
belongs to; when this cluster is decimated, the corresponding
operator renormalizes to zero.  

To calculate $[\chi_{\rm loc}]_{\rm av}(\omega)$, we run 
the RG down to energy scale $\Omega_{\rm final} = \omega/2$,
and rewrite the spectral sum in terms of the degrees of freedom 
of the renormalized problem; excitations that contribute to this 
new sum are clearly the $\tilde\sigma^x = +1$ excitations of 
the spin clusters that are being frozen into their 
$\tilde\sigma^x = -1$ states by the transverse fields 
at this scale, and the spectral sum is now easily evaluated:
\begin{eqnarray}
[\chi_{\rm loc}]_{\rm av}(\omega) & = &\frac{1}{L} 
\tilde{\sum_m} \tilde{\sum_c} \tilde\mu_c
|\langle m| \tilde\sigma^z_c |0 \rangle|^2 \delta(\omega-\tilde E_m)
\nonumber \\
& \sim & \frac{n(\Gamma_\omega)}{\omega} 
R_0(\Gamma_\omega) \overline{\mu}_0 (\Gamma_\omega) \, ,
\label{chiZZ}
\end{eqnarray}
where we used the fact that all $\tilde \mu_c$ spins that are 
active in an effective cluster $c$ contribute identically, 
and $\overline{\mu}_0(\Gamma)$ is the average magnetic moment of 
the clusters that are being eliminated at scale $\Gamma$. 
[Note that here, and in the following, we simply write 
$\Gamma_\omega$ instead of $\Gamma_{\omega/2}$ to avoid clutter 
in our notation---since we are interested only in the leading 
behavior, the difference is not important for our purposes].

{\it At criticality} ($\delta = 0$), we obtain
\begin{equation}
[\chi_{\rm loc}]_{\rm av}(\omega) 
\sim \frac{1}{\omega |\ln \omega|^{3-\phi}} \, ,
\end{equation}
for $\omega \ll \Omega_0$.  For the average spin autocorrelation 
in imaginary time $\tau \gg \Omega_0^{-1}$ we then find
\begin{equation}
[C_{\rm loc}]_{\rm av}(\tau) \sim \frac{1}{|\ln \tau|^{2-\phi}} \, .
\end{equation}

In the {\it disordered phase} ($\delta > 0$), we obtain
\begin{eqnarray}
\left[ \chi_{\rm loc} \right]_{\rm av}(\omega) 
& \sim & \delta^{4-\phi} \frac{|\ln \omega|}{\omega^{1-1/z(\delta)}} 
\, ,
\nonumber \\
\left[ C_{\rm loc} \right ]_{\rm av}(\tau)
& \sim & \delta^{4-\phi} \frac{ |\ln \tau| }{\tau^{1/z(\delta)}} \, ,
\label{chiZZdsrd} 
\end{eqnarray}
for $\omega \ll \Omega_\delta$ and 
$\tau \gg \Omega_\delta^{-1}$.  Here, we have used scaling
solutions for the off-critical flows to write the answer
for the local susceptibility and have chosen to express
the power-law in terms of the dynamical exponent $z(\delta)$. 
From the scaling solution to the RG flow equations, we have
$z^{-1} = 2|\delta|$ --- this is to be thought of as the leading 
term in a small-$\delta$ expansion for $z^{-1}$.  
Written in terms of $z(\delta)$, our result Eq.~(\ref{chiZZdsrd}) 
is valid more generally, and can be understood directly in 
the simple picture of the disordered phase given in F~Sec.~IVB:  
The average spin autocorrelation at large time $\tau$ is dominated 
by the (rare) spins that belong to the rare strongly coupled 
clusters (Griffiths regions) with low effective 
``flipping rates'' (i.e., effective transverse fields) smaller 
than $\omega \sim 1/\tau$.  The density of such clusters, 
which is also the density of the most numerous excitations 
at these low energies, is $n(\omega) \sim \omega^{1/z}$ 
(this is fixed by the relationship $\tau \sim l^z$ between 
length and time scales, that serves as the definition of $z$).
Most of these clusters have their effective 
flipping rates between $\omega$ and some fraction of $\omega$, 
and therefore effective moments of order $|\ln\omega|$ 
[since all clusters that are being eliminated at a fixed energy 
scale $\Omega$ have roughly the same magnetic moment proportional 
to $|\ln\Omega|$].  Estimating the contribution of such Griffiths 
regions clearly gives us Eq.~(\ref{chiZZdsrd}) {\it including} 
the factor of $|\ln\tau|$.

Finally, in the {\it ordered phase} ($\delta < 0$), we obtain
\begin{eqnarray}
\left[ \chi_{\rm loc} \right]_{\rm av}(\omega) 
& \sim & 
|\delta|^{4-\phi} \frac{|\ln \omega|}{\omega^{1-2/z(\delta)}} \, ,
\nonumber \\ 
\left[ C_{\rm loc} \right]_{\rm av}(\tau) 
& \sim & 
|\delta|^{4-\phi} \frac{ |\ln \tau| }{\tau^{2/z(\delta)}} \, ,
\label{chiZZord} 
\end{eqnarray}
for $\omega \ll \Omega_\delta$ and $\tau \gg \Omega_\delta^{-1}$.  
In contrast to the case of the disordered phase, 
the interpretation of Eq.~(\ref{chiZZord}) in terms of the picture 
of the ordered phase presented in F~Sec.~IVA is more subtle.  
In the ordered phase, the typical excitations at low energies 
$\omega \ll \Omega_\delta$ are {\it classical}---they are 
``domain walls'' that ``break'' large clusters apart in the 
places where the clusters are held together by weak 
(effective) bonds of strength of order $\omega$.  Such weak 
effective bonds represent the rare large regions 
(Griffiths regions) that are locally in the disordered phase.  
These domain wall excitations are the most numerous excitations 
that define the relationship between the energy and the length 
scales and determine the dynamical exponent $z(\delta)$.  
Such excitations, however, even if they are localized in 
the neighborhood of site $j$, do not contribute to 
$\chi^{zz}_{jj}(\omega)$ since they cannot be ``excited'' 
from the (classical) ground state by the action of $\sigma^z_j$.  
Excitations which do contribute involve {\it much more rare} 
ferromagnetic clusters that are are flipping back and forth 
in isolation, with flipping rates of order $\omega \sim 1/\tau$ 
or slower [of course, we exclude the macroscopic cluster flipping
at a rate of order $e^{-cL}$ as we are subtracting out the 
time-independent part of the autocorrelation]. 
In the RG language, these are precisely the clusters 
that are decimated at energy scales of order $\omega$, 
i.e., that happen to have (at these scales) an anomalously 
strong transverse fields of order $\omega$ [remember that we are 
in the ordered phase].  A simple construction, however, 
clearly shows that the density of such regions is indeed 
$\sim \omega^{2/z}$, as predicted by the scaling solution: 
For such a cluster to occur we need a ferromagnetic segment of 
length $\sim |\ln\omega|$ (which is not rare in the ordered phase) 
that is isolated (from eventually becoming a part of 
the macroscopic cluster) on each side by a disordered region 
of comparable length.  Each of the two disordered regions 
is actually a ``typical'' Griffiths region at these energy scales, 
and the two are required to occur much closer together
than their typical separation~$\omega^{-1/z}$; this explains 
the appearance of the power~$2/z$ in Eq.~(\ref{chiZZord}).  
The factor~$|\ln\tau|$ again comes from the typical
magnetic moments of such ferromagnetic droplets.

\subsubsection{Average local energy autocorrelation 
$[C^e_{\rm loc}]_{\rm av}(\tau)$}
We begin by working out the leading-order renormalizations of 
the $\sigma^x$ operators:  When a given spin~$j$ 
is combined with another spin~$k$ into a new cluster
(i.e., when the strong bond $J_{jk}$ is being eliminated)
the operator $\sigma^x_j$ renormalizes to
$(\tilde h_{(jk)}/h_j) \tilde\sigma^x_{(jk)}\,$, 
where $h_j$ is the transverse field on the spin $j$ 
before the decimation, $\tilde h_{(jk)}$ is the effective 
transverse field on the new cluster $(jk)$, and 
$\tilde\sigma^x_{(jk)}$ is the effective ``spin-flip'' operator 
of this cluster (this rule ignores a constant term proportional 
to the identity operator, which is unimportant for our purposes 
as we are not interested in the time-independent constant
piece of the energy autocorrelation function).  On the other hand, 
when the spin~$j$ is eliminated, the operator $\sigma^x_j$ 
becomes effectively zero to first order in the nearby 
interactions (we again ignore any constants).  Iterating this, 
the operator $\sigma^x_j$ is renormalized to
$(\tilde h_c/h^{(0)}_j) \tilde \sigma^x_c$ if the spin~$j$ is 
active in some cluster $c$ with the effective field $\tilde h_c$,
and is renormalized to zero if the spin is not active;
here $h^{(0)}_j$ is the original (bare) transverse field 
on the spin~$j$.

We now run the RG down to energy scale 
$\Omega_{\rm final} = \omega/2$ and rewrite the spectral sum as
\begin{equation}
[\chi^{e}_{\rm loc}]_{\rm av}(\omega) = \frac{1}{L}
\tilde{\sum_m} \tilde{\sum_c} \, \tilde g_c
|\langle m|\tilde\sigma^x_c |0 \rangle|^2 \delta(\omega-\tilde E_m)
\, ,
\label{chiXXsum}
\end{equation}
where
\begin{equation}
\tilde g_c = \sum_{j\in c} 
\left( \frac{\tilde h_c}{h^{(0)}_j} \right)^2 \, .
\end{equation}
In the last equation $\sum_{j\in c}$ is over all spins 
that are active in a given cluster $c$.  Note that at 
low energies the effective cluster field $\tilde h_c$ is 
only weakly correlated with each of the bare fields 
$h^{(0)}_j$; if the bare field distribution is not too broad, 
we can approximate $h^{(0)}_j \sim \Omega_0$, and write 
$\tilde g_c \sim \tilde\mu_c (\tilde h_c/\Omega_0)^2$,
where $\tilde\mu_c$ is the moment of the cluster $c$.  
Doing this clearly misses some nonuniversal numerical factor of 
order one that depends on the bare (high-energy) physics. 
This factor is, in principle, a random quantity that differs
from one cluster to another; however, this number is expected 
to be roughly the same for all clusters that contribute to the 
spectral sum at low frequencies due to averaging, since 
such clusters are all large, and in some sense similar. 
Thus, we expect that the low-frequency behavior is not affected.
[Note that we would have been spared this discussion if we 
were to analyze the spectral sum with matrix elements of 
$h^{(0)}_j \sigma^x_j$, which is anyway a more natural operator
to consider when thinking of the local energy fluctuations].
The excitations that contribute to the spectral sum 
Eq.~(\ref{chiXXsum}) correspond to transitions from 
the $\tilde\sigma_j^z = \tilde\sigma_k^z$ states to 
the $\tilde\sigma_j^z = - \tilde\sigma_k^z$ states
of two (effective) spins $j$ and $k$ that are being combined 
into one cluster $\tilde\sigma_{(jk)}$ by a strong bond at 
the energy scale $\Omega_{\rm final}$.  Since the log-field 
distribution is broad [we are near the critical point and at 
low energy scales], for such a pair of spins to contribute 
significantly the transverse field on at least one of 
the two spins involved must be of order $\omega$; thus, we have
\begin{equation}
[\chi^{e}_{\rm loc}]_{\rm av}(\omega) \sim
\frac{\omega}{\Omega_0^2} n(\Gamma_\omega) P_0(\Gamma_\omega) 
R_0(\Gamma_\omega) \overline{\mu}_0(\Gamma_\omega) \, ,
\label{chiXX}
\end{equation}
from which we immediately read-off our results:

{\it At criticality}, for $\omega \ll \Omega_0$ and 
$\tau \gg \Omega_0^{-1}$, we obtain
\begin{eqnarray}
\left[ \chi^{e}_{\rm loc} \right]_{\rm av}(\omega) 
& \sim & \frac{\omega}{|\ln \omega|^{4-\phi}} \, , \nonumber \\
\left[ C^{e}_{\rm loc} \right]_{\rm av}(\tau) 
& \sim & \frac{1}{ \tau^2 |\ln \tau|^{4-\phi} } \, .
\label{chiXXcrit}
\end{eqnarray}

{\it Away from the critical point}, both in the disordered phase and 
in the ordered phase, we obtain
\begin{eqnarray}
\left[ \chi^{e}_{\rm loc} \right]_{\rm av}(\omega) 
& \sim & |\delta|^{5-\phi} \omega^{1+2/z(\delta)} |\ln \omega| \, ,
\nonumber \\
\left[ C^{e}_{\rm loc} \right]_{\rm av}(\tau) 
& \sim & |\delta|^{5-\phi} \frac{ |\ln \tau| }
                                { \tau^{2+2/z(\delta)} } \, ,
\label{chiXXoff}
\end{eqnarray}
for $\omega \ll \Omega_\delta$ and $\tau \gg \Omega_\delta^{-1}$;
in the last formula we again used $z(\delta)$ as a more physical 
parameter characterizing the Griffiths phase at a given $\delta$.  
The off-critical energy autocorrelation function thus behaves 
similarly in the two phases, as expected from duality.  It is 
again possible to interpret these results in terms of the 
statistical properties of appropriate rare regions that 
dominate the average energy autocorrelation at long times.  
As the results (and their interpretation) are identical in either 
phase, we sketch only the interpretation on the ordered side:
As we have already noted, for a region to have significant 
energy fluctuations at the frequency scale of order $\omega$, 
it must contain two adjoining segments both having 
a characteristic energy of order $\omega$---a predominantly
disordered segment (in the RG language, an effective bond 
with $\tilde J \sim \omega$ across this segment) and a 
predominantly ordered segment (in the RG language, a cluster 
with $\tilde h \sim \omega$).  Clearly, the predominantly 
ordered segment with effective transverse field $\sim \omega$ 
can exist only if it is also isolated on the other side from 
the rest of the system by another predominantly disordered 
segment having the same characteristic energy scale.  
This situation has already been analyzed in the context of 
the spin autocorrelations in the ordered phase, and clearly 
one recovers precisely Eq.~(\ref{chiXXoff}) from such an analysis.

\subsubsection{Autocorrelations of the boundary spin}
So far, we have calculated average autocorrelations for 
the spins in the bulk; calculations for the first spin $\sigma_1$ 
in a semi-infinite chain $j \geq 1$ proceed analogously,
and we will simply state the results.

For the {\it spin autocorrelation}, we find
\begin{equation}
[\chi_1]_{\rm av}(\omega) \sim
\frac{ E(0|\Gamma_\omega) }{\omega} = 
\frac{ P_0(\Gamma_\omega) R_0(\Gamma_\omega)  }{\omega} \, ;
\label{chiZ1}
\end{equation}
here $E(\beta|\Gamma) d\beta$ is the probability that 
$\sigma_1$ survives to scale $\Gamma$ and is in a cluster
with $\ln(\Omega/h) = \beta$; such properties of the boundary 
spin are fully characterized in F~Sec.~V.
{\it At criticality} we obtain
\begin{eqnarray}
\left[ \chi_1 \right]_{\rm av}(\omega) 
& \sim & \frac{1}{\omega |\ln \omega|^2} \, , \nonumber \\ 
\left[ C_1 \right]_{\rm av}(\tau) 
& \sim & \frac{1}{|\ln \tau|} \, ,
\label{chiZ1crit} 
\end{eqnarray}
while {\it away from the critical point}, we get
\begin{eqnarray}
\left[ \chi_1 \right]_{\rm av}(\omega) 
& \sim & \frac{\delta^2}{\omega^{1-1/z(\delta)}} \, , \nonumber \\ 
\left[ C_1 \right]_{\rm av}(\tau) 
& \sim & \frac{\delta^2}{\tau^{1/z(\delta)}} \, .
\label{chiZ1off} 
\end{eqnarray}
As in the bulk case, we can interpret the average 
off-critical spin autocorrelation Eq.~(\ref{chiZ1off})
in terms of the rare instances that dominate this average.
In this case, the corresponding rare regions must start at 
$\sigma_1$---this explains the absence of a $|\ln\tau|$ 
factor in Eq.~(\ref{chiZ1off}) compared to the bulk results
Eqs.~(\ref{chiZZdsrd})~and~(\ref{chiZZord}).  The only other 
difference is that in the ordered phase we do not need to isolate 
the ferromagnetic droplet (containing $\sigma_1$) from the left.

For the {\it energy autocorrelation} we find
\begin{equation}
[\chi^e_1]_{\rm av}(\omega) \sim \frac{\omega}{\Omega_0^2} 
P_0^2(\Gamma_\omega) R_0(\Gamma_\omega) \, .
\label{chiE1}
\end{equation}
{\it At criticality} we obtain
\begin{eqnarray}
\left[ \chi^e_1 \right]_{\rm av}(\omega) 
& \sim & \frac{\omega}{|\ln \omega|^3} \, , \nonumber \\ 
\left[ C^e_1 \right]_{\rm av}(\tau) 
& \sim & \frac{1}{ \tau^2 |\ln \tau|^3 } \, ,
\label{chiE1crit}
\end{eqnarray}
while in the {\it disordered phase}
\begin{eqnarray}
\left[ \chi^e_1 \right]_{\rm av}(\omega) 
& \sim & \delta^3 \omega^{1+2/z(\delta)} \, , \nonumber \\ 
\left[ C^e_1 \right]_{\rm av}(\tau) 
& \sim & \frac{\delta^3}{\tau^{2+2/z(\delta)}} \, ,
\label{chiE1dsrd}
\end{eqnarray}
and in the {\it ordered phase}
\begin{eqnarray}
\left[ \chi^e_1 \right]_{\rm av}(\omega) 
& \sim & |\delta|^3 \omega^{1+1/z(\delta)} \, , \nonumber \\ 
\left[ C^e_1 \right]_{\rm av}(\tau) 
& \sim & \frac{|\delta|^3}{\tau^{2+1/z(\delta)}} \, .
\label{chiE1ord}
\end{eqnarray}
Thus, the  average off-critical energy autocorrelation of the
boundary spin differs from that of the bulk spins in exactly 
the same way (and for the same reasons) as in the case of 
the spin autocorrelation.

\subsection{Dynamic structure factor of the spins}
\label{SdynRTFIM}
Let us now briefly consider the dynamic structure factor 
$S^{zz}(k,\omega)$ defined as
\begin{equation}
S^{zz}(k,\omega) =
\frac{1}{L}\sum_m |\langle m| \sum_j e^{ikx_j} 
                                     \sigma^z_j |0\rangle|^2
\delta(\omega - E_m) \,.
\end{equation}
$S^{zz}(k,\omega)$ characterizes the spatial structure of 
the excitations at energy $\omega$.  Proceeding as before,
we find
\begin{equation}
S^{zz}(k,\omega) \sim \frac{n(\Gamma_\omega)}{\omega} 
R_0(\Gamma_\omega) \overline{|\mu_0(k)|^2}(\Gamma_\omega) \, ,
\end{equation}
where $\overline{|\mu_0(k)|^2}(\Gamma)$ is the average modulus
squared of the effective magnetic moment at wavevector $k$
for the clusters that are being eliminated at scale $\Gamma$; 
for a given cluster $c$, this effective moment is defined
as $\mu_c(k)=\sum_{j\in c} e^{ikx_j}$.  The dynamic 
structure factor can also be written in terms of the function 
$D(\beta,x|\Gamma_\omega)$ defined in F~Sec.~IIIB4; we have
\begin{equation}
S^{zz}(k,\omega) \sim \frac{ \hat{D}(0,k|\Gamma_\omega) }{\omega},
\end{equation}
where $\hat{D}(0,k|\Gamma_\omega)$ is the Fourier transform of
$D(0,x|\Gamma_\omega)$ at wavevector $k$.  We have not attempted
to analyze $D(\beta,x|\Gamma_\omega)$, even though a detailed
characterization is likely to be possible (see F~Sec.~IIIB4).
Instead, we will only analyze the behavior of the dynamic 
structure factor in some limiting cases using the scaling
picture.

First, consider the system at criticality.  Fix wavevector $k$.
Then, for $\Gamma \ll \Gamma_k \equiv 1/\sqrt{k}$ the effective 
cluster moments at wavevector $k$ ``add coherently'' 
(more precisely, the real parts of the effective moments of 
the clusters that are being combined into bigger clusters are 
of the same sign) and therefore scale as $\Gamma^\phi$.  
At scales $\Gamma \gg \Gamma_k$, the effective moments at $k$ 
``add incoherently'' (the real parts of the moments being combined 
can be of any relative sign) and therefore scale as 
$\Gamma_k^\phi (\Gamma/\Gamma_k)^{\phi_{\rm sym}}$, where 
$\phi_{\rm sym}=(1+\sqrt{5})/4$ is the growth exponent for
the cluster moments distributed symmetrically around zero
(see Appendix of Ref.~\onlinecite{fraf}).  Thus, we arrive
at the following scaling form for the dynamic structure factor
at criticality
\begin{equation}
S^{zz}(k,\omega) \sim \frac{ \Gamma_\omega^{2\phi} }
                           { \omega \Gamma_\omega^3 }
\Phi \left( k \Gamma_\omega^2 \right) \, ,
\end{equation}
where $\Phi(x) \sim {\rm const}$ for $x \ll 1$ and 
$\Phi(x) \sim 1/x^{\phi-\phi_{\rm sym}}$ for $x \gg 1$.
We cannot, however, address the regime 
$k\Gamma_\omega^2 \sim 1$ by such a scaling analysis.

Now, consider the system that is not critical, either in the
disordered or in the ordered phase, in the regime 
$\Gamma_\omega \gg \Gamma_\delta$.  The length and the 
magnetic moment of a cluster that is eliminated 
at scale $\Gamma \gg \Gamma_\delta$ are sharply defined: 
$l_0(\Gamma) = c_l (\Gamma/\Gamma_\delta) \Gamma_\delta^2 
+ O(\sqrt{\Gamma/\Gamma_\delta} \Gamma_\delta^2)$ and
$\mu_0(k=0,\Gamma) = c_\mu (\Gamma/\Gamma_\delta) \Gamma_\delta^\phi
+ O(\sqrt{\Gamma/\Gamma_\delta} \Gamma_\delta^\phi)$, where
$c_l$ and $c_\mu$ are numerical constants of order one.
Such a cluster has some internal structure on the length 
scales below the correlation length $\sim \delta^{-2}$, but
``looks'' fairly uniform on larger length scales.
Then, for the wavevectors $k \ll \delta^2$ we have 
\begin{equation}
\overline{|\mu_0(k)|^2}(\Gamma) \sim \frac{\delta^{4-2\phi}}{k^2}
\left[ 1 + \cos(c_l k \Gamma/|\delta|)
           e^{-c k^2 \Gamma/|\delta|^3} \right] \, ;
\end{equation}
note the oscillatory behavior at $k\sim |\delta|/\Gamma$ due to
the ``sharpness'' of the lengths $l_0$ of the clusters;
the gradual suppression of this oscillatory behavior at 
larger wavevectors comes from the uncertainty in $l_0$, which
is much smaller than $l_0$ itself.  To obtain the dynamic 
structure factor $S^{zz}(k,\omega)$ we simply need to multiply 
this ``cluster structure factor'' by the density $\rho(\omega)$ of 
such clusters at energy $\omega$: 
$\rho(\omega) \sim \delta^3/\omega^{1-1/z(\delta)}$ in the disordered
phase and $\rho(\omega) \sim \delta^3/\omega^{1-2/z(\delta)}$ in 
the ordered phase.

\subsection{On the validity of the results}
\label{RTFIM_ERRORS}
Our analysis in Sec.~\ref{xxzvalidity} of the effects 
on the (calculated) dynamic spin structure factor 
of higher-order corrections to the spin operator 
renormalization rules in the spin-1/2 antiferromagnetic 
chains can be carried over to the quantum Ising spin chain 
as well.  Such higher-order corrections will only renormalize 
the effective magnetic moments of the remaining clusters 
by numerical factors of order one---this is also discussed 
in detail in F~Sec.~VIA.  We do not repeat such analysis here.
Instead, we consider the effect of such corrections on
the average autocorrelations of the boundary spin $\sigma_1$. 
This is somewhat more tractable than the bulk case, and 
we can actually include these effects in an explicit calculation.  
Our results are not surprising: these next-order corrections 
only affect the values of some non-universal prefactors, 
and have no effect on the time dependence of the autocorrelations
in the long-time limit.

The first-order renormalization of the boundary spin 
operator $\sigma^z_1$ is discussed in detail in F~Sec.~VB
in the study of the end-point magnetization, and also 
in Ref.~\onlinecite{fyoung} in the study of the end-to-end
correlations.  Following these references, we write the component 
of the boundary spin on the left-most remaining 
cluster $\tilde \sigma_l$ as 
$\langle \sigma^z_1 \tilde\sigma^z_l \rangle = e^{-\Lambda}$.
Then, for the special family of scaling solutions, 
the distribution of the parameter $\Lambda$ at scale $\Gamma$ 
is described by the probability density
\begin{equation}
{\cal L}(\Lambda|\Gamma)=
\frac{P_0(\Gamma)}{P_0(\Gamma_I)} \delta(\Lambda)
+\frac{P_0(\Gamma_I)-P_0(\Gamma)}{P_0(\Gamma_I)} 
P_0(\Gamma) e^{-P_0(\Gamma) \Lambda} \, ,
\end{equation}
where the first term is simply the probability that $\sigma_1$ 
survives down to the scale $\Gamma$ from the initial scale 
$\Gamma_I$ (we already considered this probability in our 
zeroth-order calculation). The average autocorrelation is 
then given as
\begin{equation}
[\chi_1]_{\rm av}(\omega) =
\frac{ P_0(\Gamma_\omega) R_0(\Gamma_\omega)  }
     { \omega \; P_0(\Gamma_I) }
\left[ 1+ \frac{P_0(\Gamma_I)-P_0(\Gamma_\omega)}
               {2+P_0(\Gamma_\omega)}
\right] \, ,
\end{equation}
where we have kept all the numerical factors within the RG; 
the first term comes from the instances when the 
boundary spin survives down to energy scale $\Gamma_\omega$
and is precisely our zeroth-order result Eq.~(\ref{chiZ1}).
In all cases of interest $P_0(\Gamma_\omega) \ll 1$, and the
second term is thus roughly the first term times $P_0(\Gamma_I)$.  
Now, $1/P_0(\Gamma_I)$ is the width of the initial 
distribution of the logarithms of interactions, and is 
therefore some number of order one---the two contributions 
are thus of the same order.  However, the wider this initial 
distribution is, the smaller the second term is relative to 
the first.  The origin of the second term is easily understood: 
the boundary spin $\sigma_1$ may be decimated with some
finite probability at an early stage of the RG; in such a case
$\sigma_1$ will have a significant component on the nearest
surviving spin, which may now be considered as a boundary spin
in the new problem with a somewhat smaller energy cutoff;
the average large-time (low-frequency) autocorrelation
of this new boundary spin will be at least of order 
Eq.~(\ref{chiZ1}) from the instances when this new spin
survives to the energy scale $\omega$.  The above calculation
explicitly shows that the average dynamical response of the spin 
$\sigma_1$ is dominated entirely by the instances when one of 
the spins within some small (of order one) distance from 
the boundary survives down to low energies of order $\omega$.

Thus, we see that our general arguments for the effect of
higher-order terms in the operator renormalization rules
are borne out by this detailed calculation for the boundary spin
autocorrelation.  This suggests that our approach gives
asymptotically exact results (apart from non-universal prefactors)
in the bulk case as well.

%%%%%%%%%%%%%%%%%%%%%%%%%%%%%%%%%%%%%%%%%%%%%%%%%%%%%%%%%%%%%%%
%%%%%%%%%%%%%%%%%%% END QUANTUM ISING  %%%%%%%%%%%%%%%%%%%%%%%%

%%%%%%%%%%%%%%%%%%%%%%%%%%%%%%%%%%%%%%%%%%%%%%%%%%%%%%%%%%%%%%%
%%%%%%%%%%%%%%%%%%% T.NEQ.0 %%%%%%%%%%%%%%%%%%%%%%%%%%%%%%%%%%%
\section{A discussion of $T \neq 0$ properties}
\label{FiniteT}
So far we have calculated various dynamical and transport 
quantities at $T=0$.  These results are clearly valid 
even at $T \neq 0$ so long as the probe frequency 
$\omega$ satisfies $T \alt \omega$.  Unfortunately, 
it is not straightforward to generalize our calculations 
to the complementary low-frequency regime ($\omega \ll T$)
dominated by thermal effects.  There is, however, one exception. 
As mentioned earlier, the spin-1/2 XX chain is equivalent to 
a model of spinless fermions with random nearest neighbor 
hopping and zero chemical potential.  It should come as 
no surprise that the free-fermion nature of this problem 
allows us to straightforwardly calculate some dynamical 
and transport properties at small non-zero temperature.
We begin this section by formulating a fermion analog of 
the RG procedure used for the spin chains.  We will then
use this RG approach to work out the low-frequency, 
low-temperature dynamical conductivity and the $zz$ component 
of the dynamic structure factor for the spin-1/2 XX chain without
any restriction on the relative magnitudes of $\omega$ and $T$ 
(the calculation of the perpendicular component of the 
structure factor for $\omega < T$ is much more complicated, 
and we will only be able to discuss its qualitative
behavior).
Naturally, these results are not at all generic, relying as
they do on the free-fermion nature of the problem.
On the other hand, a weak $J^z$ coupling, which corresponds to 
the nearest-neighbor
interaction between the fermions, is strongly irrelevant in 
the RG sense at the free-fermion point, and the system 
flows to the non-interacting point.  Since this non-interacting 
limit is singular as far as finite temperature transport properties 
are concerned, we have here an example of a ``dangerously irrelevant
operator'', and the important physical question is how this weak, 
irrelevant interaction affects the $T \neq 0$ transport near the 
non-interacting point.~\cite{wfgc}  This is what we turn to at 
the end of this section.

\subsection{Free fermion RG}
\label{fermionRG}
The free-fermion problem ${\cal H} = 
\sum_j t_j(c^\dagger_j c_{j+1} + c^\dagger_{j+1} c_j)$
has been the subject of extensive investigation in the past 
using a variety of techniques (see, e.g., Ref.~\onlinecite{lbmpaf} 
and references therein).  For our purposes, it is most convenient 
to introduce a RG procedure analogous to the singlet RG used in 
the spin problem.  We formulate this procedure directly
in terms of the corresponding single-particle Shrodinger problem
\mbox{${\mathbf H}=\sum_j t_j(|j\rangle \langle j+1| + 
                              |j+1\rangle \langle j|)$};
this RG is, for the case of the Hamiltonian above, essentially 
just an efficient way of (approximately) diagonalizing random 
symmetric tridiagonal matrices with zeroes on the diagonal.
We begin with the observation that the particle-hole 
symmetry of the problem causes eigenstates to occur in pairs, 
with energies $\pm \epsilon$.  The strong-randomness RG proceeds 
by eliminating, at each step, such a pair of states with
energies at the top and bottom of the band:  One finds the largest 
(in absolute value) hopping amplitude in the system, 
say $t_2$ connecting sites $2$ and $3$; this defines
the bandwidth $\Omega_0 = 2 \times \max\{|t_j|\}$
of the original problem.  If the distribution of the $t_j$ 
is broad, the symmetric and antisymmetric wavefunctions living 
on these two sites will be good approximations to eigenstates 
with energies $\pm \Omega_0/2$, as $t_{1/3}$ will typically be 
much smaller in magnitude than $t_2$.  The couplings $t_{1/3}$ can 
then be treated perturbatively, and eliminating the high-energy 
states living on the sites $2$ and $3$ results in an effective 
hopping amplitude $\tilde{t}_1 = -t_1 t_3/t_2$ between the 
neighboring sites $1$ and $4$.  More precisely, in the effective 
Hamiltonian that describes the remaining $L-2$ states, the block 
$1$-$2$-$3$-$4$ is represented as 
$\tilde{\cal H}_{1-4} = \tilde{t}_1 
                       (|\tilde{1}\rangle \langle \tilde{4}| + 
                        |\tilde{4}\rangle \langle \tilde{1}|)$,
where the states $|\tilde{1}\rangle$ and $|\tilde{4}\rangle$
are essentially the original $|1\rangle$ and $|4\rangle$ states
up to $O(t_{1/3}/t_2)$ corrections. 
This rule is essentially identical to the rule for the singlet RG 
at the XX point, as the additional minus sign can be `gauged away' 
in the nearest-neighbor model in one dimension; we will, in fact,
only keep track of the absolute values of the $t_j$.
The distribution of $|t_j|$ in the renormalized problem with 
bandwidth $\Omega$ will thus be the same as the renormalized 
distribution of $J^{\perp}$ at cut-off scale $\Omega$ in 
the singlet RG for the spin problem.  The analysis of the 
asymptotic validity of this approach thus carries over 
unchanged from the singlet RG.

This procedure can therefore be iterated to reach lower and lower 
energies; at each stage we trade in our current problem for a new 
problem defined on two fewer sites.  This new problem will
have the same low-energy eigenvalues as our original problem.
However, evaluating matrix elements of operators between two 
low-energy states requires some care, as the states 
$|\tilde j\rangle$ in terms of which the renormalized problem 
is written are different from the states $|j\rangle$ of 
the original problem.  As in the singlet RG, this is best 
handled by renormalizing the {\em operators} as we go along, 
so that the matrix elements of the renormalized operators 
between the states of the new problem are the same as the 
matrix elements of the bare operators between the corresponding 
states of the original problem.  This allows us to calculate 
various dynamical properties by evaluating the corresponding 
spectral sums exactly as in the spin language.
At $T=0$, this amounts to nothing more than a restatement 
in terms of the fermions of our previous calculations. 
The new language, however, has one important advantage: 
thermal effects are easily incorporated into this framework, 
essentially because the non-interacting nature of the problem 
is made explicit.  [We emphasize again that the RG finds
{\it all} eigenstates of the free-fermion problem.
The corresponding statement can also be made in terms of 
the singlet RG on the XX spin chain: when eliminating 
a pair of spins $2$ and $3$ the effective Hamiltonians in
{\it all} sectors (corresponding to the states $|s_0\rangle$, 
$|t_0\rangle$, and $|t_{\pm 1}\rangle$, of the pair) 
are {\it identical} up to a sign of $\tilde J^\perp_{14}$ 
in the $|t_{\pm 1}\rangle$ sector.]
Moreover, it is now possible to address questions specific to
the free-fermion problem that do not have a natural analog in
the spin language (see Appendix~\ref{fermionApp}).

Finally, we note in passing that this RG procedure can be 
generalized to analyze other particle-hole symmetric free-fermion 
problems in one and two dimensions (which are not immediately 
equivalent to any quantum spin problem) as well as analyze the
general properties of the Bogoliubov-de Gennes equation for 
quasiparticles in a one dimensional superconducting wire in 
the absence of spin-rotation symmetry (the results of such 
an analysis will be published separately).

\subsection{$T \neq 0$ dynamics and transport at the XX point}
\label{dynfiniteT}
Let us begin by working out the full $T$ and $\omega$ dependence 
of the dynamical conductivity, Eq.~(\ref{freekubo}), at the free 
fermion point.  Our first task is to work out the rules 
that govern the renormalization of the current operators 
${\bf T}(j)$.  Assume once again that the hopping element 
$t_2$ has maximum magnitude.  We wish to work out what operators 
we should use in place of ${\bf T}(1)$, ${\bf T}(2)$
and ${\bf T}(3)$ when we renormalize down to lower energies
by eliminating the corresponding two states at the top and
bottom of the band (the other current operators to the left 
and right of this segment will be unchanged to leading order 
by this elimination).  An explicit perturbative calculation 
immediately yields 
\mbox{$\tilde{\mathbf T}(2) = \tilde{\mathbf T}(1/3) = 
i \tilde{t}_1 (|\tilde 1\rangle \langle \tilde 4| - 
               |\tilde 4\rangle \langle \tilde 1|)$};
this is completely analogous to the rule obtained in the spin 
representation, and as before, we will call this operator
$\tilde{\bf T}(1)$ for consistency of notation.

\narrowtext
\begin{figure}[!b]
\epsfxsize=\columnwidth
\centerline{\epsffile{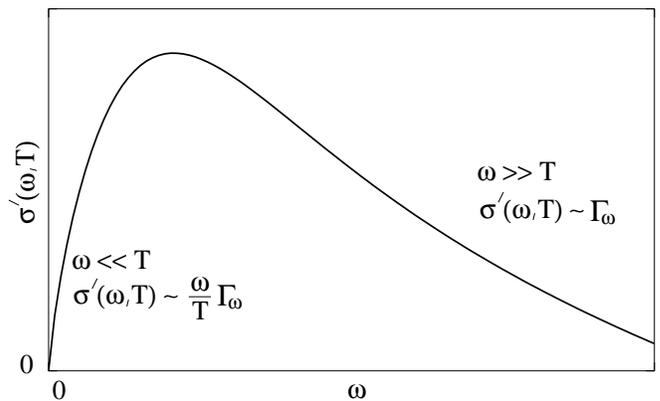}}
\vspace{0.15in}
\caption{A plot of the frequency dependence of 
$\sigma^{\prime}(\omega)$ at low $T \neq 0$ at the XX point.
Note that this result is expected to break down below
a frequency scale $1/\tau_{tr} \ll T$ when a small but 
non-zero $J^z$ interaction is turned on (see 
Sec.~{\protect \ref{smallint}}).}
\label{figTnz}
\end{figure}

As we carry out the RG and reduce the bandwidth by removing
states from the top and bottom of the band, the above
result implies that $\sum_j {\bf T}(j)$ renormalizes to
$\tilde{\sum}_j \tilde l_j \tilde{\bf T}(j)$, where $j$
now labels the sites of the renormalized problem with
bandwidth $\Omega$, and the $\tilde l_j$ are the
lengths of the renormalized bonds in this problem.
With this in hand, we run the RG until the bandwidth 
is reduced to $\Omega_{\rm final}=\omega$ and rewrite 
the spectral sum Eq.~(\ref{freekubo}) as
\begin{eqnarray}
\sigma'(\omega) = \frac{1}{\omega L} &&
\tilde{\sum_{\mu_1 ,\mu_2}} 
|\langle \tilde\phi_{\mu_2}
|\tilde{\sum_j} \tilde l_j \tilde{\bf T}(j)|
 \tilde\phi_{\mu_1} \rangle|^2 \, \times \nonumber \\
&&
  \times \, [f(\epsilon_{\mu_1}) - f(\epsilon_{\mu_2})]
  \delta(\omega - \epsilon_{\mu_2} + \epsilon_{\mu_1}) \, .
\label{freekubo1}
\end{eqnarray}
Because of the extremely broad distribution of the $\tilde{t}_j$,
the dominant contribution to the sum Eq.~(\ref{freekubo1}) comes 
from transitions between the two members, one at the bottom and
the other at the top of the renormalized band, of each
pair of states that is being eliminated at this energy scale.
The matrix element for this transition is just 
$\tilde{l}|\omega|/2$, where $\tilde l$ is the length of 
the hop in question.  In the thermodynamic limit, we thus have
\begin{eqnarray}
\sigma'(\omega) & = & [f(-\omega/2) - f(\omega/2)] 
\times \nonumber \\
&& \times \frac{n(\Gamma_\omega)}{\omega}
\int \!\! dl \, d\zeta \, \omega^2 l^2 \, P(\zeta,l | \Gamma_\omega)
\delta(\omega - \omega e^{-\zeta}) \nonumber \\
& =  &  \frac{\sinh(\omega/2T)}{2\cosh^2(\omega/4T)}
\sigma'_{T=0} (\omega) \, , 
\label{sigmafree}
\end{eqnarray}
which is the leading behavior for $\omega, T \ll \Omega_0$.
This result smoothly interpolates between the logarithmic
frequency dependence seen earlier for $\omega \gg T$ and
the limiting form 
$\sigma'(\omega) \sim \omega \ln(\Omega_0/\omega)/T$ 
valid for $\omega \ll T$---a plot of this frequency dependence
is shown in Fig.~\ref{figTnz}.

Let us now turn to the $T \neq 0$ spin dynamic structure
factor at low frequencies in the vicinity of $k=\pi/a$.  
In the single-particle language, the spectral representation 
for $S^{zz}(k,\omega)$ reads
\begin{eqnarray}
S^{zz}(k,\omega) & = & \frac{1}{(1-e^{-\omega/T}) L}
\sum_{\mu_1, \mu_2} |\langle \phi_{\mu_2}|\hat {\bf S}^z(k)
|\phi_{\mu_1} \rangle|^2 \times \nonumber \\
&& \times [f(\epsilon_{\mu_1}) - f(\epsilon_{\mu_2})]
\delta(\omega -\epsilon_{\mu_2} + \epsilon_{\mu_1}) \, .
\end{eqnarray}
Here $\hat{\bf S}^{z}(k) \equiv \sum_j {\bf S}^z (j)e^{ikx_j}$ 
is the Fourier transform of the position-dependent matrix 
operator ${\bf S}^z (j)$; in the real-space basis 
${\bf S}^z (j) = {\bf n}(j) - 1/2$, where 
${\bf n}(j) = |j\rangle \langle j|$.  This spectral sum can 
also be evaluated within our RG approach.  The leading-order 
operator renormalizations in this case are, in complete 
analogy with the spin problem, very simple: each 
${\bf S}^z (j)$ remains unchanged unless a state living 
on site $j$ is eliminated, in which case ${\bf S}^z (j)$
renormalizes to a multiple of the identity.  As before, we run 
the RG till the bandwidth is reduced to 
$\Omega_{\rm final}=\omega$ and do the spectral sum with
the renormalized operators in the new problem. 
This renormalized sum may be evaluated by again recognizing 
that it is dominated by transitions between pairs of
states with energies $\pm \omega/2$ that live on pairs
of sites connected by `strong' hopping amplitudes (of magnitude
$\omega/2$) in the renormalized problem.  The corresponding 
matrix element is just $(1-e^{ik\tilde l})/2$, where $\tilde l$ 
is the length of the hop in question.  Counting the contributions 
exactly as in our zero-temperature calculations, we thus get
\begin{eqnarray}
S^{zz}(k,\omega) = \frac{1}{(1+e^{-\omega/2T})^2} 
\times S^{zz}_{T = 0}(k,\omega) \, .
\label{strucfree}
\end{eqnarray}
Thus, we see that $S^{zz}$ is essentially unaffected by thermal 
fluctuations at the XX point; in particular, the low-frequency 
divergence is {\it not} cut-off by temperature effects even 
when $\omega \ll T$.

A similar analysis can clearly be performed in the XX-RD phase.
Again, both $\sigma'(\omega)$ and $S^{zz}(k,\omega)$ at
$T>0$ are simply given by the corresponding expressions at
$T=0$ multiplied by simple functions of $\omega/T$, 
exactly as in Eqs.~(\ref{sigmafree})~and~(\ref{strucfree}).
Thus, though temperature effects are simple to work out at the
XX point and in the XX-RD phase, the results are rather special 
due to the free-fermion character of the problem.

\subsection{Going beyond the free-fermion results}
\label{smallint}
What happens when we turn on the nearest-neighbor interaction? 
This is the question we need to address next.

Let us first consider the effects of small $J^z$ couplings 
added on to the XX model.  The analysis of Ref.~\onlinecite{fraf} 
shows that this term is irrelevant in the RG sense; 
the typical value of $\tilde J^z/\tilde J^\perp$ at
log-cutoff $\Gamma$ scales as 
$(\tilde J^z/\tilde J^\perp) \sim u_0 \exp(-c\Gamma^\phi)$,
where $c$ is an $O(1)$ constant, $\phi$ is the golden mean 
$(1+\sqrt{5})/2$, and $u_0$ is the typical value of 
$J^z/J^\perp$ in the microscopic model.  A useful way of thinking 
about the low-frequency behavior of the conductivity is as follows:
Imagine running the RG till the cutoff $\Omega \sim T$. 
In this renormalized problem the typical 
$(\tilde J^z/\tilde J^\perp) \sim u_0 \exp(-c\Gamma_T^\phi)$,
where $\Gamma_T \equiv \ln(\Omega_0/T)$. 
In the fermion language, this is the typical value of the ratio of 
the nearest neighbor interactions to the hopping amplitudes. 
In this renormalized problem, a naive Fermi's Golden Rule
estimate of the corresponding inelastic collision rate due to
interactions gives
$1/\tau_{\rm coll} \sim u_0^2 T \exp(-2c\Gamma_T^\phi)$.
This gives us a frequency scale below which our $T \neq 0$
free-fermion results are expected to break down as a result of
the residual interaction effects.

Unfortunately, we are unable to do a controlled calculation that
determines the transport properties in the frequency regime
$\omega < 1/\tau_{\rm coll}$.  The best we can do is to
work out what a naive scaling argument would predict for the
d.c.~limit of the conductivity.  The basic idea is as follows:
The collision rate may be converted into a corresponding 
dephasing length $L_{\rm coll}$ by appealing to the activated 
scaling that is a characteristic of our problem.  This gives 
$L_{\rm coll} \sim \ln^2(\tau_{\rm coll}) \sim \Gamma_T^{2\phi}$.  
This is the length scale beyond which quantum coherence is lost 
due to inelastic collisions.  Now, we can imagine breaking up 
the system into blocks of length $L_{\rm coll}$.  A d.c.~current 
$I$ passing through the system will see a chain of resistors 
corresponding to these blocks---the resistance values of each of 
these blocks is simply given by the $T=0$ Landauer resistance of 
the corresponding system of length $L_{\rm coll}$.
The voltage developed across a system of total length $L$ will
therefore be $V = I \sum_{j=1}^{L/L_{\rm coll}} R_j 
= I L R_{\rm av}(L_{\rm coll})/L_{\rm coll}$.
Since $R_{\rm av}(L_{\rm coll}) \sim e^{c_1 L_{\rm coll}}$ 
(where $c_1$ is an $O(1)$ scale factor), the d.c.~conductivity 
works out to be
$\sigma_{\rm d.c.} \sim L_{\rm coll} e^{-c_1 L_{\rm coll}} 
\sim \ln^{2 \phi}(\Omega_0/T)e^{-c' \ln^{2 \phi}(\Omega_0/T)}$.
Note that in the absence of interactions, we had earlier
found $\sigma'(\omega) \to 0$ as $\omega \to 0$ at $T>0$---our 
scaling argument implies that interactions render this conclusion 
invalid.  Unfortunately, while this scaling argument is certainly 
plausible, the question of the true low-frequency limit can only 
be settled by a controlled calculation in the regime 
$\omega \ll 1/\tau_{\rm coll}$, which is beyond our current 
capabilities.

The above arguments also suggest that $S^{zz}(k,\omega)$ 
will deviate from the $T \neq 0$ free-fermion result for 
$\omega < 1/\tau_{\rm coll}$.  In particular, one expects 
that the $\omega \to 0$ divergence of $S^{zz}$ would
be cut-off below this frequency scale.  Similar behavior is also
expected of $S^{+-}$, but again, what is really needed is a
controlled calculation as opposed to a scaling argument.
Note also that we expect something different at the XXX and XXZC 
quantum critical points: since the theory at these critical points 
already includes interactions, one expects that 
$1/\tau_{\rm coll} \sim T$, and the relaxational behavior 
characteristic of an interacting system at finite temperature 
will set in for $\omega \sim T$, in contrast to the behavior 
in the vicinity of the XX point.
%%%%%%%%%%%%%%%%%%%%%%% END T.NEQ.0 %%%%%%%%%%%%%%%%%%%%%%%%%%%%%%

%%%%%%%%%%%%%%%%%%%%%%%%%%%%%%%%%%%%%%%%%%%%%%%%%%%%%%%%%%%%%%%%%%
%%%%%%%%%%%%%%%%%%%%%%% EXPERIMENT? %%%%%%%%%%%%%%%%%%%%%%%%%%%%%%
\section{Prospects for experimental tests}
\label{Experiments}
Previous experimental work on one-dimensional random-exchange
Heisenberg antiferromagnetic spin chains has characterized 
the dynamics of these systems in terms of the observed NMR
$1/T_1$ relaxation rate,~\cite{protonrelax} and ESR relaxation 
rates and linewidths.~\cite{rafm1}

As far as the NMR measurements are concerned, our calculations 
are unfortunately not directly relevant to the experimental 
measurements of $1/T_1$.  This may be seen as follows: In the 
usual case of a pure, translationally invariant system, $1/T_1$
is directly related, by Fermi's Golden Rule, to the local 
dynamic structure factor $S_{\rm loc}$ evaluated at frequency 
$\omega$ equal to the nuclear resonance frequency $\gamma_N H$, 
where $\gamma_N$ is the nuclear magnetic moment and $H$ is 
the external field.  In a random system, with a broad 
variation in the value of $S_{\rm loc}(\omega)$, the following 
question immediately arises: what measure of the distribution 
of $S_{\rm loc}(\omega)$ does the experimentally measured 
$1/T_1(H)$ reflect?

Now, we have seen that the average $S_{\rm loc}(\omega)$ 
diverges strongly as $\omega \to 0$ at $T=0$.  Naively, one 
might have thought that this would imply a corresponding 
divergence in $1/T_1$ at small $H$, at least when $T \ll H$.  
However, the divergence in $S_{\rm loc}(\omega)$ comes from 
a few very rare sites which give a very large contribution.  
Clearly, the observed $1/T_1$ will be completely insensitive 
to this effect, since all that will happen is that a tiny 
fraction of nuclear spins (in the neighborhoods of those rare 
electron spins that have significant spin fluctuations at 
the frequency $\omega=\gamma_N H$) will flip almost 
instantaneously, while the rest of the nuclear spins 
will have an extremely small probability to flip, and this is 
what will be reflected in the spin relaxation experiments.
In this sense, it is the typical value of $S_{\rm loc}(\omega)$
that is more relevant for comparisons with NMR $1/T_1$ data.
A typical nuclear spin will in fact have essentially no spin 
fluctuations to couple to at frequency $\omega = \gamma_N H$---it 
can therefore relax only by paying an activation energy
that is set by $\gamma_e H$ (where $\gamma_e \gg \gamma_N$ is
the electron magnetic moment) since the external field acts
to freeze out all modes below this energy scale in most of the system
(with the exception of the rare regions alluded to above).
The experiments actually do see activated behavior for $1/T_1$
at finite temperature. However, the activation gap seems to
scale as $\Delta \sim H^{1.6}$---the rough argument above
of course cannot explain this non-trivial $H$ dependence of the 
observed activation energy.

Our second remark relates to the ESR linewidth measurements
of Clark and Tippie.\cite{rafm1}  Here, again, our results 
do not address the experimentally relevant questions.  
This is because all our calculations for the XXX case are done 
within the context of the simple Heisenberg exchange Hamiltonian,
while the observed linewidth in the experiments is determined
by other effects such as dipolar interactions or anisotropy.
%With full Heisenberg symmetry, there can be no linewidth
%to the ESR line.  The observed linewidth is thus entirely 
%due to effects such as dipolar interactions or anisotropy,
%with the much stronger exchange interactions only counter-acting 
%to narrow the line.  It may be possible to treat this exchange
%narrowing within the random-exchange model, but this is outside
%the scope of this paper.

Inelastic neutron scattering experiments, on the other hand, 
if they can be done on these systems, provide a direct testing 
ground for our predictions.  We conclude with some remarks on 
the relevance of our calculations of the dynamic structure 
factor to such experiments.  First of all, note that we 
considered randomness in the exchanges only, with the spins 
themselves assumed positioned on regular lattice sites; 
thus, our results are restricted to compounds with 
{\it chemical} disorder in exchanges.  It is clear that small 
randomness in the positions of spins (e.g., due to thermal 
fluctuations) will result only in some further suppression 
(by the standard Debye-Waller factor at wavevector $k$) of 
the features relative to an overall background.  In the dimer 
phase, a possible difference in the lengths of even and odd 
bonds will result only in some phase factor in the cosine 
of~Eq.~(\ref{SkoRD}).  Also note that the non-magnetic neutron 
scattering from such spin chains will actually be suppressed 
near $k=\pi/a$, and this may facilitate a possible experimental 
observation of the predicted features.  We caution, however, 
that while it would be extremely interesting to see the sharp 
oscillatory structure predicted in the Griffiths phases, this 
may be difficult to achieve without going to very low 
temperatures and energy transfers.  Regarding transport,
we hope that our results will motivate experiments to probe
the spin conductivity in these systems.
%%%%%%%%%%%%%%%%%%%%%%% END EXPERIMENT? %%%%%%%%%%%%%%%%%%%%%%%%%%

\section{Acknowledgements}
\label{ack}
We thank P.~W.~Anderson, R.~N.~Bhatt, D.~Dhar, D.~S.~Fisher, 
F.~D.~M~Haldane, M.~Hastings, A.~Millis, A.~Madhav, R.~Moessner,
S.~Sachdev, T.~Senthil, S.~L.~Sondhi, and A.~Vishwanath for 
useful discussions. This work was supported by NSF grants 
DMR-9809483 and DMR-9802468.

\appendix
\section{Finite-size scaling function for the conductivity}
\label{finite_size}
Consider a finite chain with an even number of sites $N_I = L+1$, 
where $L$ is the length of the chain, and with free boundary 
conditions (a similar analysis can be carried out for chains 
with an odd number of sites, and also for chains with periodic 
boundary conditions).  We want to calculate the real part of 
the dynamical conductivity averaged over the distribution of bond 
strengths in the limit of low frequencies and large $L$. 
We work this out for the XX chain; the result in the presence of 
$J^z$ couplings will differ only in the values of some non-universal 
scale factors so long as the system does not develop Ising 
antiferromagnetic order in the thermodynamic limit.
To proceed, we need to keep track of the joint distribution at 
scale $\Gamma$ of the number of remaining spins $N$, 
the $N-1$ couplings $\zeta_i$, and the corresponding bond 
lengths $l_i$.  In a finite system, the couplings become 
correlated due to the constraint imposed by the finite length of 
the system.  However, following Fisher and Young,~\cite{fyoung}
we note that the couplings remain `quasi-independent', and can be 
described in terms of the infinite-chain distribution 
$P(\zeta,l|\Gamma)$ exactly as in Ref.~\onlinecite{fyoung}.  
More precisely, if we also keep track of the lengths $l_F$ and 
$l_R$ of the `dead' regions (consisting of singlet pairs formed at 
higher energy scales) at the left and right ends of the chain, 
then a distribution of the form
\begin{eqnarray}
d{\rm Prob}&& \!\!\!\!\!\!\!\!
[N; \zeta_1,l_1 \dots \zeta_{N-1},l_{N-1}; l_F,l_R | L,\Gamma]
\;=~~~~~~~~~~~~\nonumber \\
&=& a_N(L|\Gamma) 
P(\zeta_1,l_1)d\zeta_1 \dots
P(\zeta_{N-1},l_{N-1})d\zeta_{N-1} \nonumber \\
&& \times \, {\cal L}(l_F) {\cal L}(l_R) 
\delta_{l_1+\dots+l_{N-1}+l_F+l_R,L}
\end{eqnarray} 
for even $N \geq 2$ has its from preserved under renormalization 
if $a_N(L|\Gamma)$ is independent of $N$ with
\begin{equation}
\frac{1}{a}\frac{da}{d\Gamma}=2P_0(\Gamma)=2\int dl P(0,l) \, .
\end{equation}
Here, $P(\zeta,l|\Gamma)$ satisfies the same flow equation as 
in the infinite chain, and ${\cal L}(l|\Gamma)$ satisfies
\begin{equation}
\frac{\partial {\cal L}}{\partial \Gamma}={\cal L}(\cdot) *_l 
P(0,\cdot) *_l 
\int_0^\infty P(\zeta,\cdot) d\zeta - P_0{\cal L} \, .
\end{equation}
In the above, the $\Gamma$ dependence is left implicit, and
$f(\cdot) *_l g(\cdot)$ is used to denote a (discrete)
convolution in the length variables.  For clarity, we work 
explicitly with discrete lengths, with $l_F$ and $l_R$ even, 
and $l_i$ odd integers; this is clearly preserved under the RG.

We start the RG with $\Omega_0 = 1$, $\Gamma_I=0$, the initial 
bond distribution $P(\zeta|\Gamma_I)=e^{-\zeta}$ 
(this corresponds simply to choosing the initial $J^\perp$ to be 
uniformly distributed in the interval $[0,1]$), $l_i=1$, 
$l_F=l_R=0$, and $N_I\equiv L+1$; with initial distributions 
$P(\zeta,l|\Gamma_I)$ and ${\cal L}(l|\Gamma)$ normalized to 
unity, the normalization factor is $a(\Gamma_I)=1$.
The dynamical conductivity is now given by
\begin{equation} 
\sigma'(\omega,L)=\frac{1}{4}
\frac{a(\Gamma_\omega)}{L} A(L|\Gamma_\omega) \, , 
\end{equation}
with $a(\Gamma) = (\Gamma +1)^2$ (for our specific choice of
initial conditions), and
\begin{eqnarray}
A(L|\Gamma) & = &\sum_{N=2}^{L+1}\!\!' \;\; (N-1) 
\sum_{l_1,l_2,\dots l_{N-1},l_F,l_R} P(0,l_1)l_1^2 \nonumber \\
&& \times \int_0^\infty P(\zeta_2,l_2)d\zeta_2 \dots 
\int_0^\infty P(\zeta_{N-1},l_{N-1})d\zeta_{N-1}\nonumber \\
&& \times \; {\cal L}(l_F) {\cal L}(l_R) 
\delta_{l_1+l_2+\dots+l_{N-1}+l_F + l_R,L} \, ,
\end{eqnarray}
where the sum is over even $N$.
 
Now, multiplying $A(L|\Gamma)$ by $e^{-yL}$ and summing over 
odd $L\geq 1$---i.e., doing a (discrete) Laplace transform 
in $L$---removes the constraint on the lengths, and we find 
\begin{equation}
A(y|\Gamma)={\cal L}^2(y) Q(y) \frac{1+T^2(y)}{(1-T^2(y))^2},
\end{equation} 
where $Q(y)$ and $T(y)$ are respectively the Laplace transforms 
of $P(0,l)l^2$ and $\int_0^\infty P(\zeta,l)d\zeta$.
Thus, we can straightforwardly work out $A(y)$, given
$P(\zeta,y)$ and ${\cal L}(y)$.  Using the results of 
Refs.~\onlinecite{fyoung}~and~\onlinecite{fising},
we can write the following explicit expressions for these
two functions:
\begin{eqnarray}
P(\zeta,y|\Gamma) &= & Y(y|\Gamma) e^{-\zeta u(y|\Gamma)} \, , \\
{\cal L}(y|\Gamma) &= & \frac{u(0|\Gamma)u(y|\Gamma_I)}
{u(y|\Gamma)u(0|\Gamma_I)} {\cal L}(y|\Gamma_I) \, ,
\end{eqnarray}
where $u(y|\Gamma)=D(y)\coth[D(y)(\Gamma+C(y))]$ and 
$Y(y|\Gamma)=D(y)/\sinh[D(y)(\Gamma+C(y))]$.  The functions 
$D(y)$ and $C(y)$ depend on the initial distribution 
$P(\zeta,y|\Gamma_I)$, and in our case are given by 
$D(y)=\sqrt{1-e^{-2y}}$ and $D(y)C(y)=y+\ln(1+\sqrt{1-e^{-2y}})$.
Also, ${\cal L}(y|\Gamma_I)=1$.

With this in hand, it is a relatively simple matter to work out
$A(L|\Gamma)$, $L$ odd, by performing the inverse Laplace transform:
\begin{equation}
A(L)
%=\frac{1}{2\pi i} \int_{c-i\pi}^{c+i\pi} A(y)e^{yL}dy
=\frac{1}{\pi i} \int_{c-i\pi/2}^{c+i\pi/2} A(y)e^{yL}dy \, .
\label{InvLaplace}
\end{equation}
In Sec.~\ref{numeric_sigma}, we evaluated this integral numerically
to compare the RG predictions with the results of the 
exact-diagonalization studies (note that in the main body of the
paper we didn't make a distinction between $N_I$ and $L$, since
it is irrelevant in the thermodynamic limit; in the more detailed
notation of this Appendix, our numerical results of 
Sec.~\ref{numeric_sigma} are for system sizes $N_I=128$ and $256$).

In the scaling limit $\Gamma \gg 1$, $L \gg 1$, the integral 
Eq.~(\ref{InvLaplace}) for $A(L|\Gamma)$ is dominated by small 
$y$ and can be approximated by 
\begin{equation}
A(L|\Gamma)=2 \,{\mathrm LT}^{-1}\, A(y|\Gamma) \, ,
\end{equation}
where ${\mathrm LT}^{-1}$ denotes the inverse of the continuous 
Laplace transform. Moreover, in this limit, $A(y)$ may be worked 
out using the following scaling forms for ${\cal L}(y)$ and 
$P(\zeta,y)$:
\begin{eqnarray}
{\cal L}(y|\Gamma) & = & \frac{1}
{\Gamma \sqrt{2y} \coth[\Gamma \sqrt{2y}]} \, , \\
P(\zeta,y|\Gamma) & = & \frac{\sqrt{2y}}{\sinh[\Gamma \sqrt{2y}]}
e^{-\zeta \sqrt{2y} \coth[\Gamma \sqrt{2y}]} \, .
\end{eqnarray}
Putting everything together, we can now write 
$A(L,\Gamma)=\Gamma f(\Gamma^2/L)$, which immediately implies 
a scaling form for the conductivity: $\sigma'(\omega, L) = 
\Gamma_\omega \Theta(\Gamma_\omega^2/L)$.
Thus, we see that the dynamical conductivity in a finite
system satisfies a scaling form that reflects the activated 
dynamical scaling at the XX fixed point.  Note that while 
the scaling form holds more generally, the values of the 
non-universal scale factors that we have used are specific 
to our choice of initial distribution.  Analyzing the behavior 
starting with an arbitrary initial distribution allows one 
to relate these non-universal scale factors to the properties 
of the initial distribution {\em under the assumption} that 
`bad decimations' early in the RG do not affect these values. 
Such an analysis allows us to write 
$\sigma'(\omega, L) = l_v \ln(\Omega_0/\omega)
\Theta\left(l_v \ln^2(\Omega_0/\omega)/L \right)$, with the
microscopic energy scale $\Omega_0$ and the microscopic
length scale $\l_v$ precisely as defined in the main text.
Moreover, it is clear that the same scaling function also
describes the low-frequency dynamical conductivity in a large
but finite system  even in the presence of $J^z$ interactions 
as long as the system is in a Random Singlet state---only 
the values of the non-universal scale factors are expected 
to change.

While it is possible to calculate the full scaling function 
$\Theta(x)$ by a detailed analysis of the inverse Laplace 
transform, we will confine ourselves here to working out 
$\Theta(x)$ in two limiting cases: For $x \ll 1$, 
$\Theta(x) = 7/180$ (which correctly reproduces the 
infinite-size result, as it must), while in the limit 
$x \gg 1$ we have $\Theta(x) = e^{-x/2}/\sqrt{2\pi x}$.
This is the result used in the Griffiths argument of 
Sec.~\ref{SigmaXXZ}.

%%%%%%%%%%%%%%%%%%%% fermionApp %%%%%%%%%%%%%%%%%%%%%%%%%%%%%%%%

\section{Low-energy eigenfunctions of the one-dimensional random 
hopping Hamiltonian}
\label{fermionApp}
We have already noted on several occasions that subleading terms
in our evaluation of low-energy properties via the
strong-randomness RG can be treated consistently within
the RG picture, and are naturally related to the {\em ground state 
correlations} in the system.  Here we discuss more explicitly
one example of this connection, for the case of the XX model; 
our discussion is couched completely in the free-fermion language 
and uses the fermionic RG introduced in Sect.~\ref{fermionRG}---this
will allow us to work in a more familiar context and to connect with 
the known results for the fermionic problem.

Our example relates to the structure of the low-energy eigenstates 
of the one-dimensional random hopping Hamiltonian 
\mbox{${\mathbf H}=\sum_{j=1}^{L-1} t_j
(|j+1\rangle \langle j|+|j\rangle \langle j+1|)$};
for simplicity, we work with free boundary conditions.
We note at the very outset that the average Greens functions 
$[G(x,x';E)]_{\rm av}$ have been calculated exactly in 
Ref.~\onlinecite{lbmpaf} using SUSY techniques.  Moreover, 
a great deal is known about the statistics of the low-energy 
wavefunctions, via a sophisticated real space 
analysis.~\cite{dsfunpub}  Here, we present a simple (if somewhat 
rough) picture of the wavefunctions which allows one to easily 
reproduce the previously published SUSY results, as well as 
provides us with the following rough picture of a typical pair of 
wavefunctions at energies $\pm \epsilon$: such a pair is typically 
``localized'' on two sites $i$ and $j$ separated by a distance of 
order $\Gamma_\epsilon^2$, $\Psi_{\pm \epsilon}={\cal N}
(|i\rangle \pm |j\rangle) + \dots$ with ${\cal N}$ of order 1,
with the wavefunction amplitudes decaying as $e^{-c\sqrt{r}}$
away from the two sites for $r \alt \Gamma_\epsilon^2$ and as 
$e^{-c'r/\Gamma_\epsilon}$ for $r \agt \Gamma_\epsilon^2$.
Moreover, as we shall see below, the above picture is intimately
related to corresponding structure in the {\em zero-energy} 
wavefunction.

\begin{figure}[!b]
\epsfxsize=\columnwidth
\centerline{\epsffile{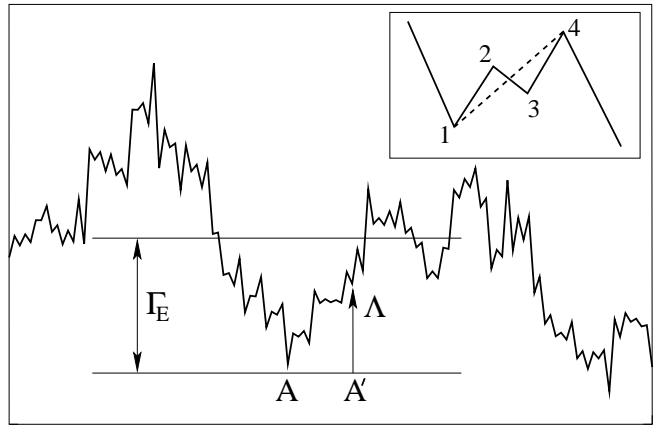}}
\vspace{0.15in}
\caption{Pictorial representation of the bare distribution
of the hopping elements: ``random walker'' steps up by 
$\ln(\Omega_0/t_{2n+1})$ from every odd site $2n+1$ and steps 
down by $\ln(\Omega_0/t_{2n+2})$ from every even site $2n+2$.
Inset shows the basic RG transformation when a pair of 
strongly coupled sites $2$ and $3$ is eliminated.
Site $A$ survives down to energy scale $\Gamma_E$, while
site $A'$ is eliminated before this scale is reached. 
To leading order, the $A'$  has component only onto
the $A$, with the amplitude given by $e^{-\Lambda}$. 
}
\label{walk}
\end{figure}

We start by considering this zero-energy wavefunction.
If the number of sites $L$ is odd, the Hamiltonian~$\mathbf H$
has a (unique) zero-energy eigenstate $\psi$, and this eigenstate
can be written down immediately as $\psi_{2n} \equiv 0$ and 
$\psi_{2n+1}=\Pi_{j=1}^{n} (-t_{2j-1}/t_{2j}) \psi_1$.
Thus, the logarithm of the wavefunction $\ln|\psi_{2n+1}|$
performs a random walk (see Fig.~\ref{walk}).  While the relative
amplitudes of the wavefunction are thus understood directly,
we are interested in the properties of the {\it normalized} 
zero-energy wavefunction $\Psi$, that is,  we also need to 
understand normalization of the eigenvector $\psi$.  
Such properties of the normalized wavefunction $\Psi$ were 
analyzed by Balents and Fisher\cite{lbmpaf} by reducing 
the calculation of correlators to the Liouville quantum 
mechanics.  Here we suggest a more direct approach by noting
that normalization of the $\psi$ for a given realization can be 
achieved by simply setting the largest amplitude
$\psi_{\max}=\max \{\psi_{2n+1}\}$ to a number of order one. 
More precisely, if we set $\psi_{\max}=1$, then for any $q>0$ 
the norm ${\cal N}_{2q}=\sum_{j=1}^L |\psi_j|^{2q}$ is a random
quantity of order one with a well-behaved (``narrow'') 
distribution of width of order one.  This can be seen 
directly by, for example, calculating first moments of the
distribution of ${\cal N}_q$ using the properties of such
a random walk near its global maximum (absorbing boundary).  
With this in hand, we can characterize the zero-energy 
wavefunction essentially quantitatively even without the 
knowledge of the right normalization constant, which is
always of order one.

\narrowtext

Let us now characterize the pairs of wavefunctions at energies 
$\pm \epsilon$.  Consider running the fermion RG on a given system; 
the concrete disorder realization can viewed as a random-walk 
pattern, while the basic RG step is a transformation
on this pattern that removes the smallest-scale structure 
and keeps the rest of the structure unchanged---this
is shown in Fig.~\ref{walk} (note that this is similar
to the RG approach of Ref.~\onlinecite{fmld} to the problem of
random walks in random environments).  In this picture, the sites
that remain down to the energy scale $\Gamma_\epsilon$ are 
completely specified from the initial random-walk pattern 
as follows: an even site $A$ survives down to this energy if
and only if the ``walks'' to the right and to the left
from the site $A$ do not return the same level (of the site $A$) 
until after they crossed the level $\Gamma_\epsilon$ higher 
than the level of the site $A$---see Fig.~\ref{walk}; similar 
conditions can be formulated for odd sites.  Moreover, this
also gives a corresponding criterion for the pairs of sites
on which our wavefunctions at energies $\pm \epsilon$ are localized
(within the zeroth order RG picture for these wavefunctions). 
Thus, as far as the zeroth order RG picture of these wavefunctions 
is concerned, one is faced with the problem of characterizing such 
extremal properties of random walks, and the RG itself can be 
viewed as a powerful tool for doing this, even though many 
questions like the number density of remaining sites or 
the distribution of the effective time steps (i.e., bond lengths) 
and heights (i.e., bond log-couplings) can be answered directly 
in the random-walk language by appealing to the properties of 
a random walk near its minima (absorbing walls).  

So far, we have only reformulated our leading (zeroth) order
result for the wavefunctions at $\pm \epsilon$ in more visual 
terms.  The real usefulness of this picture appears when we go
beyond the leading order, and keep track of operators
``measuring'' the wavefunction amplitude throughout the system.
Consider going beyond the leading order for the renormalization of
the ``participation'' operators 
${\bf n}_j = |j\rangle \langle j|$.  When we eliminate 
sites $2$ and $3$, the renormalization of the corresponding 
operators {\em including the leading second-order correction to 
the zeroth order result} is given as
\begin{eqnarray*}
({\bf n}_2)^{\rm eff} &=& (t_3/t_2)^2 {\bf n}_{\tilde 4} \, , \\
({\bf n}_3)^{\rm eff} &=& (t_1/t_2)^2 {\bf n}_{\tilde 1}  \, .
\end{eqnarray*} 
It is now easy to see that if at a given stage of the RG
the site $A'$ is no longer active, it will have component
only onto the nearest active site $A$ on the same sublattice,
with the magnitude of the ``participation'' given by $e^{-2\Lambda}$,
where $\Lambda$ is the absolute ``height'' difference of the 
random walker at ``times'' corresponding to the two sites.
The connection to the statistics of the zero energy wavefunction
is now explicit.
Moreover, our picture of the wavefunctions at energy $\epsilon$ 
now follows:  the dominant feature of such wavefunctions 
$\psi_{\pm \epsilon}$ are two sites, with order one amplitudes 
of the $\psi_{\pm \epsilon}$, separated by a distance of order 
$\Gamma_\epsilon^2$; the wavefunctions will have significant weight 
only near the two ``peaks'' and, to this order, only on the sites of 
the same sublattice as the nearby peak, i.e., the wavefunction
can be thought of as being composed of two bumps centered on
the two sites.  As we move away from the two peaks by a distance
$r \alt \Gamma_\epsilon^2$, the amplitude of the wavefunction
is given by the above random-walk prescription and is typically
decreasing as $e^{-c\sqrt{r}}$; for still larger distances
the amplitudes are given by high-order perturbation theory, with
the perturbative expression for $\psi(r)$ necessarily ``stepping''
through the ``active'' segments of lengths $\sim \Gamma_\epsilon^2$
separating $r$ from the two bumps and acquiring a suppression 
of order $e^{-c\Gamma_\epsilon}$ from each such segment---thus, 
for $r \agt \Gamma_\epsilon^2$ we will 
typically have the decay $\psi(r) \sim e^{-c'r/\Gamma_\epsilon}$
(note that this gives us an interpretation, in terms of the typical
wavefunction at energy $\epsilon$, of the two length scales---one 
proportional to $\Gamma_\epsilon$ and the other to 
$\Gamma_\epsilon^2$---that diverge in the limit of low $\epsilon$, 
and are usually identified with the typical and average 
localization length respectively).

%%%%%%%%%%%%%%%%%%%%%%%%%%%%%%%%%%%%%%%%%%%%%%%%%%%%%%%%%%%%%%%
%%%%%%%%%%%%%%%   REFERENCES AND COMMENTS   %%%%%%%%%%%%%%%%%%%


\begin{references}
\bibitem{foot2dI} Recent work~Ref.~\protect{\onlinecite{RTFIM2d}} 
has shown that the quantum Ising model in two dimensions can 
also be described by a strong-disorder effective theory at 
low energies.  Low-frequency dynamical properties for
the two-dimensional case can be readily obtained from these 
results using our approach; see also Ref.~\protect{\onlinecite{ss}}
for the calculation of average dynamical susceptibilities
in the dilute quantum Ising system.
\bibitem{saddle} The fact that the volumes and moments of the
Griffiths regions of a given characteristic energy are sharply 
defined can also be made plausible by reasoning as follows: 
These regions can be thought of as being selected by a saddle-point
criterion that balances the probability of their occurrence 
against the strength of their contribution to, say, the 
susceptibility at a given low frequency.  In the limit of very low 
frequencies, the corresponding saddle point becomes very sharp, 
leading to a sharp definition of the lengths and moments of these
regions.
\bibitem{worry1} We need to be critical about evaluating 
$\sigma'(\omega)$ using the finite-size result Eq.~\ref{Lsigma}.
In the regime considered, the main contribution to the (average) 
dynamical conductivity at frequency $\omega$ of a finite segment of 
length $L$ comes from the cases when the last bond decimated has 
$\tilde J^\perp = \omega$, i.e., precisely from the cases 
when the lowest gap of the segment is $\omega$; in almost all of 
these cases the bond stretches essentially across the whole 
length $L$ and the conductivity is simply related to the 
distribution of gaps in such segments of length $L$.  Thus, 
two spins in a typical pair contributing to Eq.~(\ref{Lsigma}) 
(i.e., ``resonating'' at $\omega$) are near the two respective 
ends of such region and are likely to have strong 
$\tilde J^z > \omega$ couplings with the rest of the system, 
so the pair will not contribute to the dynamical conductivity 
at all!  Clearly, we need to consider cases when a pair with 
$\tilde J^\perp = \omega$ occurs deep inside our RS region and 
is thus well isolated from the rest of the system---in some sense, 
we need to consider distribution of the second energy gap in such 
finite segments.  One can analyze these situations along the lines 
of a similar analysis of Ref.~\onlinecite{fyoung} for 
the distribution of the first gap, with the result that 
the contributions to the dynamical conductivity from such events, 
which will give the actual $\sigma'(\omega)$, are further 
suppressed as compared to Eq.~(\ref{Lsigma}) by an additional
factor $\sim \exp(-c_2 |\ln\omega|^2/L)$; this will modify 
the exponent $\alpha$ in Eq.~(\ref{IAFsigma}) 
(which is unknown to us in any case from these arguments), 
but not the factor $|\ln\omega|^2$.  Our scaling
arguments in the main text provide a more consistent constructive 
approach for addressing these issues.
\bibitem{wrongsigma}
There is also a calculation~{\protect\cite{gogolin}}, 
referred to in literature~{\protect\cite{lifshitc}},
of the dynamical conductivity at the free-fermion point
that relies on a diagrammatic resummation technique---this 
predicts that $\sigma'(\omega) \to {\rm const}$ as $\omega \to 0$, 
which is a weaker signature of the unusual localization 
properties of this problem than predicted by us.  We have been 
unable to pinpoint the error in this diagrammatic calculation;
numerical simulations of Sec.~\ref{numeric_sigma}, however, seem 
to agree with our predictions.
\end{references}
\end{document}